\documentclass[12pt,a4paper]{amsart}
\usepackage[T1]{fontenc}
\usepackage[utf8]{inputenc}
\usepackage{lmodern,microtype}
\usepackage{amsmath,amssymb,amsthm,mathtools,mathrsfs,esint}
\usepackage[a4paper,scale=0.8]{geometry}
\usepackage{booktabs,array,enumitem,xcolor,listings,needspace,placeins}
\usepackage[hypertexnames=false,colorlinks=true,linkcolor=blue!45!black,citecolor=blue!45!black,urlcolor=blue!45!black]{hyperref}
\usepackage{bookmark,tikz,graphicx}
\usetikzlibrary{arrows.meta,calc}
\hypersetup{
 pdftitle={Anderson localisation in infinite and finite resonator crystals with long-range interactions},
 pdfauthor={Habib Ammari, Xin Fu, Wenjia Jing, Jiayu Qiu, Alexander Uhlmann},
 pdfsubject={Anderson localisation with long-range capacitance interactions and finite-crystal limits},
 pdfkeywords={Anderson localisation, capacitance operator, long-range interactions, subwavelength resonators, finite-volume estimates}}

\makeatletter
\let\ams@adminfootnotes\@adminfootnotes
\def\@adminfootnotes{\begingroup\ams@adminfootnotes\endgroup}
\let\ams@setauthors\@setauthors
\def\@setauthors{%
  \begingroup\scshape
  \let\MakeUppercase\@firstofone
  \ams@setauthors
  \endgroup}
\makeatother
\numberwithin{equation}{section}

\newtheorem{theorem}{Theorem}[section]
\newtheorem{proposition}[theorem]{Proposition}
\newtheorem{lemma}[theorem]{Lemma}
\newtheorem{corollary}[theorem]{Corollary}
\theoremstyle{definition}
\newtheorem{definition}[theorem]{Definition}

\theoremstyle{remark}
\newtheorem{remark}[theorem]{Remark}
\newcommand{\R}{\mathbb R}\newcommand{\C}{\mathbb C}
\newcommand{\Z}{\mathbb Z}
\newcommand{\EE}{\mathbb E}\newcommand{\PP}{\mathbb P}

\newcommand{\dd}{\,\mathrm d}\newcommand{\one}{\mathbf 1}
\newcommand{\Id}{\mathrm{Id}}\newcommand{\ac}{\mathrm{ac}}
\DeclareMathOperator{\supp}{supp}\DeclareMathOperator{\dist}{dist}
\DeclareMathOperator{\diag}{diag}\DeclareMathOperator{\Crit}{Crit}

\DeclareMathOperator{\Tr}{Tr}\DeclareMathOperator{\rank}{rank}
\DeclareMathOperator{\Lip}{Lip}
\setlist[enumerate]{itemsep=3pt,topsep=5pt}
\allowdisplaybreaks[2]\emergencystretch=2em
\newcommand{\Lap}{\mathsf L}
\newcommand{\ip}[2]{\langle #1,#2\rangle}

\AddToHook{cmd/section/before}{\Needspace{8\baselineskip}}
\newcommand{\CrystalFigure}{%
\begin{figure}[htbp]\centering
\begin{tikzpicture}[
 x={(0.92cm,-0.28cm)},y={(0.65cm,0.42cm)},z={(0cm,0.95cm)},
 line cap=round,line join=round,>=Stealth]
 \begin{scope}[xshift=-3.496cm]
  \draw[gray!65,dashed]
   (-.5,.5,-.5)--(-.5,.5,.5)
   (-.5,-.5,-.5)--(-.5,.5,-.5)--(.5,.5,-.5);
  \draw[gray!65]
   (-.5,-.5,-.5)--(.5,-.5,-.5)--(.5,.5,-.5)
   (-.5,-.5,-.5)--(-.5,-.5,.5)
   (.5,.5,-.5)--(.5,.5,.5);
  \shade[ball color=blue!40] (0,0,0) circle[radius=.24cm];
  \draw[blue!65!black]
   (-.5,-.5,.5)--(.5,-.5,.5)--(.5,.5,.5)--(-.5,.5,.5)--cycle;
  \draw[->] (0,0,0)--(1.1,0,0) node[right] {$x_1$};
  \draw[->] (0,0,0)--(0,1.05,0) node[right] {$x_2$};
  \draw[->] (0,0,.3)--(0,0,1.3) node[right] {$x_3$};
  \node[left] at (-.25,0,0) {$D$};
  \draw[gray!65] (.5,-.5,-.5)--(.5,-.5,.5);
 \end{scope}

 \begin{scope}[xshift=1.84cm,y={(0.45cm,0.62cm)}]
  \draw[white,line width=1.2pt,dashed] (-1,1,-1)--(0,1,-1);
  \draw[gray!40,dashed] (-1,1,-1)--(0,1,-1);
  \draw[white,line width=1.2pt,dashed] (-1,1,-1)--(-1,1,0);
  \draw[gray!40,dashed] (-1,1,-1)--(-1,1,0);
  \draw[white,line width=1.2pt,dashed] (-1,0,-1)--(-1,1,-1);
  \draw[gray!40,dashed] (-1,0,-1)--(-1,1,-1);
  \fill[white] (-1,1,-1) circle[radius=0.17cm];
  \shade[ball color=blue!15] (-1,1,-1) circle[radius=0.15cm];
  \draw[white,line width=1.2pt,dashed] (0,1,-1)--(1,1,-1);
  \draw[gray!40,dashed] (0,1,-1)--(1,1,-1);
  \draw[white,line width=1.2pt,dashed] (0,1,-1)--(0,1,0);
  \draw[gray!40,dashed] (0,1,-1)--(0,1,0);
  \draw[white,line width=1.2pt,dashed] (0,0,-1)--(0,1,-1);
  \draw[gray!40,dashed] (0,0,-1)--(0,1,-1);
  \fill[white] (0,1,-1) circle[radius=0.17cm];
  \shade[ball color=blue!15] (0,1,-1) circle[radius=0.15cm];
  \draw[white,line width=1.2pt,dashed] (-1,1,0)--(0,1,0);
  \draw[gray!40,dashed] (-1,1,0)--(0,1,0);
  \draw[white,line width=1.2pt,dashed] (-1,1,0)--(-1,1,1);
  \draw[gray!40,dashed] (-1,1,0)--(-1,1,1);
  \draw[white,line width=1.2pt,dashed] (-1,0,-1)--(0,0,-1);
  \draw[gray!40,dashed] (-1,0,-1)--(0,0,-1);
  \draw[white,line width=1.2pt,dashed] (-1,0,0)--(-1,1,0);
  \draw[gray!40,dashed] (-1,0,0)--(-1,1,0);
  \fill[white] (-1,1,0) circle[radius=0.17cm];
  \shade[ball color=blue!15] (-1,1,0) circle[radius=0.15cm];
  \draw[white,line width=1.2pt] (1,1,-1)--(1,1,0);
  \draw[gray!25] (1,1,-1)--(1,1,0);
  \draw[white,line width=1.6pt,line cap=butt] (0.10709,0.5,-0.5)--(0.440488,0.5,-0.5);
  \draw[orange!80!black,thick] (-0.5,0.5,-0.5)--(0.5,0.5,-0.5);
  \draw[white,line width=1.2pt,dashed] (-1,0,-1)--(-1,0,0);
  \draw[gray!40,dashed] (-1,0,-1)--(-1,0,0);
  \draw[white,line width=1.2pt] (1,0,-1)--(1,1,-1);
  \draw[gray!25] (1,0,-1)--(1,1,-1);
  \fill[white] (1,1,-1) circle[radius=0.17cm];
  \shade[ball color=blue!38] (1,1,-1) circle[radius=0.15cm];
  \draw[white,line width=1.2pt,dashed] (-1,-1,-1)--(-1,0,-1);
  \draw[gray!40,dashed] (-1,-1,-1)--(-1,0,-1);
  \fill[white] (-1,0,-1) circle[radius=0.17cm];
  \shade[ball color=blue!15] (-1,0,-1) circle[radius=0.15cm];
  \draw[white,line width=1.2pt,dashed] (0,1,0)--(1,1,0);
  \draw[gray!40,dashed] (0,1,0)--(1,1,0);
  \draw[white,line width=1.2pt,dashed] (0,0.5,0)--(0,1,0);
  \draw[gray!40,dashed] (0,0.5,0)--(0,1,0);
  \draw[white,line width=1.6pt,line cap=butt] (-0.5,0.5,0.324407)--(-0.5,0.5,0.47239);
  \draw[orange!80!black,thick] (-0.5,0.5,-0.5)--(-0.5,0.5,0.5);
  \draw[white,line width=1.6pt,line cap=butt] (-0.5,0.280733,-0.5)--(-0.5,0.464241,-0.5);
  \draw[orange!80!black,thick] (-0.5,-0.5,-0.5)--(-0.5,0.5,-0.5);
  \draw[white,line width=1.2pt,dashed] (0,1,0)--(0,1,1);
  \draw[gray!40,dashed] (0,1,0)--(0,1,1);
  \fill[white] (0,1,0) circle[radius=0.17cm];
  \shade[ball color=blue!15] (0,1,0) circle[radius=0.15cm];
  \draw[white,line width=1.2pt,dashed] (0,0,-1)--(0,0,-0.5);
  \draw[gray!40,dashed] (0,0,-1)--(0,0,-0.5);
  \draw[white,line width=1.2pt,dashed] (0,0,-1)--(1,0,-1);
  \draw[gray!40,dashed] (0,0,-1)--(1,0,-1);
  \draw[white,line width=1.2pt] (-1,1,1)--(0,1,1);
  \draw[gray!25] (-1,1,1)--(0,1,1);
  \draw[white,line width=1.2pt,dashed] (-1,0,0)--(-0.5,0,0);
  \draw[gray!40,dashed] (-1,0,0)--(-0.5,0,0);
  \draw[white,line width=1.2pt,dashed] (0,-1,-1)--(0,0,-1);
  \draw[gray!40,dashed] (0,-1,-1)--(0,0,-1);
  \fill[white] (0,0,-1) circle[radius=0.17cm];
  \shade[ball color=blue!15] (0,0,-1) circle[radius=0.15cm];
  \draw[white,line width=1.6pt,line cap=butt] (0.5,0.5,0.324407)--(0.5,0.5,0.47239);
  \draw[orange!80!black,thick] (0.5,0.5,-0.5)--(0.5,0.5,0.5);
  \draw[white,line width=1.2pt,dashed] (0,0,0)--(0,0.5,0);
  \draw[gray!40,dashed] (0,0,0)--(0,0.5,0);
  \draw[white,line width=1.2pt,dashed] (0,0,-0.5)--(0,0,0);
  \draw[gray!40,dashed] (0,0,-0.5)--(0,0,0);
  \draw[white,line width=1.2pt] (-1,0,1)--(-1,1,1);
  \draw[gray!25] (-1,0,1)--(-1,1,1);
  \fill[white] (-1,1,1) circle[radius=0.17cm];
  \shade[ball color=blue!38] (-1,1,1) circle[radius=0.15cm];
  \draw[white,line width=1.6pt,line cap=butt] (0.5,0.280733,-0.5)--(0.5,0.464241,-0.5);
  \draw[orange!80!black,thick] (0.5,-0.5,-0.5)--(0.5,0.5,-0.5);
  \draw[white,line width=1.2pt,dashed] (-0.5,0,0)--(0,0,0);
  \draw[gray!40,dashed] (-0.5,0,0)--(0,0,0);
  \draw[white,line width=1.2pt] (1,1,0)--(1,1,1);
  \draw[gray!25] (1,1,0)--(1,1,1);
  \draw[white,line width=1.6pt,line cap=butt] (0.10709,0.5,0.5)--(0.440488,0.5,0.5);
  \draw[orange!80!black,thick] (-0.5,0.5,0.5)--(0.5,0.5,0.5);
  \draw[white,line width=1.2pt,dashed] (-1,0,0)--(-1,0,1);
  \draw[gray!40,dashed] (-1,0,0)--(-1,0,1);
  \draw[white,line width=1.2pt] (-1,-1,-1)--(0,-1,-1);
  \draw[gray!25] (-1,-1,-1)--(0,-1,-1);
  \draw[white,line width=1.2pt] (1,0,0)--(1,1,0);
  \draw[gray!25] (1,0,0)--(1,1,0);
  \fill[white] (1,1,0) circle[radius=0.17cm];
  \shade[ball color=blue!38] (1,1,0) circle[radius=0.15cm];
  \draw[white,line width=1.2pt,dashed] (-1,-1,0)--(-1,0,0);
  \draw[gray!40,dashed] (-1,-1,0)--(-1,0,0);
  \fill[white] (-1,0,0) circle[radius=0.17cm];
  \shade[ball color=blue!15] (-1,0,0) circle[radius=0.15cm];
  \draw[white,line width=1.2pt] (0,1,1)--(1,1,1);
  \draw[gray!25] (0,1,1)--(1,1,1);
  \draw[white,line width=1.2pt] (1,0,-1)--(1,0,0);
  \draw[gray!25] (1,0,-1)--(1,0,0);
  \draw[white,line width=1.6pt,line cap=butt] (0.10709,-0.5,-0.5)--(0.440488,-0.5,-0.5);
  \draw[orange!80!black,thick] (-0.5,-0.5,-0.5)--(0.5,-0.5,-0.5);
  \draw[white,line width=1.2pt] (-1,-1,-1)--(-1,-1,0);
  \draw[gray!25] (-1,-1,-1)--(-1,-1,0);
  \fill[white] (-1,-1,-1) circle[radius=0.17cm];
  \shade[ball color=blue!38] (-1,-1,-1) circle[radius=0.15cm];
  \draw[white,line width=1.2pt,dashed] (0,0,0)--(0.5,0,0);
  \draw[gray!40,dashed] (0,0,0)--(0.5,0,0);
  \draw[white,line width=1.6pt,line cap=butt] (-0.5,-0.113976,0.5)--(-0.5,0.0695315,0.5);
  \draw[white,line width=1.6pt,line cap=butt] (-0.5,0.280733,0.5)--(-0.5,0.464241,0.5);
  \draw[orange!80!black,thick] (-0.5,-0.5,0.5)--(-0.5,0.5,0.5);
  \draw[white,line width=1.2pt] (1,-1,-1)--(1,0,-1);
  \draw[gray!25] (1,-1,-1)--(1,0,-1);
  \fill[white] (1,0,-1) circle[radius=0.17cm];
  \shade[ball color=blue!38] (1,0,-1) circle[radius=0.15cm];
  \draw[white,line width=1.2pt,dashed] (0,0,0)--(0,0,0.5);
  \draw[gray!40,dashed] (0,0,0)--(0,0,0.5);
  \draw[white,line width=1.2pt,dashed] (0,-0.5,0)--(0,0,0);
  \draw[gray!40,dashed] (0,-0.5,0)--(0,0,0);
  \fill[white] (0,0,0) circle[radius=0.17cm];
  \shade[ball color=orange!75] (0,0,0) circle[radius=0.15cm];
  \draw[white,line width=1.6pt,line cap=butt] (-0.5,-0.5,-0.259489)--(-0.5,-0.5,-0.111505);
  \draw[white,line width=1.6pt,line cap=butt] (-0.5,-0.5,0.324407)--(-0.5,-0.5,0.47239);
  \draw[orange!80!black,thick] (-0.5,-0.5,-0.5)--(-0.5,-0.5,0.5);
  \draw[white,line width=1.2pt] (0,0,1)--(0,1,1);
  \draw[gray!25] (0,0,1)--(0,1,1);
  \fill[white] (0,1,1) circle[radius=0.17cm];
  \shade[ball color=blue!38] (0,1,1) circle[radius=0.15cm];
  \draw[white,line width=1.2pt,dashed] (0.5,0,0)--(1,0,0);
  \draw[gray!40,dashed] (0.5,0,0)--(1,0,0);
  \draw[white,line width=1.2pt] (0,-1,-1)--(1,-1,-1);
  \draw[gray!25] (0,-1,-1)--(1,-1,-1);
  \draw[white,line width=1.2pt] (-1,0,1)--(0,0,1);
  \draw[gray!25] (-1,0,1)--(0,0,1);
  \draw[white,line width=1.2pt,dashed] (0,0,0.5)--(0,0,1);
  \draw[gray!40,dashed] (0,0,0.5)--(0,0,1);
  \draw[white,line width=1.2pt] (0,-1,-1)--(0,-1,0);
  \draw[gray!25] (0,-1,-1)--(0,-1,0);
  \fill[white] (0,-1,-1) circle[radius=0.17cm];
  \shade[ball color=blue!38] (0,-1,-1) circle[radius=0.15cm];
  \draw[white,line width=1.6pt,line cap=butt] (0.5,-0.113976,0.5)--(0.5,0.0695315,0.5);
  \draw[white,line width=1.6pt,line cap=butt] (0.5,0.280733,0.5)--(0.5,0.464241,0.5);
  \draw[orange!80!black,thick] (0.5,-0.5,0.5)--(0.5,0.5,0.5);
  \draw[white,line width=1.6pt,line cap=butt] (0.5,-0.5,-0.259489)--(0.5,-0.5,-0.111505);
  \draw[white,line width=1.6pt,line cap=butt] (0.5,-0.5,0.121203)--(0.5,-0.5,0.47239);
  \draw[orange!80!black,thick] (0.5,-0.5,-0.5)--(0.5,-0.5,0.5);
  \draw[white,line width=1.2pt,dashed] (0,-1,0)--(0,-0.5,0);
  \draw[gray!40,dashed] (0,-1,0)--(0,-0.5,0);
  \draw[white,line width=1.2pt] (-1,-1,0)--(0,-1,0);
  \draw[gray!25] (-1,-1,0)--(0,-1,0);
  \draw[white,line width=1.2pt] (1,0,1)--(1,1,1);
  \draw[gray!25] (1,0,1)--(1,1,1);
  \fill[white] (1,1,1) circle[radius=0.17cm];
  \shade[ball color=blue!38] (1,1,1) circle[radius=0.15cm];
  \draw[white,line width=1.2pt] (-1,-1,1)--(-1,0,1);
  \draw[gray!25] (-1,-1,1)--(-1,0,1);
  \fill[white] (-1,0,1) circle[radius=0.17cm];
  \shade[ball color=blue!38] (-1,0,1) circle[radius=0.15cm];
  \draw[white,line width=1.2pt] (1,0,0)--(1,0,1);
  \draw[gray!25] (1,0,0)--(1,0,1);
  \draw[white,line width=1.6pt,line cap=butt] (-0.339398,-0.5,0.5)--(-0.19321,-0.5,0.5);
  \draw[white,line width=1.6pt,line cap=butt] (-0.0839636,-0.5,0.5)--(0.0622245,-0.5,0.5);
  \draw[white,line width=1.6pt,line cap=butt] (0.10709,-0.5,0.5)--(0.440488,-0.5,0.5);
  \draw[orange!80!black,thick] (-0.5,-0.5,0.5)--(0.5,-0.5,0.5);
  \draw[white,line width=1.2pt] (-1,-1,0)--(-1,-1,1);
  \draw[gray!25] (-1,-1,0)--(-1,-1,1);
  \fill[white] (-1,-1,0) circle[radius=0.17cm];
  \shade[ball color=blue!38] (-1,-1,0) circle[radius=0.15cm];
  \draw[white,line width=1.2pt] (1,-1,0)--(1,0,0);
  \draw[gray!25] (1,-1,0)--(1,0,0);
  \fill[white] (1,0,0) circle[radius=0.17cm];
  \shade[ball color=blue!38] (1,0,0) circle[radius=0.15cm];
  \draw[white,line width=1.2pt] (0,0,1)--(1,0,1);
  \draw[gray!25] (0,0,1)--(1,0,1);
  \draw[white,line width=1.2pt] (1,-1,-1)--(1,-1,0);
  \draw[gray!25] (1,-1,-1)--(1,-1,0);
  \fill[white] (1,-1,-1) circle[radius=0.17cm];
  \shade[ball color=blue!38] (1,-1,-1) circle[radius=0.15cm];
  \draw[white,line width=1.2pt] (0,-1,0)--(1,-1,0);
  \draw[gray!25] (0,-1,0)--(1,-1,0);
  \draw[white,line width=1.2pt] (0,-1,1)--(0,0,1);
  \draw[gray!25] (0,-1,1)--(0,0,1);
  \fill[white] (0,0,1) circle[radius=0.17cm];
  \shade[ball color=blue!38] (0,0,1) circle[radius=0.15cm];
  \draw[white,line width=1.2pt] (0,-1,0)--(0,-1,1);
  \draw[gray!25] (0,-1,0)--(0,-1,1);
  \fill[white] (0,-1,0) circle[radius=0.17cm];
  \shade[ball color=blue!38] (0,-1,0) circle[radius=0.15cm];
  \draw[white,line width=1.2pt] (-1,-1,1)--(0,-1,1);
  \draw[gray!25] (-1,-1,1)--(0,-1,1);
  \fill[white] (-1,-1,1) circle[radius=0.17cm];
  \shade[ball color=blue!38] (-1,-1,1) circle[radius=0.15cm];
  \draw[white,line width=1.2pt] (1,-1,1)--(1,0,1);
  \draw[gray!25] (1,-1,1)--(1,0,1);
  \fill[white] (1,0,1) circle[radius=0.17cm];
  \shade[ball color=blue!38] (1,0,1) circle[radius=0.15cm];
  \draw[white,line width=1.2pt] (1,-1,0)--(1,-1,1);
  \draw[gray!25] (1,-1,0)--(1,-1,1);
  \fill[white] (1,-1,0) circle[radius=0.17cm];
  \shade[ball color=blue!38] (1,-1,0) circle[radius=0.15cm];
  \draw[white,line width=1.2pt] (0,-1,1)--(1,-1,1);
  \draw[gray!25] (0,-1,1)--(1,-1,1);
  \fill[white] (0,-1,1) circle[radius=0.17cm];
  \shade[ball color=blue!38] (0,-1,1) circle[radius=0.15cm];
  \fill[white] (1,-1,1) circle[radius=0.17cm];
  \shade[ball color=blue!38] (1,-1,1) circle[radius=0.15cm];
  \draw[->,blue!70!black]
   (1,1,1)--(1.8,1.5,1.25) node[right] {$D+n$};
 \end{scope}

 \node[anchor=base] at (-3.496cm,-2.45cm)
  {$Y=(-\tfrac12,\tfrac12)^3$};
 \node[anchor=base] at (1.84cm,-2.45cm)
  {$\mathcal D=\displaystyle\bigcup_{n\in\Z^3}(D+n)$};
\end{tikzpicture}
\caption{Left: The reference cell. Right: A simple cubic sphere crystal, shown schematically. The orange
box marks the reference cell in the repeated structure. The model is periodic in all three directions.}
\label{fig:crystal}\end{figure}}

\newcommand{\CellFigure}{%
\begin{figure}[htbp]\centering
\begin{tikzpicture}[>=Stealth,font=\small]
 \begin{scope}[scale=3.4]
  \draw[thick] (-.5,-.5) rectangle (.5,.5);
  \filldraw[fill=blue!12,draw=blue!60!black] (0,0) ellipse[x radius=.29,y radius=.2];
  \draw[dashed,orange!80!black] (0,-.62)--(0,.64);
  \draw[dashed,orange!80!black] (-.64,0)--(.64,0);
  \fill (0,0) circle[radius=.008] node[below right] {$0$};
  \node at (0,.11) {$D$};
  \draw[<->,thick,blue!70!black] (-.26,.32)--(.26,.32) node[midway,above] {$R_1$};
  \draw[<->,thick,blue!70!black] (.36,-.2)--(.36,.2) node[midway,right] {$R_2$};
  \node[left] at (-.5,-.34) {$F_1^-$};
  \node[right] at (.5,-.34) {$F_1^+$};
  \node[above] at (-.24,.5) {$F_2^+$};
  \node[below] at (-.24,-.5) {$F_2^-$};
  \node at (0,-.77) {$\gamma_j^+u=e^{i\alpha_j}\gamma_j^-u$};
 \end{scope}
 \begin{scope}[shift={(6.2,-1.7)},scale=1.1]
  \draw[->] (-.25,0)--(3.5,0) node[right] {$\alpha_1$};
  \draw[->] (0,-.25)--(0,3.5) node[above] {$\alpha_2$};
  \draw[gray!60] (0,0) rectangle (3,3);
  \foreach \x/\y/\lab/\pos in {0/0/\Gamma/below left,3/0/X/below,0/3/Y/left,3/3/M/above right}
    \fill[blue!70!black] (\x,\y) circle[radius=.055] node[\pos] {$\lab$};
  \foreach \y in {.8,1.5,2.2}\draw[->,orange!90!black] (.6,\y)--(2.35,\y);
  \foreach \x in {.8,1.5,2.2}\draw[->,orange!90!black] (\x,.5)--(\x,2.55);
  \node[align=center] at (1.5,-.85) {$\partial_{\alpha_j}c>0$ for $0<\alpha_j<\pi$};
 \end{scope}
\end{tikzpicture}
\caption{Left: a two-dimensional section illustrating the coordinate
reflections and the opposite-face trace convention, with
$F_j^\pm=\{x\in\overline Y:x_j=\pm1/2\}$. In three dimensions
all three coordinate reflections are required. Right: the positive quadrant
of the two-dimensional Brillouin torus, with $\Gamma=(0,0)$,
$X=(\pi,0)$, $Y=(0,\pi)$, and $M=(\pi,\pi)$. All four vertices are
critical; only $M$ is the maximum. The arrows indicate coordinate
monotonicity.}
\label{fig:cell}\end{figure}}

\newcommand{\FiniteFigure}{%
\begin{figure}[htbp]\centering
\begin{tikzpicture}[scale=.7,font=\small]
 \foreach \panel in {0,1}{
  \begin{scope}[shift={(8*\panel,0)}]
   \foreach \i in {-2,-1,0,1,2}\foreach \j in {-2,-1,0,1,2}{
    \ifnum\panel=0
      \filldraw[fill=gray!12,draw=gray!45] (\i,\j) circle[radius=.23];
    \fi
   }
   \foreach \i in {-1,0,1}\foreach \j in {-1,0,1}
     \filldraw[fill=blue!22,draw=blue!70!black] (\i,\j) circle[radius=.23];
   \draw[dashed,thick,orange!80!black] (-1.5,-1.5) rectangle (1.5,1.5);
   \node[above] at (0,2.65) {\ifnum\panel=0 $C_{\Lambda_L}^t$\else $C_{\Lambda_L}^f$\fi};
   \node[align=center] at (0,-3.1) {\ifnum\panel=0 Outside resonators retained\\with zero prescribed potential\else Outside resonators removed\\unconstrained exterior\fi};
  \end{scope}
 }
\end{tikzpicture}
\caption{Truncated and finite capacitance matrices. The dashed
line encloses the retained indices, represented here by a square sample.
The variational constraints give $0\le C_{\Lambda_L}^f\le C_{\Lambda_L}^t$.
Both sequences recover the infinite operator as $L\to\infty$.}
\label{fig:finite}\end{figure}}

\newcommand{\LocalisationFigure}{%
\begin{figure}[htbp]\centering
\begin{tikzpicture}[font=\small,>=Stealth]
 \draw[->,thick] (.3,5.7)--(12.4,5.7) node[right] {$E$};
 \draw (.65,5.61)--(.65,5.79) node[above=3pt] {$0$};
 \draw[blue!65!black,line width=4pt] (8.65,5.7)--(11.55,5.7);
 \draw (8.65,5.6)--(8.65,5.8) node[above=3pt] {$c_M-\Delta_W$};
 \draw (11.55,5.6)--(11.55,5.8) node[above=3pt] {$c_M$};
 \node[blue!65!black] at (10.1,5.26) {$I_W$};
 \node[anchor=west] at (.65,5.12) {Fixed $W>0$ and fixed edge interval};
 \begin{scope}[shift={(.8,.4)}]
  \fill[blue!2] (0,0) rectangle (4.2,4.2);
  \draw[step=.42,gray!30,thin] (0,0) grid (4.2,4.2);
  \foreach \i in {0,...,9}\foreach \j in {0,...,9}{
   \pgfmathsetmacro{\tone}{8+80*exp(-((\i-4)^2+(\j-5)^2)/2.4)}
   \fill[blue!\tone!white] ({.21+.42*\i},{.21+.42*\j}) circle[radius=.045];
  }
  \draw[thick] (0,0) rectangle (4.2,4.2);
  \draw[dashed,thick,orange!85!black] (1.89,2.31) circle[radius=1.13];
  \fill[blue!65!black] (1.89,2.31) circle[radius=.045];
  \draw[->,orange!85!black] (1.89,2.31)--(2.7,3.1);
  \node[orange!85!black,fill=white,inner sep=1pt] at (3.15,3.52) {$r_\varepsilon$};
  \node[fill=white,inner sep=2pt] at (3.8,.37) {$\Omega$};
  \draw (.84,-.14)--(.84,-.26) (.84,-.2)--(1.26,-.2)
        (1.26,-.14)--(1.26,-.26);
  \node[below] at (1.05,-.25) {$\varepsilon$};
  \node at (2.1,-1.1) {Cell size $\varepsilon$};
 \end{scope}
 \draw[->,thick,gray!75] (5.45,2.5)--(6.65,2.5);
 \node[align=center] at (6.05,3.1) {smaller\\cells};
 \begin{scope}[shift={(7.2,.4)}]
  \fill[blue!2] (0,0) rectangle (4.2,4.2);
  \draw[step=.21,gray!30,thin] (0,0) grid (4.2,4.2);
  \foreach \i in {0,...,19}\foreach \j in {0,...,19}{
   \pgfmathsetmacro{\tone}{8+80*exp(-((\i-12)^2+(\j-8)^2)/3.2)}
   \fill[blue!\tone!white] ({.105+.21*\i},{.105+.21*\j}) circle[radius=.028];
  }
  \draw[thick] (0,0) rectangle (4.2,4.2);
  \draw[dashed,thick,orange!85!black] (2.625,1.785) circle[radius=.72];
  \fill[blue!65!black] (2.625,1.785) circle[radius=.035];
  \draw[->,orange!85!black] (2.625,1.785)--(3.13,2.29);
  \node[orange!85!black,fill=white,inner sep=1pt] at (3.12,2.76) {$r_{\varepsilon/2}$};
  \node[fill=white,inner sep=2pt] at (3.8,.37) {$\Omega$};
  \draw (.84,-.14)--(.84,-.26) (.84,-.2)--(1.05,-.2)
        (1.05,-.14)--(1.05,-.26);
  \node[below] at (.945,-.25) {$\varepsilon/2$};
  \node at (2.1,-1.1) {Cell size $\varepsilon/2$};
 \end{scope}
\end{tikzpicture}
\caption{Localisation in a fixed spectral window and a bounded physical
sample. The dots schematically indicate the mass of one mode. For finite capacitance matrices, with probability tending to one,
every mode in $I_W$ outside the excluded family of boundary-order size
has negligible mass outside a ball of radius
$r_\varepsilon=O_{W,p}(\varepsilon\log(1/\varepsilon))$.
For truncated matrices the conclusion holds for all modes in the window.
The dashed circles illustrate this shrinking bound. The two centres are
different because the theorem does not identify modes or centres across
sample sizes. The spectral window stays fixed as the cells shrink.}
\label{fig:localisation}\end{figure}}

\title[Anderson localisation with long-range interactions]
{Anderson localisation in infinite and finite resonator crystals with long-range interactions}

\author[H. Ammari, X. Fu, W. Jing, J. Qiu, and A. Uhlmann]
{Habib Ammari\footnote{\raggedright ETH Z\"urich, Department of Mathematics, Rämistrasse 101, 8092 Z\"urich, Switzerland and 
  Hong Kong Institute for Advanced Study, City University of Hong Kong, Kowloon Tong, Hong Kong Special Administrative Region. Email: habib.ammari@math.ethz.ch.} \and 
Xin Fu\footnote{\raggedright School of Science, Institute for Theoretical Sciences, Westlake University, Hangzhou,
Zhejiang Province, 310030, P.R. China. Email: 1550862645cf@gmail.com.}
\and Wenjia Jing\footnote{\raggedright Yau Mathematical Sciences Center, Tsinghua University, Beijing 100084 and Beijing Institute of Mathematical Sciences and Applications, Beijing 101408, P.R. China. Email: wjjing@tsinghua.edu.cn.}
\and Jiayu Qiu\footnote{\raggedright ETH Z\"urich, Department of Mathematics, Rämistrasse 101, 8092 Z\"urich, Switzerland. Email: jiayu.qiu@math.ethz.ch.}
\and Alexander Uhlmann\footnote{\raggedright Department of Applied Mathematics and Theoretical Physics,
University of Cambridge. Email: aju27@cam.ac.uk.}}
\date{}

\begin{document}
\begin{abstract}
We establish Anderson localisation for random capacitance operators and quantify concentration of individual modes of finite capacitance matrices for clusters of subwavelength high-contrast resonators. The underlying geometry is fully periodic in dimensions \(d=2,3\), with one resonator per cell, and the inverse material factors are independent and uniformly distributed. We retain the full capacitance interaction, whose
coefficients decay exponentially with lattice separation. Under
coordinate reflection symmetry, spectral and dynamical localisation
hold near the upper spectral edge for every fixed positive disorder
strength. At weak disorder $W$, we obtain an explicit localisation
window of energy width proportional to $W$; prescribed positive
intervals are localised at sufficiently large disorder. A variational
boundary comparison and a material-variable eigenvalue-spacing
estimate transfer localisation from truncated matrices to individual
modes of the finite capacitance matrix. In a bounded sample of cell
size $\varepsilon$, with probability tending to one, all except for $O(\varepsilon^{1-d}\log(1/\varepsilon))$ modes in the localisation
interval concentrate within radii $O(\varepsilon\log(1/\varepsilon))$.
The estimates are uniform for the contrast-dependent capacitance family
at small contrast. We also prove a common limiting density of states
for the truncated and finite matrices and illustrate concentration
numerically. 
\end{abstract}
\maketitle
\setcounter{footnote}{5}
{\small\tableofcontents}
\medskip

\section{Introduction}\label{sec:introduction}
High-contrast resonators exhibit resonances at wavelengths much larger
than their characteristic size. Their collective response is described,
to leading order in the subwavelength regime, by a capacitance matrix or
operator; see \cite{ADH}. This operator describes long-range coupling
between the resonators and retains their geometry. We study Anderson
localisation caused by independent random perturbations of the material
parameters, keeping the full capacitance interaction both in the infinite
crystal and in finite clusters containing an increasing number of
resonators.

For Anderson models with short-range hopping, the mathematical theory
of localisation is based on multiscale analysis \cite{FS,DK} and
fractional-moment estimates \cite{AM,ASFH}; see also \cite{Stolz}.
Band-edge localisation and associated Wegner and density of states
estimates for classes of random acoustic and electromagnetic
operators were established in \cite{FK1994,CHT1999}.
Recent OpenAI preprints \cite{OAIAC,OAI2D} report pure-point spectrum
at every positive disorder in two dimensions and an absolutely
continuous spectral interval at weak disorder in dimensions at least
three, for nearest-neighbour models with independent uniform site
potentials. 

Several recent works on resonator crystals provide the starting point of the present analysis that gives probabilistic localisation estimates for the infinite random operator and simultaneous concentration estimates for individual modes of the finite matrix corresponding to a finite cluster of resonators. Ammari, Davies, and Hiltunen \cite{ADH2024Anderson} develop a generalised capacitance formulation for resonator arrays with random material parameters. Their capacitance formulation describes the coupling of localised modes and relates mode hybridisation to level repulsion and changes in localisation. The material perturbations enter through a diagonal multiplication operator, while the capacitance operator describes the geometric interactions. Ammari, Du, Fu, Jing, and
Melcher \cite{EM} establish a variational capacitance framework,
band-edge homogenisation, and the nonresonant sound-soft limit for
finite crystals of subwavelength high-contrast resonators. Their global
analyticity and nondegenerate upper band edge provide the deterministic inputs for the infinite-volume
localisation argument. Their positive boundary comparison between
truncated and finite capacitance matrices is essential for our
finite-cluster result. For chains embedded in three dimensions,
Ammari, Barandun, Qiu, and Uhlmann \cite{ABQU} establish critical
long-range decay of the capacitance coefficients, prove convergence
of finite operators and their spectra, and give numerical evidence
of localisation at arbitrary disorder strength. Ammari and Qiu
\cite{AQchain} prove absence of
absolutely continuous spectrum for one-dimensional capacitance operators
with multiplicative disorder and critical long-range interactions.
That result extends the one-dimensional long-range approach of
\cite{JM} and motivates the present spectral question. Full periodicity
gives exponential decay of the capacitance coefficients. This decay
allows fractional-moment methods \cite{AM,ASFH} to establish spectral
and dynamical localisation while retaining the long-range interaction.

Let $d\in\{2,3\}$, $Y=(-1/2,1/2)^d$, and let $D\Subset Y$ be
simply-connected with $C^2$ boundary. We also assume that \(Y\setminus\overline D\) is connected. The unperturbed crystal consists of $D_n:=D+n$, $n\in\Z^d$, all with the same wave speed. We denote by $\alpha\in Y^*=\R^d/(2\pi\Z)^d$ the quasimomentum associated with the crystal when performing the Floquet transform. Let $C$ be its 
static capacitance operator, introduced in Section \ref{sec:model}, and $c(\alpha)$ its scalar Bloch symbol.
Thus $c(\alpha)=\operatorname{Cap}_{\alpha,D}/|D|$ and
$C_{nm}=c_{n-m}$, where $\operatorname{Cap}_{\alpha,D}$ is the quasiperiodic capacity of $D$ (see \cite{ADH}) and $c_r$ are the Fourier coefficients of $c$. See Section~\ref{sec:model} for the precise variational definitions.

To fix the physical normalisation, let $\rho$ and $\kappa_{\rm out}$
be the background density and bulk modulus, and set
$v^2=\kappa_{\rm out}/\rho$. Inside the resonators the density is
$\rho_r=\delta\rho$, $0<\delta\ll1$, and the squared wave speeds
are $v_r^2b_n$, with $v_r>0$ fixed and $b_n>0$. We denote the
material configuration by $\varpi$ and reserve $\omega\ge0$ for physical
angular frequency. At unit period the transmission
problem is
\begin{equation}
 \begin{cases}
 (\Delta+\omega^2/v^2)u=0&
       \text{in }\R^d\setminus\overline{\bigcup_nD_n},\\
 (\Delta+\omega^2/(v_r^2b_n))u=0&\text{in }D_n,\\
 u|_+=u|_-,\quad\delta\partial_\nu u|_+=\partial_\nu u|_-
       &\text{on }\partial D_n .
 \end{cases}\label{eq:physical}
\end{equation}
The normal $\nu$ points out of the resonator; the signs $+$ and $-$
denote exterior and interior traces, respectively. With $q_n$ denoting
the leading resonator amplitude, integration of the interior equation
and matching of the fluxes give
\begin{equation}
 Cq=EM_\varpi q,\qquad B_\varpi Cq=Eq,\qquad
 E=\frac{\omega^2}{\delta v_r^2},\quad
 M_\varpi=\diag(b_n^{-1}),\quad B_\varpi=M_\varpi^{-1}.
 \label{eq:physical-reduction}
\end{equation}
For the periodic material configuration, the first Bloch frequency has
the asymptotic expansion
\begin{equation}
 \omega_1^\alpha(\delta)
 =v_r\sqrt{\delta\,c(\alpha)}+O(\delta^{3/2})
 \label{eq:Bloch-asymptotic}
\end{equation}
for fixed nonzero $\alpha$; see \cite[Theorem~4.4]{ADH}.
At physical period $0<\varepsilon \ll 1$, the spectral conversion is instead
$E=\varepsilon^2\omega^2/(\delta v_r^2)$. In particular,
$\delta=\varepsilon^2$ gives $E=\omega^2/v_r^2$.
Throughout, energy means this reduced spectral parameter, proportional
to squared frequency, rather than the physical energy of a wave.
Unless explicitly stated otherwise, spectral intervals and their
widths are expressed in $E$.

We prescribe the inverse material factors by
\[
 m_n=b_n^{-1}=1+W\xi_n,\qquad
 \xi_n\stackrel{\rm iid}{\sim}\operatorname{Unif}[0,1],\qquad W>0,
\]
where ``iid'' means independent and identically distributed. The reduced random operator \(B_\varpi C\) has the self-adjoint representation
on $\ell^2(\Z^d)$ given by
\[
 H_\varpi=M_\varpi^{-1/2}CM_\varpi^{-1/2},\qquad
 \psi=M_\varpi^{1/2}q.
\]
Our spectral results concern this reduced problem and its finite-cluster
analogue. Transfer to the full transmission field requires an
approximation uniform in the number of resonators.

For the infinite crystal, the variational cell problem gives an
analytic symbol and exponentially decaying coefficients \cite{EM,AQ}.
Under coordinate reflection symmetry, the symbol $c$ is strictly increasing in
each coordinate on $(0,\pi)$ for arbitrary transverse phases. Its
critical points are the vertices $\{0,\pi\}^d$, and its unique maximum
$c_M=c(\alpha_M)$, with $\alpha_M=(\pi,\ldots,\pi)$, is nondegenerate.
For every fixed $W>0$, Theorem~\ref{thm:localisation} proves spectral
and dynamical localisation on
$I_W=(c_M-\Delta_W,c_M)$ for some $\Delta_W\in(0,c_M)$.
Theorem~\ref{thm:strong-main} and Corollary~\ref{cor:strong-localisation}
give the same conclusion on prescribed positive intervals under an
explicit large-disorder condition. For each fixed $0<a<1/2$ and
sufficiently small $W$, Theorem~\ref{im:thm:app-linear-window} gives
the explicit interval
\[
 I_{W,a}=\left(\frac{c_M}{1+aW},c_M\right).
\]
This bounds the upper endpoint of any absolutely continuous spectral
component, without establishing an extended interior regime.

For a finite set of retained indices $\Lambda$, we distinguish the
truncated capacitance matrix $C_\Lambda^t=(C_{nm})_{n,m\in\Lambda}$
from the finite capacitance matrix $C_\Lambda^f$, defined for the
retained resonators after all others are removed.
With $M_{\varpi,\Lambda}=\diag(m_n:n\in\Lambda)$, their self-adjoint
representations are
$H_{\varpi,\Lambda}^{a}=M_{\varpi,\Lambda}^{-1/2}
C_\Lambda^{a}M_{\varpi,\Lambda}^{-1/2}$, $a\in \{t,f\}$.
The asymptotic-equivalence method of \cite{ADHconv} extends to this
diagonal weighting. The two matrix sequences have the same limiting
density of states; we also obtain a surface-order trace comparison
and convergence of their spectra.

To control eigenvectors, we combine the positive boundary estimate of
\cite[Lemma~5.1]{EM} with uniform localisation for truncated matrices
and an eigenvalue-spacing estimate for material disorder.
The boundary correction is exponentially small on vectors supported
far from the boundary, while the spacing estimate separates the
nearby finite eigenvalues. These bounds match modes of the truncated matrix centred in the
interior to individual eigenvectors of the finite matrix.

More precisely, let $\Omega\subset\R^d$ be a bounded Lipschitz sample
containing $N_\varepsilon\asymp\varepsilon^{-d}$ cells. For every
fixed $p>0$, with probability at least $1-C_{W,p}N_\varepsilon^{-p}$,
all but $O_{W,p}(\varepsilon^{1-d}\log(1/\varepsilon))$ modes in
the localised interval concentrate in balls of radius
$O_{W,p}(\varepsilon\log(1/\varepsilon))$, with squared tail at
most $N_\varepsilon^{-p}$; see Theorem~\ref{thm:cluster-main}.
The retained modes are close to exponentially localised modes of
the truncated matrix. The unmatched family is controlled in cardinality, but its
elements are not necessarily boundary-supported. The estimates are
uniform for the contrast-dependent family $C_\delta$ defined in
Section~\ref{sec:uniform-family}. They apply at $\delta=\varepsilon^2$
on an energy interval of fixed disorder-dependent width below its upper
edge $c_{M,\delta}$. Under $\delta=\varepsilon^2$, its frequency width
also has a positive limit. This differs from the $O(\varepsilon^2)$
frequency detuning, and hence $O(\varepsilon^2)$ reduced-energy detuning,
used in deterministic band-edge homogenisation \cite{EM}.

At regular interior energies, deterministic estimates give lower
bounds of order $W^{-1}$ on eigenfunction extent, compatible with
localisation on larger scales. For the unperturbed crystal,
\cite[Appendix~B]{EM} derives a formal local transport equation for
a selected regular Bloch mode; one-dimensional finite-sample resonance
effects are analysed in \cite[Appendix~A.5]{EM}. Boundary matching is
needed to obtain a finite-sample scattering model. This deterministic
modulation description does not determine the spectral type of the
randomly perturbed operator.

Section~\ref{sec:numerics} treats square arrays of $L^2$ disks for
$L=36,51,70$, using finite capacitance matrices and $500$
realisations at each positive disorder strength. Centred intensity
profiles and spectral participation show concentration near the top
of the spectrum and, at stronger disorder, in the interior of the
unperturbed band. The selected disordered modes occupy a decreasing
fraction of the sample as $L$ increases.

The paper is organised into three parts. Part~\ref{part:infinite} studies the infinite operator, its symbol,
and localisation. Part~\ref{part:truncated} establishes the
estimates for truncated matrices and eigenvalue spacing needed for
the finite-cluster problem. Part~\ref{part:finite} treats finite capacitance matrices,
spectral convergence, and numerical illustrations.
Appendix~\ref{app:heuristic} discusses the three-dimensional
weak-scattering prediction, and Appendix~\ref{sec:other-laws}
extends the band-edge absence result to other material distributions.

\part{The infinite crystal}\label{part:infinite}

\section{The random capacitance operator}\label{sec:model}
We use the Euclidean norm $|x|=|x|_2$, the lattice norms
$|x|_1=\sum_j|x_j|$ and $|x|_\infty=\max_j|x_j|$, and the
standard coordinate vectors $e_j\in\R^d$. The vector $\delta_n$
is the coordinate vector at $n$ in $\ell^2(\Z^d)$.
Inner products are conjugate-linear in the first argument, $T^*$
denotes the adjoint, and $[S,T]=ST-TS$. We write $\Id$ for the
identity, $\one_A$ for an indicator or its multiplication operator,
and $r_+=\max\{r,0\}$. For a self-adjoint operator $H$, $\one_I(H)$ is
its spectral projection onto an interval $I$. 
Throughout, the operator norm refers to the induced norm on the corresponding Hilbert space, and we will omit the subscript when it is clear from the context. In particular, the Hilbert-Schmidt norm satisfies $\|T\|_{\rm HS}^2=\Tr(T^*T)$. Positive constants in estimates
may change from line to line; their parameter dependence is stated
when uniformity is required.

\subsection{Geometry and capacitance normalisation} \label{sec:geo_cap_normalize}
Recall that $d\in\{2,3\}$, $Y=(-1/2,1/2)^d$, and
denote by $\mathcal D=\bigcup_{n\in\Z^d}(D+n)$. Assume that $D\Subset Y$ is simply-connected with $C^2$ boundary. We also assume that \(Y\setminus\overline D\) is connected. The reflection symmetry is described as follows, which will be imposed in particular results of this paper, e.g., the band-edge localisation,
\begin{equation}
 R_jD=D\quad(1\le j\le d),\qquad
 R_j(x_1,\ldots,x_d)=(x_1,\ldots,-x_j,\ldots,x_d).
 \tag{R}\label{eq:reflection}
\end{equation}
The geometry is illustrated in Figures~\ref{fig:crystal} and~\ref{fig:cell}.
Corresponding to the lattice structure, the Brillouin torus associated with the Floquet transform is
$Y^*=\R^d/(2\pi\Z)^d$, represented by $[-\pi,\pi]^d$ with opposite
faces identified. 

For $\alpha \in Y^*$, let $H^1_\alpha(Y)$ denote the Sobolev functions in $H^1(Y)$ whose face traces satisfy
$\gamma_j^+u=e^{i\alpha_j}\gamma_j^-u$, where $\gamma_j^\pm$ is the
trace on $x_j=\pm1/2$. Define
\begin{equation}
 c(\alpha)=\widehat C^\alpha
 =\frac1{|D|}\min_{u\in H^1_\alpha(Y),\ u=1\ {\rm in}\ D}
                         \int_Y|\nabla u|^2.
 \label{eq:symbol-definition}
\end{equation}
Section~\ref{sec:deterministic} proves the existence and uniqueness of the minimiser, together with its
operator interpretation. With the discrete Floquet convention of \cite[Section~4.8]{ADH}, set
\begin{equation}
 \widehat q(\alpha)=\sum_{n\in\Z^d}q_ne^{in\cdot\alpha},\qquad
 c_r=\frac1{(2\pi)^d}\int_{Y^*}c(\alpha)e^{-ir\cdot\alpha}\dd\alpha,
 \qquad (Cq)_n=\sum_m c_{n-m}q_m.
 \label{eq:notation-Fourier}
\end{equation}
For $q\in\ell^2(\mathbb{Z}^d)$, the Fourier series is understood in $L^2(Y^*)$.
We use $q_n$ and $q(n)$ interchangeably. With these notions, $C_{nm}=c_{n-m}$ and
$\widehat{Cq}=c\widehat q$.
Write $c_M=\max_{Y^*}c>0$. Under \eqref{eq:reflection},
$c_M=c(\alpha_M)$ with $\alpha_M=(\pi,\ldots,\pi)$.
The geometric capacity of \cite{ADH} is $\operatorname{Cap}_{\alpha,D}
=|D|c(\alpha)$. Its generalised one-resonator symbol is
$\delta v_r^2c(\alpha)$. The factor $\delta v_r^2$ has been removed
in the unit-period spectral variable $E$ of \eqref{eq:physical-reduction}.
For a physical period $\varepsilon>0$ and contrast $\delta>0$, the
conversion between reduced spectral energy and frequency is
\begin{equation}
 E=\frac{\varepsilon^2\omega^2}{\delta v_r^2},\qquad
 \omega=\omega_{\varepsilon,\delta}(E)
 :=\frac{v_r\sqrt\delta}{\varepsilon}\sqrt E.
 \label{eq:frequency-conversion}
\end{equation}
Thus $\delta=\varepsilon^2$ gives $E=\omega^2/v_r^2$. This convention
is used for the shrinking-cell limit; formulas retaining both
$\varepsilon$ and $\delta$ allow other scalings.

For the static reduced model, set
\begin{equation}
 \omega_M:=\omega_{\varepsilon,\delta}(c_M)
       =\frac{v_r\sqrt{\delta c_M}}{\varepsilon},\qquad
 \frac E{c_M}=\left(\frac\omega{\omega_M}\right)^2.
 \label{eq:normalised-frequency}
\end{equation}
Here $\omega_M$ is the reduced edge frequency. By rescaling
\eqref{eq:Bloch-asymptotic}, the full periodic Bloch frequency at
$M$ differs from it by $O(\varepsilon^{-1}\delta^{3/2})$, which is
$O(\varepsilon^2)$ when $\delta=\varepsilon^2$.
The frequency conversions below are exact for the
reduced model; their interpretation for the transmission problem
requires a uniform reduction estimate.

\CrystalFigure

\subsection{Random material factors}
The main operator family under consideration has independent variables $\xi_n$, uniform on $[0,1]$,
and, for each fixed $W>0$,
\begin{equation}
 m_n=1+W\xi_n,\quad M_\varpi=\diag(m_n),\quad
 B_\varpi=M_\varpi^{-1},\quad
 A_\varpi=B_\varpi C,\quad
 H_\varpi=M_\varpi^{-1/2}CM_\varpi^{-1/2}.
 \label{eq:random-model}
\end{equation}
In the sequel, $\PP$ and $\EE$ denote the probability and expectation with respect to the product measure generated by $\xi_{n}$, respectively, and
$\mu_W:=\EE m_n=1+W/2$ is the mean inverse material factor.
Uniform inverse factors
$m_n$ correspond to direct factors $b_n=m_n^{-1}$ with density
$(Wb^2)^{-1}\one_{[(1+W)^{-1},1]}(b)$. Appendix~\ref{sec:other-laws}
also treats bounded densities and uniform direct factors.

The operator $A_\varpi$ is self-adjoint on
$\mathcal H_\varpi=\ell^2(\Z^d;m_\varpi)$, with norm
$\|q\|_\varpi^2=\sum_nm_n|q_n|^2$. 
The map $\mathcal U_\varpi:\mathcal H_\varpi\to\ell^2(\Z^d)$ given by
\begin{equation}
 \mathcal U_\varpi q=M_\varpi^{1/2}q,\qquad
 \mathcal U_\varpi A_\varpi\mathcal U_\varpi^{-1}=H_\varpi
 \label{eq:unitary}
\end{equation}
is unitary. We formulate spectral statements for
$H_\varpi$. This convention is equivalent to the $BC$ model in
\cite[Section~4.8]{ADH}. If a model is written as $VCV$, as in
\cite{AQchain}, its diagonal $V$ corresponds to our $B_\varpi^{1/2}$.

For finite $\Lambda\subset\Z^d$, let
$E_\Lambda:\ell^2(\Lambda)\to\ell^2(\Z^d)$ be the extension by zero;
$E_\Lambda^*$ is the restriction. Write
$P_\Lambda=E_\Lambda E_\Lambda^*$ and
$M_{\varpi,\Lambda}=E_\Lambda^*M_\varpi E_\Lambda$. Put
\begin{equation}
 C_\Lambda^t=E_\Lambda^*CE_\Lambda,\qquad
 H_{\varpi,\Lambda}^t=M_{\varpi,\Lambda}^{-1/2}
 C_\Lambda^tM_{\varpi,\Lambda}^{-1/2},\qquad
 \Lambda_L=[-L,L]^d\cap\Z^d,\quad L\in\mathbb N.
 \label{eq:finite}
\end{equation}
We call $C_\Lambda^t$ the \emph{truncated capacitance matrix}.
The superscript $f$ denotes the \emph{finite capacitance matrix}
$C_\Lambda^f$ of the retained resonators, defined variationally in
Section~\ref{sec:finite-crystals}. The superscripts on $H$ indicate
the corresponding random matrices. For centred cubes, write $N_L=|\Lambda_L|=(2L+1)^d$
and $P_L=P_{\Lambda_L}$.

\begin{table}[ht]
\centering\small
\begin{tabular}{@{}ll@{}}\toprule
Notation & Meaning\\\midrule
$\alpha\in Y^*,\ \alpha_M$ & Quasimomentum and upper-edge corner\\
$c(\alpha)=\widehat C^\alpha$ & 
scalar capacitance symbol\\
$C,\ c_r,\ c_M$ & Lattice operator, convolution coefficients, upper edge\\
$\gamma_0,\ A_0$ & Decay constants: $|c_r|\le A_0e^{-\gamma_0|r|_1}$\\
$\varpi,\ \xi_n,\ W$ & Configuration, independent variables, disorder width\\
$m_n,\ M_\varpi,\ B_\varpi$ & Inverse material factors, their diagonal, its inverse\\
$H_\varpi,\ H_{\varpi,\Lambda}^{t,f}$ & Self-adjoint infinite and finite operators\\
$\omega,\ \delta,\ E$ & Angular frequency, contrast, reduced spectral energy\\
$\omega_M$ & Reduced edge frequency $v_r\sqrt{\delta c_M}/\varepsilon$\\
$\Delta_W,\ I_W$ & Fixed disorder-dependent edge width and interval\\
$\nu_W$ & Positive decay rate in the localisation bound\\
$\mu_W=1+W/2,\ \lambda=W/\mu_W$ & Mean inverse material factor and centred disorder\\
$\widetilde H_\varpi=\mu_WH_\varpi$ & Centred operator in Section~\ref{sec:interior}\\
$s=\mu_WE$ & Centred energy in the interior and mobility-edge analysis\\
$E_a(W),\ I_{W,a}$ & Quantitative threshold and edge interval\\
$\varepsilon,\ I_\varepsilon$ & Cell size and indices in a bounded sample\\
\bottomrule\end{tabular}
\vspace{2mm}
\caption{Principal notation. Spectral energies are denoted by $E$ or
by the centred variable $s=\mu_WE$. Fractional-moment exponents are
denoted by $\vartheta$.}
\label{tab:notation}
\end{table}

\subsection{Spectral measures and correlators}
For a finite set $\Lambda$ and a Borel interval $I\subset\R$, define
\begin{equation}
 Q_{\varpi,\Lambda}^t(x,y;I)
 =\sup_{\substack{f:\R\to\C\ {\rm Borel}\\\|f\|_\infty\le1}}
 \left|\langle\delta_x,f(H_{\varpi,\Lambda}^t)
                \one_I(H_{\varpi,\Lambda}^t)\delta_y\rangle\right|.
 \label{eq:Q-definition}
\end{equation}
Write $Q_\varpi$ for the infinite-volume analogue. These correlators
are the total variations of the spectral measures on $I$ and are
bounded by one. On an open interval they can equivalently be defined
using a countable dense family in the unit ball of $C_0(I)$, which
also proves measurability.

\begin{remark}[Energy and frequency statements]\label{rem:energy-frequency}
Resolvents of $H_\varpi$ and its finite versions use a spectral
parameter $z=E+i\eta$ in energy units. The positive-frequency
generator of the reduced model is
\[
 \mathcal G_{\varepsilon,\delta,\varpi}
   :=\omega_{\varepsilon,\delta}(H_\varpi)
   =\frac{v_r\sqrt\delta}{\varepsilon}H_\varpi^{1/2}.
\]
It has the same eigenvectors as $H_\varpi$. For an energy interval
$I=(E_-,E_+)\Subset(0,\infty)$, its corresponding frequency interval is
$(\omega_{\varepsilon,\delta}(E_-),\omega_{\varepsilon,\delta}(E_+))$,
and the two spectral projections coincide. Since the conversion and
its inverse are smooth on positive intervals, pure point spectrum
and absence of absolutely continuous spectrum transfer between them.
The time parameter in $e^{-itH_\varpi}$ is conjugate to reduced
energy. The correlator also bounds
$e^{-it\mathcal G_{\varepsilon,\delta,\varpi}}\one_I(H_\varpi)$,
by taking $f(E)=e^{-it\omega_{\varepsilon,\delta}(E)}$ on $I$ in
\eqref{eq:Q-definition}. Thus the same spatial dynamical-localisation
bound holds for the reduced positive-frequency evolution. These
observations also apply to the finite matrices and to $C_\delta$.
\end{remark}

\subsection{The almost-sure spectrum}
We determine the almost-sure spectrum, thereby fixing the spectral edges used in the localisation estimates below.

\begin{proposition}\label{prop:spectrum-new}
For every fixed $W>0$, $\sigma(H_\varpi)=[0,c_M]$ almost surely.
\end{proposition}
\begin{proof}
        Since $0\leq C\leq c_M \mathrm{Id}$ and $0<B_{\varpi} \leq \mathrm{Id}$, $0\leq H_{\varpi} \leq c_M \mathrm{Id}$. Thus, $\sigma(H_{\varpi})\subset [0,c_M]$ for every configuration $\varpi$. Conversely, by Weyl's criterion for the spectrum, to prove that $[0,c_M] \subset \sigma(H_{\varpi}) $, it remains to show that, for every $E\in [0,c_M]$, there are unit vectors $u_j$ satisfying
        \begin{equation}
            \| (H_{\varpi} - E) u_j \| \rightarrow 0 \qquad \mathrm{as} \ j\rightarrow \infty.
        \end{equation}
    The remainder of the proof is divided into several steps. 

    \textit{Step 1: construct the approximate eigenvectors for $C$.} Since $c(Y^*)=[0,c_M]$, for every $E\in [0,c_M]$, there is an $\alpha_E$ such that $c(\alpha_E)=E$. Concentrate the Fourier transform of a unit vector $g_j$ near $\alpha_E$ such that 
    \begin{equation}
        \max_{\alpha \in \mathrm{supp}\,\widehat{g_j}}  |c(\alpha) - E| \rightarrow 0 \qquad \mathrm{as} \ j\rightarrow \infty. 
    \end{equation}
    Since the symbol of $C$ is $c$, we have
    \begin{equation}
        \| (C-E)g_j \|^2 = \frac{1}{(2\pi)^d}\int_{Y^*} |c(\alpha )- E|^2 |\widehat{g_j}(\alpha)|^2\,\mathrm{d}\alpha \rightarrow 0 \qquad \mathrm{as} \ j\rightarrow \infty. 
    \end{equation}
    Although these $g_j$ need not have finite support,  finitely supported vectors are dense in $\ell^2(\mathbb{Z}^d)$, and $C-E$ is bounded. By truncating and normalising, we obtain finitely supported unit vectors $f_j$ such that
    \begin{equation}
        \| (C - E) f_j \| \rightarrow 0 \qquad \mathrm{as} \ j\rightarrow \infty.
    \end{equation}
    Since $Cf_j\in \ell^2(\mathbb{Z}^d)$, we may choose a finite set $B_j$, containing $\mathrm{supp}\,f_j$, large enough such that $\| \mathbf{1}_{B_j^c} Cf_j \| \rightarrow 0$ as $j\rightarrow \infty$.

    \textit{Step 2: construct the good event.} Choose a countable dense subset $\mathcal{E}\subset [0,c_M]$. For any $E\in \mathcal{E}$, by Step 1, after passing to a subsequence, there is a sequence $\{\varepsilon_j\}_{j\in \mathbb{N}}$ converging to zero such that 
    \begin{equation}\label{estimateforCEfj}
        \|(C-E)f_j\| + \| \mathbf{1}_{B_j^c} Cf_j \|\leq \varepsilon_j, \qquad j\in \mathbb{N}.
    \end{equation}
    For each $j\in \mathbb{N}$, since $B_j$ is finite, there is a sequence $\{z_j^k \}_{k\in \mathbb{N}} \subset \mathbb{Z}^d$ such that the translated sets
    \begin{equation}
        B_j^k := B_j + z_j^k
    \end{equation}
    are mutually disjoint. Define the event
    \begin{equation}
        \mathcal{A}_j^k := \bigcap_{n\in B_j^k}\{ \varpi: \xi_n(\varpi) \leq a_j \}, \qquad a_j:=\min \left\{ \frac{1}{2} , \frac{2\varepsilon_j}{W} \right\} \in (0,1).
    \end{equation}
    Denote $M^{-1/2}_{\varpi}$ by $D_{\varpi}$. By the choice of $a_j$, we see $\xi_n(\varpi) \leq a_j$ implies $|D_{\varpi}(n,n)-1|\leq \varepsilon_j$. Since $\{\xi_n\}$ are independent and uniformly distributed in $[0,1]$, 
    \begin{equation}
        \PP[\mathcal{A}_j^k]=a_j^{|B_j|}>0.
    \end{equation}
    For fixed $j$, the events $\{\mathcal{A}_j^k\}$ are independent. Therefore,
    \begin{equation}
        \PP\left[ \bigcap_{k=1}^\infty (\mathcal{A}_j^k)^c \right] = \lim_{K\rightarrow \infty} (1- a_j^{|B_j|})^K =0.
    \end{equation}
    This implies that the event
    \begin{equation}
        \mathcal{G}:=\bigcap_{E\in \mathcal{E}} \bigcap_{j=1}^\infty \bigcup_{k=1}^\infty \mathcal{A}_j^k 
    \end{equation}
    has probability one. 
    
    \textit{Step 3: construct the approximate eigenvectors for $H_{\varpi}$.} Fix $\varpi \in \mathcal{G}$ and $E\in \mathcal{E}$. For each $j$, choose an index $k_j$ such that $\varpi \in \mathcal{A}_j^{k_j}$. Set $u_j $ by
    \begin{equation}
        u_j(n):= f_j(n-z_j^{k_j}) , \qquad n\in \mathbb{Z}^d.
    \end{equation}
    Then, $u_j$ is still a unit vector and supported on $B_j^{k_j}$. Translation invariance and \eqref{estimateforCEfj} give
    \begin{equation}\label{estimateCEuj}
        \|(C-E)u_j\| + \| \mathbf{1}_{(B_j^{k_j})^c} Cu_j \|\leq \varepsilon_j, \qquad j\in \mathbb{N}.
    \end{equation}
    Furthermore, since $\varpi \in \mathcal{A}_j^{k_j}$, 
    \begin{equation}\label{estimateDpinn}
        |D_{\varpi}(n,n)-1|\leq \varepsilon_j, \qquad n \in B_j^{k_j}.
    \end{equation}
    We rewrite
    \begin{equation}\label{identityHEu}
    \begin{aligned}
        (H_{\varpi} - E) u_j &= (C-E)u_j + D_{\varpi} C(D_{\varpi} - \mathrm{Id}) u_j \\
        & \quad + (D_{\varpi}-\mathrm{Id})\mathbf{1}_{B_j^{k_j}} Cu_j+ (D_{\varpi}-\mathrm{Id})\mathbf{1}_{(B_j^{k_j})^c} Cu_j.
    \end{aligned}
    \end{equation}
    For the first and last terms in the right-hand side of \eqref{identityHEu}, we use \eqref{estimateCEuj} and $\| D_{\varpi}- \mathrm{Id}\| \leq 1$; for the second and third terms we use $\mathrm{supp} \,u_j \subset B_j^{k_j}$ and \eqref{estimateDpinn}. As a result, we get $\| (H_{\varpi} - E) u_j \|\rightarrow 0$ as $j\rightarrow \infty$. By Weyl’s criterion, $E \in \sigma(H_{\varpi})$. Since $E\in\mathcal E$ is arbitrary, and the spectrum $\sigma(H_{\varpi})$ is closed, we obtain $[0,c_M]=\overline{\mathcal{E}}\subset \sigma(H_{\varpi})$. The proof is complete.
    \end{proof}

\section{Deterministic properties of the capacitance operator}\label{sec:deterministic}
We establish the deterministic properties used in the localisation arguments: analyticity of the Bloch symbol, exponential decay of the capacitance coefficients, and, under coordinate reflection symmetry, uniqueness and nondegeneracy of the maximum of the symbol. Write $\mathcal O=Y\setminus\overline D$ for the connected cell exterior.
\subsection{A cell Poincar\'e inequality}
We begin with a Poincaré inequality that gives coercivity of the cell energy uniformly in the Bloch phase. We use complex-valued Sobolev spaces. Put
\begin{equation}
 C_P=\frac{\sqrt{1+|D|^{-1}}}{\pi}.
 \label{eq:CP}
\end{equation}
\begin{lemma}\label{lem:CP}
Every $h\in H^1(Y)$ that vanishes almost everywhere in $D$ satisfies
\begin{equation}
 \|h\|_{L^2(Y)}\le C_P\|\nabla h\|_{L^2(Y)}.
 \label{eq:CP-bound}
\end{equation}
No boundary condition on $\partial Y$ is required.
\end{lemma}
\begin{proof}
Let $\bar h=\int_Yh$, since $|Y|=1$. The Neumann Poincar\'e inequality on
the unit cube is $\|h-\bar h\|_2\le\pi^{-1}\|\nabla h\|_2$; it follows,
for example, from the cosine expansion and the first nonzero Neumann
eigenvalue $\pi^2$. Since $h=0$ in $D$,
$|D||\bar h|^2\le\|h-\bar h\|_2^2$. Therefore
\[
 \|h\|_2^2=\|h-\bar h\|_2^2+|\bar h|^2
 \le(1+|D|^{-1})\pi^{-2}\|\nabla h\|_2^2.
\]
The same argument applies to complex functions.
\end{proof}

\subsection{Bloch cell problem and the infinite operator}
Using the trace convention of Section~\ref{sec:model}, set
\begin{equation}
 \mathscr U_\alpha=\{u\in H^1(Y;\C):u=1\text{ in }D,
 \ \gamma_j^+u=e^{i\alpha_j}\gamma_j^-u\ (1\le j\le d)\}.
 \label{eq:Utheta}
\end{equation}
Choose a real $H^1$ cutoff $\chi$, equal to one in a neighbourhood of
$\overline D$ and zero in a neighbourhood of $\partial Y$. 
Set $E_\chi=|D|^{-1}\int_Y|\nabla\chi|^2$. We see that $\mathscr{U}_\alpha$ is nonempty since $\chi \in \mathscr{U}_\alpha$. 

The minimiser in \eqref{eq:symbol-definition} is denoted by
\[
 u_\alpha=\mathop{\rm argmin}_{u\in\mathscr U_\alpha}
                         \int_Y|\nabla u|^2.
\]
Differences
of functions in $\mathscr{U}_\alpha$ vanish in $D$, so Lemma~\ref{lem:CP} gives
coercivity. Lax--Milgram, or strict convexity of the energy, proves existence
and uniqueness of the minimiser. The Euler equation is
\begin{equation}
 \int_Y\nabla u_\alpha\cdot\overline{\nabla h}=0
 \quad(h=0\text{ in }D,\ h\text{ has Bloch phase }\alpha).
 \label{eq:Euler-cell}
\end{equation}
The minimiser is harmonic in $\mathcal O$. Its Bloch extension is weakly
harmonic across the cell faces: test the Euler equation with the
Bloch periodisation of a compactly supported exterior test function.
In particular it is smooth near any open face patch away from its edges and its partial derivatives $\partial_j u_\alpha$ satisfy
\begin{equation}
 \gamma_j^+(\partial_j u_\alpha)
 =e^{i\alpha_j}\gamma_j^-(\partial_j u_\alpha), \qquad j=1,\dots,d
 \label{eq:flux-matching}
\end{equation}
at the cell faces. Integration by parts cancels opposite face terms and gives
\begin{equation}
 c(\alpha)=-\frac1{|D|}\int_{\partial D}\partial_{\nu_D}^+u_\alpha\dd S = \frac{1}{|D|} \int_Y |\nabla u_\alpha|^2,
 \label{eq:cap-flux}
\end{equation}
where $\nu_D$ points out of $D$ and the superscript means the exterior trace.

For finitely supported $q=(q_n)$, minimise the global energy
\begin{equation}
 Q(q)=\frac1{|D|}\min\left\{\int_{\R^d}|\nabla U|^2:
 U\in H^1(\R^d),\quad U=q_n\text{ in }D+n\right\}.
 \label{eq:global-cap}
\end{equation}
A sum of translated versions of the cutoff $\chi$, given below \eqref{eq:Utheta}, gives an admissible lift with norm bounded
by a constant times $\|q\|_{\ell^2}$. Summing
\eqref{eq:CP-bound} over cells gives coercivity on the homogeneous space
of functions vanishing in every resonator. Thus the minimiser exists
uniquely. Polarisation of $Q$ defines a bounded, nonnegative,
self-adjoint operator $ C$ on $\ell^2(\Z^d)$.

With
\[
 (\mathcal BU)(\alpha,x)=\sum_n U(x+n)e^{in\cdot\alpha},
\]
the transformed minimiser is $\widehat q(\alpha)u_{-\alpha}(x)$. Indeed,
the transform has phase $-\alpha$ on opposite faces and value
$\widehat q(\alpha)$ in $D$, and the transformed Euler equation has a
unique solution. Complex conjugation gives $c(-\alpha)=c(\alpha)$.
Parseval's identity therefore gives
\begin{equation}
 Q(q)=\frac1{(2\pi)^d}\int_{[-\pi,\pi]^d}
 c(\alpha)|\widehat q(\alpha)|^2\dd\alpha.
 \label{eq:Fourier-Q}
\end{equation}
This proves the convolution representation \eqref{eq:notation-Fourier}.
Real coefficients of the exterior
problem and self-adjointness give $c_r\in\R$ and $c_{-r}=c_r$.
At phase zero the constant function one is admissible, so $c(0)=0$.
If $c(\alpha)=0$, its minimiser must be constant and equal to one;
then all phases equal one. Thus $c(\alpha)>0$ for $\alpha\ne0$.

\subsection{Analyticity and exponential decay}
\label{sec:decay}

Exponential off-diagonal decay is established for the crystal capacitance
operator in \cite[Proposition~2.5]{AQ}. The variational formulation
in \cite[Proposition~3.5 and Remark~2]{EM} proves analyticity of the Floquet symbol associated with the capacitance operator on the
whole Brillouin torus and derives the same decay, including a uniform
analytic extension for the contrast-dependent family.
This is the decay input used in our finite-volume localisation proof.
We give a self-contained proof for the static operator, with explicit
geometric constants, in both dimensions.

\begin{proposition}\label{prop:exp}
The symbol is real analytic on the whole Brillouin torus, including zero.
With $C_P$ and $E_\chi$ above, set
\begin{equation}
 \gamma_0=\frac1{8\sqrt d\,C_P},\qquad
 A_0=3E_\chi\exp\left(\frac{\sqrt d}{8C_P}\right).
 \label{eq:explicit-A-alpha}
\end{equation}
Then
\begin{equation}
 |c_r|\le A_0 e^{-\gamma_0|r|_1}, \quad r\in\Z^d.
 \label{eq:exponential}
\end{equation}
In particular all polynomially weighted coefficient sums are finite,
and $\sum_rc_r=c(0)=0$.
\end{proposition}
\begin{proof}
Use the fixed space
\[
 \mathscr H=\{w\in H^1(Y):w=0\text{ in }D,
 \ \gamma_j^+w=\gamma_j^-w\text{ for all }j\}.
\]
For real $\alpha$, write $u_\alpha=\chi+e^{i\alpha\cdot x}w_\alpha$.
The equation for $w_\alpha$ is $a_\alpha(w_\alpha,v)=b_\alpha(v)$ for $v\in \mathscr H$, where
the following formulas also define holomorphic dependence on a complex
vector $z$:
\begin{align}
 a_z(w,v)&=\int_Y(\nabla+iz)w\cdot(\nabla-iz)\overline v,
 \label{eq:az}\\
 b_z(v)&=-\int_Y e^{-iz\cdot x}\nabla\chi
                   \cdot(\nabla-iz)\overline v.
 \label{eq:bz}
\end{align}
Notice that $z$, rather than $\bar z$, occurs in both factors of
\eqref{eq:az}; it is the holomorphic continuation of the real-parameter
form. For real $\alpha$, $a_\alpha$ is a positive Hermitian form.

Put $D_\alpha=\nabla+i\alpha$. Applying Lemma~\ref{lem:CP} to
$e^{i\alpha\cdot x}w$ gives
\begin{equation}
 \|w\|_2\le C_P\|D_\alpha w\|_2.
 \label{eq:cov-P}
\end{equation}
Thus, for fixed $\alpha$, $\|w\|_\alpha=\|D_\alpha w\|_2$ is a norm on
$\mathscr H$ equivalent to its $H^1$ norm.
Let $z=\alpha+i\beta$ with real $\alpha,\beta$, and put
$\tau=C_P|\beta|_2$. Expanding the factors in \eqref{eq:az} gives
\begin{equation}
 |a_z(w,v)-a_\alpha(w,v)|
 \le(2\tau+\tau^2)\|w\|_\alpha\|v\|_\alpha.
 \label{eq:perturb-a}
\end{equation}
If $|\beta_j|\le\gamma_0$ for every $j$, then $\tau\le1/8$, so
$2\tau+\tau^2\le17/64<1$. The operator associated with $a_z$
is invertible by a Neumann series relative to $a_\alpha$, with inverse
norm at most $(1-2\tau-\tau^2)^{-1}$. The solution $w_z$ depends
holomorphically on $z$ in a neighbourhood of this closed strip.

Since $|x|_2\le\sqrt d/2$ in $Y$, let
$b_\beta=\sqrt d\,|\beta|_2/2$. From \eqref{eq:bz} and
\eqref{eq:cov-P},
\[
 \|w_z\|_\alpha\le
 \frac{e^{b_\beta}(1+\tau)}{1-2\tau-\tau^2}(|D|E_\chi)^{1/2}.
\]
For real phases, using the test function $u_\alpha-\chi$ in \eqref{eq:Euler-cell} shows that $c(\alpha)=|D|^{-1}\int_Y\nabla u_\alpha\cdot\nabla\chi$.
Its holomorphic continuation is therefore
\begin{equation}
 c(z)=\frac1{|D|}\int_Y\nabla(\chi+e^{iz\cdot x}w_z)\cdot\nabla\chi.
 \label{eq:hol-cap}
\end{equation}
It is periodic by uniqueness of analytic continuation from the real
torus. The preceding estimates give
\begin{align*}
 |c(z)|
 &\le E_\chi\left(1+
 e^{2b_\beta}\frac{(1+\tau)^2}{1-2\tau-\tau^2}\right)\\
 &\le E_\chi\left(1+\frac{81}{47}
                  e^{\sqrt d/(8C_P)}\right)
 \le A_0.
\end{align*}
Shift each Fourier integration contour in \eqref{eq:notation-Fourier} by
$-i\gamma_0\operatorname{sgn}(r_j)$; periodicity cancels the vertical
sides. The exponential factor has modulus $e^{-\gamma_0|r|_1}$.
This proves \eqref{eq:exponential}. The remaining assertions follow
by absolute convergence of the Fourier series and its derivatives.
\end{proof}

\begin{remark}
Reflection symmetry is not needed for Proposition~\ref{prop:exp}. Moreover, the constants in Proposition~\ref{prop:exp} are deliberately conservative. We point out that the full spatial periodicity
and the Dirichlet boundary condition on the resonator in every cell are essential to this argument. 
\end{remark}

\CellFigure

\subsection{Critical points and coordinate monotonicity}
\label{sec:gradient}

In this subsection, we assume the reflection symmetry \eqref{eq:reflection}.
Centred balls, and centred axis-aligned ellipsoids contained in the
cell, satisfy these assumptions and the connected-exterior condition.
The reflection method is related to the half-cell variational argument
for a resonator chain in \cite[Appendix~A]{AQchain}. Here we obtain a coordinate derivative identity
for the fully periodic cell problem.

\begin{theorem}\label{thm:gradient}
Let $j\in\{1,\dots,d\}$ and $\alpha\in Y^*$ with $\alpha_j\notin\{0,\pi\}$. Then
\begin{equation}
 \partial_{\alpha_j}c(\alpha)
 =\frac{1}{2|D|\sin\alpha_j}\int_Y|\nabla (u_\alpha-u_\alpha\circ R_j)|^2,
 \label{eq:gradient-identity}
\end{equation}
and the integral is strictly positive. Hence $\partial_{\alpha_j}c$ is
positive for $0<\alpha_j<\pi$ and negative for $-\pi<\alpha_j<0$, and it
vanishes on $\alpha_j\in\{0,\pi\}$ because $c$ is even and $2\pi$-periodic
in $\alpha_j$. Consequently
\begin{equation}
 \Crit(c):=\{\alpha\in Y^*:\nabla c(\alpha)=0\}
       =\mathcal V:=\{0,\pi\}^d.
 \label{eq:vertices}
\end{equation}
The unique minimum is $c(0)=0$ and the unique maximum is
$c_M=c(\alpha_M)$, $\alpha_M=(\pi,\ldots,\pi)$. Thus
$\sigma( C)=[0,c_M]$. If $J$ is a nonempty compact interval in
$(0,c_M)$ avoiding $\mathcal T:=\{c(\nu):\nu\in\mathcal V\}$, then
\begin{equation}
 \min_{\alpha:\ c(\alpha)\in J}|\nabla c(\alpha)|_2>0.
 \label{eq:regular-window}
\end{equation}
\end{theorem}
\begin{proof}
Reflection $R_j$ preserves Dirichlet energy and reverses only the $j$-th
phase. Hence $c$ is even in each coordinate separately. Fix the
transverse phases $\alpha_1,\cdots,\alpha_{j-1},\alpha_{j+1},\cdots,\alpha_d$ and let $t=\tan(\alpha_j/2)$. For an arbitrary
admissible $u$, put
\[
 v=\frac{u+u\circ R_j}{2},\qquad
 z=\frac{u-u\circ R_j}{2it}.
\]
Then $v$ is even, $z$ is odd, $v=1$ and $z=0$ in $D$, and both
retain all transverse phases. Their traces satisfy
$\gamma_j^+z=\gamma_j^+v$. Conversely these constraints, which do
not involve $t$, imply that $u=v+it z$ has the required $j$th phase:
its two traces are $(1+it)b$ and $(1-it)b$, whose ratio is
$e^{i\alpha_j}$. Thus the correspondence parametrises the entire
admissible space. The functions remain complex-valued; taking real
and imaginary parts would not preserve arbitrary transverse phases.

Parity and change of variables under $R_j$ give
$\int_Y\nabla v\cdot\overline{\nabla z}=0$. Therefore, on a fixed
affine space $\mathscr A$ of pairs,
\begin{equation}
 c(\alpha)=F(t^2),\qquad
 F(a)=\frac1{|D|}\min_{(v,z)\in\mathscr A}
 \left\{\int_Y|\nabla v|^2+a\int_Y|\nabla z|^2\right\}.
 \label{eq:fixed-pair}
\end{equation}
This affine space is nonempty: average a cutoff with its reflection
to obtain an even cutoff, and pair it with zero. Its homogeneous
space consists of pairs whose two components vanish in $D$.
Lemma~\ref{lem:CP} makes the form coercive, locally uniformly for
$a>0$. The minimising pair $(v_a,z_a)$ depends continuously in
$H^1\times H^1$ on $a$. To see this directly, subtract the two
Euler equations at $a$ and $a'$ and test by the difference. Coercivity
and the bound from the fixed cutoff competitor yield
$\|(v_a,z_a)-(v_{a'},z_{a'})\|_{H^1\times H^1}\le C|a-a'|$
on every compact subinterval of $(0,\infty)$.

Using each minimiser as a competitor for the other parameter gives,
for $a_2>a_1$,
\[
 \frac{a_2-a_1}{|D|}\|\nabla z_{a_2}\|_2^2
 \le F(a_2)-F(a_1)
 \le\frac{a_2-a_1}{|D|}\|\nabla z_{a_1}\|_2^2.
\]
Consequently $F'(a)=|D|^{-1}\|\nabla z_a\|_2^2$.

This derivative is strictly positive. Otherwise $z_a$ is constant
and zero in $D$, hence zero. The corresponding $u_\alpha$ is even.
Its equal face traces and its Bloch condition imply zero trace on
the $j$th face because $e^{i\alpha_j}\ne1$. The coordinate
derivative is odd under reflection. Combining this with
\eqref{eq:flux-matching} gives zero normal derivative on that face
because $e^{i\alpha_j}\ne-1$. On an open face patch, the harmonic
Bloch extension has both zero value and zero normal derivative.
Tangential differentiation and the Laplace equation make every
derivative zero there; harmonic analyticity makes the solution zero
in a neighbourhood. Unique continuation in the connected exterior
$\mathcal O$ then gives zero everywhere in $\mathcal O$, contradicting its
trace one on $\partial D$. The Sobolev trace and harmonic regularity
facts used here are standard; see \cite{Evans}.

Now $\mathrm d(t^2)/\mathrm d\alpha_j=t(1+t^2)$ and $(1+t^2)/t=2/\sin\alpha_j$,
\eqref{eq:gradient-identity} follows by the chain rule. Evenness gives the negative
derivative on the negative half-period and zero derivative at zero.
Evenness together with periodicity gives zero derivative at $\pi$.
Every point outside $\mathcal V$ has at least one coordinate with
nonzero derivative, proving \eqref{eq:vertices}.

The extrema follow from strict coordinate monotonicity and from
$c(0)=0$. Continuity on the connected torus and the multiplier
representation give $\sigma( C)=[0,c_M]$. Finally, the inverse
image of $J$ is compact and contains no critical point. The positive
continuous function $|\nabla c|_2$ has a positive minimum there.
\end{proof}

\begin{remark}
All $2^d$ reflection vertices are critical, including mixed vertices;
$\alpha_M$ is not the only critical quasimomentum. On compact energy
intervals avoiding their finitely many values, the level surfaces are
regular analytic hypersurfaces. This does not assert nonzero Gaussian
curvature. The nondegeneracy at the upper vertex, needed below, is
proved next.
\end{remark}

\subsection{Nondegeneracy of the upper edge}\label{sec:hessian}
The uniqueness and nondegeneracy of the upper maximum are proved in
\cite[Lemmas~3.6--3.7]{EM}. We recall the nondegeneracy proof with our
normalisation. This result is the geometric input that allows the
upper-edge deficit to be compared with a discrete Laplacian in the
localisation argument.

\begin{proposition}\label{prop:Hessian}
Under \eqref{eq:reflection}, the matrix $-\nabla_\alpha^2c(\alpha_M)$ is
positive definite. In particular, there are $a_*,a^*>0$ and $r_*>0$ such that
\begin{equation}
 a_*|\beta|_2^2\le c_M-c(\alpha_M+\beta)
       \le a^*|\beta|_2^2\qquad(|\beta|_2\le r_*).
 \label{eq:quadratic-edge}
\end{equation}
\end{proposition}
\begin{proof}
Uniqueness and reflection show that $u_{\alpha_M}$ is even in every
coordinate: reflection changes $\pi$ into the equivalent phase $-\pi$.
Antiperiodicity therefore gives zero trace on every face of $Y$. Complex
conjugation and uniqueness show that $u_{\alpha_M}$ is real. In the
connected exterior it solves the Dirichlet problem with value one on
$\partial D$ and zero on $\partial Y$. The maximum principle and the
Hopf lemma give
\[
 0<u_{\alpha_M}<1\quad\text{in }Y\setminus\overline D,
 \qquad \partial_\nu u_{\alpha_M}<0
 \quad\text{on the interior of each cell face}.
\]
No regularity at the edges of the cube is needed.

Fix $\xi\in\R^d$ and put
$X_M=\{h\in H^1_{\alpha_M}(Y):h=0\text{ in }D\}$.
Since $u_{\alpha_M}$ vanishes on the cell boundary, for real small $t$
we can write uniquely
\[
 u_{\alpha_M+t\xi}=u_{\alpha_M}+e^{it\xi\cdot x}w_t,
 \qquad w_t\in X_M,\quad w_0=0.
\]
The gauged variational equation on the fixed space $X_M$ is coercive
near zero. The analytic inverse argument from Proposition~\ref{prop:exp}
therefore gives analytic $H^1$ dependence of $w_t$ on $t$.
The Euler equation, tested by the difference from $u_{\alpha_M}$,
gives the exact energy identity
\[
 |D|\{c_M-c(\alpha_M+t\xi)\}
       =\int_Y|\nabla(e^{it\xi\cdot x}w_t)|^2.
\]
Consequently
\begin{equation}
 \xi\cdot \nabla_\alpha^2c(\alpha_M)\xi
       =-\frac2{|D|}\int_Y|\nabla w'_0|^2.
 \label{eq:Hessian-energy}
\end{equation}
If this is zero, the cell Poincar\'e inequality gives $w'_0=0$.
Differentiate the Euler equation against $e^{it\xi\cdot x}h$ at zero.
With $w'_0=0$ the result is
\[
 \int_{Y\setminus\overline D}\nabla u_{\alpha_M}\cdot
            \nabla\bigl((\xi\cdot x)\overline h\bigr)=0
 \qquad(h\in X_M).
\]
Integration by parts and pairing opposite faces yield
\begin{equation}
 \sum_{j=1}^d\xi_j\int_{F_j^+}
         \partial_\nu u_{\alpha_M}\,\overline h\dd S=0,
 \qquad F_j^+=\{x\in\partial Y:x_j=1/2\}.
 \label{eq:face-Hessian}
\end{equation}
Here the outward normal derivatives agree on the two faces by reflection,
the traces of $h$ have opposite signs, and the values of $\xi\cdot x$
differ by $\xi_j$. Choose $h$ with a nonzero nonnegative smooth trace
supported in the interior of $F_j^+$, the opposite trace on $F_j^-$,
and zero trace on other faces and on $D$. Such a function is obtained
by an $H^1$ extension in collars of the two faces, which are separated
from $D$. Equation~\eqref{eq:face-Hessian} and the strict Hopf inequality
force $\xi_j=0$. This holds for each $j$, so $\xi=0$.
Thus the Hessian is negative definite. Compactness of the unit sphere
and Taylor's theorem give \eqref{eq:quadratic-edge}.
\end{proof}

\begin{corollary}\label{cor:periodic-ac}
Under \eqref{eq:reflection}, the periodic operator $C$ has purely absolutely continuous spectrum.
\end{corollary}
\begin{proof}
Its spectral measure at $q\in\ell^2(\mathbb{Z}^d)$ is the push-forward under $c$ of
$(2\pi)^{-d}|\widehat q(\alpha)|^2\dd\alpha$. Away from the finite
critical set, a local coordinate change can use $c(\alpha)$ as one
coordinate. The inverse image of a Lebesgue-null energy set therefore
has zero $d$-dimensional measure in each such chart, by Fubini's theorem.
A countable chart cover and the nullity of the finite critical set
prove absolute continuity of every spectral measure.
\end{proof}

\subsection{The contrast-dependent family}
\label{sec:uniform-family}
For $\delta>0$, let $\fint_Du=|D|^{-1}\int_Du$ and define
the family of \cite{EM} by
\begin{equation}
 c_\delta(\alpha)=\widehat C_\delta^\alpha
 =\frac1{|D|}\min_{\substack{u\in H^1_\alpha(Y)\\ \fint_Du=1}}
 \int_Y(\delta^{-1}\one_D+\one_{Y\setminus D})|\nabla u|^2,
 \qquad
 (C_\delta q)_n=\sum_m c_{\delta,n-m}q_m .
 \label{eq:uniform-cell}
\end{equation}
The Fourier coefficients use \eqref{eq:notation-Fourier}.
Set $C_0=C$, $c_0(\alpha)=c(\alpha)$, and
$c_{M,\delta}=c_\delta(\alpha_M)$ under \eqref{eq:reflection}.
Here $c_0(\alpha)$ denotes an element of the symbol family; the coefficient
$c_0$ without an argument retains its meaning in
\eqref{eq:notation-Fourier}. We use $c_{\delta,r}$ for family coefficients.
Also put
$H_{\delta,\varpi}=M_\varpi^{-1/2}C_\delta M_\varpi^{-1/2}$ and
$C_{\delta,\Lambda}^t=E_\Lambda^*C_\delta E_\Lambda$, with
$H_{\delta,\varpi,\Lambda}^t=M_{\varpi,\Lambda}^{-1/2}
C_{\delta,\Lambda}^tM_{\varpi,\Lambda}^{-1/2}$.

\begin{proposition}[Uniform deterministic estimates]\label{prop:uniform-family}
There are $\delta_0>0$ and constants $A_*,\gamma_*,a_*,c_*,c^*>0$
such that, for $0\le\delta\le\delta_0$ and every fixed nonnegative
integer $k$,
\begin{equation}
 |c_{\delta,r}|\le A_*e^{-\gamma_*|r|_1},\qquad
 c_\delta\longrightarrow c\quad\hbox{in }C^k(Y^*)\quad(\delta\downarrow0).
 \label{eq:uniform-decay}
\end{equation}
Under \eqref{eq:reflection}, the unique maximum is at $\alpha_M$,
$c_*\le c_{M,\delta}\le c^*$, and
\begin{equation}
 c_{M,\delta}-c_\delta(\alpha_M+\beta)
       \ge a_*\,2\sum_{j=1}^d(1-\cos\beta_j)
       \quad(\beta\in Y^*).
 \label{eq:uniform-deficit}
\end{equation}
\end{proposition}
\begin{proof}
Proposition~3.5 of \cite{EM} gives a bounded holomorphic extension
to a common neighbourhood of the whole real torus, and convergence
in each $C^k$ norm. Shrink the complex neighbourhood to a common strip
and shift the Fourier contours as in Proposition~\ref{prop:exp};
this gives \eqref{eq:uniform-decay}. The cell Poincar\'e estimate also holds when $\fint_D h=0$:
indeed $|D||\bar h|^2\le\int_D|h-\bar h|^2$, by
Cauchy--Schwarz, and the proof of Lemma~\ref{lem:CP} applies.
This is the coercivity used in the contrast-dependent construction.

Each symbol is even in every coordinate by reflection of the
variational problem, so $\alpha_M$ is a critical point. The negative
definite Hessian in Proposition~\ref{prop:Hessian} and $C^2$ convergence
give strict concavity and a common quadratic deficit in a fixed
neighbourhood of $\alpha_M$. Away from that neighbourhood, the unique
maximum of $c$ and uniform convergence give a common positive gap.
Thus the maximum remains unique and at $\alpha_M$. The positive
upper and lower bounds follow by convergence of its value.
Comparison with $2\sum_j(1-\cos\beta_j)$ near zero and on the
remaining compact set proves \eqref{eq:uniform-deficit}.
This also recovers the edge properties in \cite[Lemmas~3.6--3.7]{EM}.
\end{proof}

We use this family only where uniformity in the simultaneous period
and contrast limit is needed. Statements without a $\delta$ subscript
refer to the static operator. The family is variational and real
self-adjoint. It is distinct from the nonzero-frequency outgoing
capacitance matrix of \cite[Definition~3.9 and Proposition~3.16]{ALLSU},
which is generally non-Hermitian and describes resonance shifts.
Its spectral parameter is still $E$ in
\eqref{eq:frequency-conversion}. Its reduced edge frequency is
$\omega_{M,\delta}:=\omega_{\varepsilon,\delta}(c_{M,\delta})$.
The notation $C_0=C$ labels a static limit and does not set the
physical contrast to zero in the frequency conversion.

\section{Localisation estimates on prescribed intervals}\label{sec:strong}
We derive explicit sufficient conditions for exponential fractional-moment bounds on prescribed positive energy intervals. For the uniform inverse-material law, these conditions give quantitative thresholds on the disorder strength.

Let $P_{\ac}(H)$ be the projection onto the absolutely continuous spectral
subspace of a self-adjoint operator $H$. For $0<\vartheta<1$ and $0<\mu<\vartheta\gamma_0$, define
\begin{equation}
 T_\vartheta(\mu)=\sum_{r\ne0}|c_r|^{\vartheta}e^{\mu|r|_1}<\infty.
 \label{eq:Ts-main}
\end{equation}
The constants $\gamma_0,A_0$ are established in
Proposition~\ref{prop:exp}. The fixed-interval estimate does not require reflection symmetry.

\begin{theorem}[A fixed interval at sufficiently large disorder]
\label{thm:strong-main}
Let $I=(E_0,E_1)$ with $0<E_0<E_1<c_M$. Fix $\vartheta\in(0,1)$ and
$\mu\in(0,\vartheta\gamma_0)$. If
\begin{equation}
 W>W_*(I;\vartheta,\mu):=\frac1{E_0}
       \left(\frac{2^{\vartheta}T_\vartheta(\mu)}{1-\vartheta}\right)^{1/\vartheta},
 \label{eq:large-threshold}
\end{equation}
then, for every finite $\Lambda$ and every $y\in\Lambda$,
\begin{equation}
 \sup_{E\in I,\ \eta>0}\sum_{x\in\Lambda}e^{\mu|x-y|_1}
 \EE\bigl|\langle\delta_x,(H_{\varpi,\Lambda}^t-E-i\eta)^{-1}
                                      \delta_y\rangle\bigr|^\vartheta<\infty,
 \label{eq:strong-main-bound}
\end{equation}
with one bound independent of $\Lambda$ and $y$. Consequently
$\one_I(H_\varpi)P_{\ac}(H_\varpi)=0$ almost surely.
An improved two-step sufficient condition is given in
Theorem~\ref{thm:FM}.
\end{theorem}

We first allow general iid inverse factors $m_n\in[m_-,m_+]$,
where $0<m_-\le m_+<\infty$. The uniform family of
\eqref{eq:random-model} will then follow explicitly. Write the Green's function
\[
 G_{\varpi,\Lambda}(x,y;z)=
 \langle\delta_x,(H_{\varpi,\Lambda}^t-z)^{-1}\delta_y\rangle.
\]
\begin{definition}[Condition (F)]\label{def:F}
For an open bounded interval $I=(E_0,E_1)\Subset(0,\infty)$,
condition (F) means that there is $\vartheta\in(0,1)$ such that, for every
fixed $y$, there is a finite $K_y^{F}$ with
\begin{equation}
 \sup_{L:\,y\in\Lambda_L}\ 
 \sup_{\substack{E\in I\\0<\eta\le1}}
 \sum_{x\in\Lambda_L}\EE
 |G_{\varpi,\Lambda_L}(x,y;E+i\eta)|^\vartheta\le K_y^{F}.
 \tag{F}\label{eq:F}
\end{equation}
A stronger sufficient condition is
\begin{equation}
 \EE|G_{\varpi,\Lambda}(x,y;E+i\eta)|^\vartheta
 \le B e^{-\mu|x-y|_1},
 \tag{$F_{\exp}$}\label{eq:Fexp}
\end{equation}
uniformly in finite volume, sites, $E\in I$, and $\eta>0$.
\end{definition}

\begin{remark}
Uniformity both in volume and as $\eta\downarrow0$ is indispensable.
For a discrete random law, even one finite-volume eigenvalue occurring
with positive probability can make the corresponding supremum
diverge. In the sufficient tests below we impose a fractional
integrability condition on the inverse-material distribution.
\end{remark}

\subsection{The fractional-moment tests}\label{sec:criteria}

Define
\begin{equation}
 K_\vartheta=\sup_{a\in\R}\EE|m_n-a|^{-\vartheta}<\infty,\qquad
 \kappa=K_{\vartheta}E_0^{-\vartheta}.
 \label{eq:Ks}
\end{equation}
If $m_n$ has density bounded by $R_m$, then
\begin{equation}
 K_\vartheta\le\frac{(2R_m)^\vartheta}{1-\vartheta}.
 \label{eq:density-bound}
\end{equation}
Indeed, $\PP\{|m_n-a|<r\}\le\min(1,2R_mr)$; integrate the
distribution function of $|m_n-a|^{-\vartheta}$ over its positive levels.
For $0<\mu<\vartheta\gamma_0$, put
\begin{equation}
 T_\vartheta=\sum_{r\ne0}|c_r|^{\vartheta}e^{\mu|r|_1},\qquad
 U_\vartheta=\sum_{r\ne0}|c_r|^{2\vartheta}e^{2\mu|r|_1}.
 \label{eq:TU}
\end{equation}
Their dependence on $\mu$ is suppressed only within this section.

\begin{theorem}[Uniform fractional-moment estimates]\label{thm:FM}
Either of the following conditions implies (F) and $(F_{\exp})$:
\begin{align}
 q_1&=\kappa T_\vartheta<1, \tag{C1}\label{eq:C1}\\
 q_2&=\kappa^2(T_\vartheta^2-U_\vartheta)<1. \tag{C2}\label{eq:C2}
\end{align}
More precisely, for every finite $\Lambda$ and every $y\in\Lambda$,
\begin{equation}
 \sup_{\substack{E\in I\\\eta>0}}
 \sum_{x\in\Lambda}e^{\mu|x-y|_1}
 \EE|G_{\varpi,\Lambda}(x,y;E+i\eta)|^\vartheta\le B,
 \label{eq:weighted-F}
\end{equation}
where one may take, respectively,
\begin{equation}
 B_1=m_+^\vartheta\frac{\kappa}{1-q_1},\qquad
 B_2=m_+^\vartheta\frac{\kappa(1+\kappa T_\vartheta)}{1-q_2}.
 \label{eq:Bconstants}
\end{equation}
Condition (C1) implies (C2), but the converse need not hold.
\end{theorem}

\subsection{Conditional averaging for the capacitance operator}
Define
\begin{equation}
 R_\Lambda(z)=( C_\Lambda^t-zM_{\varpi,\Lambda})^{-1}.
 \label{eq:pencil-resolvent}
\end{equation}
For $\Im z>0$ this inverse exists, since
\begin{equation}
 (H_{\varpi,\Lambda}^t-z)^{-1}
 =M_{\varpi,\Lambda}^{1/2}R_\Lambda(z)M_{\varpi,\Lambda}^{1/2}.
 \label{eq:resolvent-relation}
\end{equation}
The advantage is that, in \eqref{eq:pencil-resolvent}, the randomness appears only in the diagonal entries. Fix $y\in\Lambda$ and $D_y=\Lambda\setminus\{y\}$. The Schur
complement gives
\begin{equation}
 R_\Lambda(y,y;z)
 =\bigl(c_0-zm_y-\Sigma_y(z)\bigr)^{-1},
 \quad
 \Sigma_y(z)=\sum_{u,v\in D_y}c_{y-u}R_{D_y}(u,v;z)c_{v-y}.
 \label{eq:Schur}
\end{equation}
The self-energy $\Sigma_y(z)$ is independent of $m_y$. For the possibly complex
$a=(c_0-\Sigma_y(z))/z$, conditioning on all other variables yields
\begin{equation}
 \EE_{m_y}|R_\Lambda(y,y;z)|^\vartheta
 =|z|^{-\vartheta}\EE_{m_y}|m_y-a|^{-\vartheta}
 \le |z|^{-\vartheta}K_\vartheta\le\kappa.
 \label{eq:conditional}
\end{equation}
The supremum over real centres suffices because
$|m-a|\ge|m-\Re a|$. This estimate is uniform in the conditioned
variables and in every depleted finite set.

For $x\ne y$, block inversion gives
\begin{equation}
 R_\Lambda(x,y;z)
 =-R_\Lambda(y,y;z)\sum_{u\in D_y}R_{D_y}(x,u;z)c_{u-y}.
 \label{eq:offdiag}
\end{equation}
Subadditivity of $|\cdot|^\vartheta$, followed by conditional averaging in
$m_y$, proves
\begin{equation}
 \EE|R_\Lambda(x,y;z)|^\vartheta
 \le\kappa\sum_{u\in D_y}|c_{u-y}|^\vartheta
                    \EE|R_{D_y}(x,u;z)|^\vartheta.
 \label{eq:depletion}
\end{equation}
This is the site-deletion fractional-moment argument associated with
\cite{AM,Stolz}, applied directly to the multiplicative model.

\subsection{Closing the one-step and two-step recursions}
Put $w_r=|c_r|^{\vartheta}e^{\mu|r|_1}$ for $r\ne0$, and
\[
 S_\Lambda(y)=\sum_{x\in\Lambda}e^{\mu|x-y|_1}
                         \EE|R_\Lambda(x,y;z)|^\vartheta.
\]
Equations \eqref{eq:conditional}--\eqref{eq:depletion} and the triangle
inequality imply
\begin{equation}
 S_\Lambda(y)\le\kappa+
       \kappa\sum_{u\in\Lambda\setminus\{y\}}w_{u-y}
                      S_{\Lambda\setminus\{y\}}(u).
 \label{eq:one-recursion}
\end{equation}
If (C1) holds, induction on $|\Lambda|$ gives
$S_\Lambda(y)\le\kappa/(1-q_1)$. The singleton case follows from
\eqref{eq:conditional}; inserting the induction hypothesis into
\eqref{eq:one-recursion} gives the same bound.

For the improved estimate apply \eqref{eq:one-recursion} again to
the depleted set:
\begin{align}
 S_\Lambda(y)\le{}&\kappa+\kappa^2T_\vartheta\nonumber\\
 &+\kappa^2\sum_{u\in\Lambda\setminus\{y\}}w_{u-y}
 \sum_{v\in\Lambda\setminus\{y,u\}}w_{v-u}
                      S_{\Lambda\setminus\{y,u\}}(v).
 \label{eq:two-recursion}
\end{align}
The second step cannot return to $y$, which has been removed.
Since $w_{-r}=w_r$, the full-lattice sum of two-step weights is
\begin{equation}
 \sum_{u\ne y}w_{u-y}\sum_{v\ne u,y}w_{v-u}
 =T_\vartheta^2-\sum_{r\ne0}w_rw_{-r}=T_\vartheta^2-U_\vartheta.
 \label{eq:no-return}
\end{equation}
Induction on the cardinality now gives
$S_\Lambda(y)\le\kappa(1+\kappa T_\vartheta)/(1-q_2)$ under (C2):
the empty-set remainder is zero, the singleton bound is immediate,
and \eqref{eq:two-recursion} closes with precisely this constant.

Finally,
\[
 |G_{\varpi,\Lambda}(x,y;z)|^\vartheta
 =(m_xm_y)^{\vartheta/2}|R_\Lambda(x,y;z)|^\vartheta
 \le m_+^\vartheta|R_\Lambda(x,y;z)|^\vartheta.
\]
This proves Theorem~\ref{thm:FM}. Retaining a single summand gives
$(F_{\exp})$, and dropping the exponential weight gives (F).

\begin{remark}[Longer paths]
The same deletion argument can retain paths of length $k$ with no
repeated vertices. Let $Z_0=1$ and let $Z_k(\mu)$ be the sum of
$\prod_{j=1}^k w_{x_j-x_{j-1}}$ over distinct vertices
$x_0=0,x_1,\ldots,x_k$. If $\kappa^kZ_k(\mu)<1$, then
\eqref{eq:weighted-F} holds with
\[
 B=m_+^\vartheta\kappa
 \frac{\sum_{j=0}^{k-1}\kappa^jZ_j(\mu)}{1-\kappa^kZ_k(\mu)}.
\]
Indeed the iterated Schur expansion terminates on every finite set,
and $Z_{n+k}\le Z_nZ_k$ by discarding restrictions between a prefix
and its translated suffix. Group the resulting nonnegative path
series in blocks of length $k$. The cases $k=1,2$ are the estimates
just proved. Related self-avoiding-walk refinements for additive
Anderson models are discussed in \cite{Schenker}.
\end{remark}

\subsection{From (F) to absence of absolutely continuous spectrum}
\label{sec:noac}

\begin{theorem}\label{thm:noac}
If (F) holds for the truncated matrices of $H_\varpi$, then
\begin{equation}
 \one_I(H_\varpi)P_{\ac}(H_\varpi)=0
 \quad\text{almost surely}.
 \label{eq:noac}
\end{equation}
Equivalently, $A_\varpi=B_\varpi C$, considered on
$\mathcal H_\varpi$, has no absolutely continuous spectral component
in $I$. 
\end{theorem}

\begin{proof}[Proof of Theorem~\ref{thm:noac}]
\emph{Infinite-volume resolvent.}
Let $P_L=E_{\Lambda_L}E_{\Lambda_L}^*$. Extend the finite restriction
by zero outside $\Lambda_L$:
$\widehat H_{\varpi,L}=P_LH_\varpi P_L$ on $\ell^2(\Z^d)$.
The uniform bound $\|H_\varpi\|\le m_-^{-1}\| C\|$ and
$P_L\to\Id$ strongly imply
$\widehat H_{\varpi,L}\to H_\varpi$ strongly. For $\Im z>0$,
the resolvent identity and the bound $\|(\widehat H_{\varpi,L}-z)^{-1}\|
\le(\Im z)^{-1}$ give strong resolvent convergence. For fixed sites
and sufficiently large $L$, the matrix elements of this extended
resolvent equal those of the truncated matrix. Thus
\[
 G_{\varpi,\Lambda_L}(x,y;z)\longrightarrow
 G_\varpi(x,y;z)
 :=\langle\delta_x,(H_\varpi-z)^{-1}\delta_y\rangle.
\]
Extend finite-volume summands by zero when $x\notin\Lambda_L$.
Fatou's lemma over counting measure and probability gives, for each
$E\in I$ and $0<\eta\le1$,
\begin{equation}
 \EE\sum_{x\in\Z^d}|G_\varpi(x,y;E+i\eta)|^\vartheta\le K_y^F.
 \label{eq:infinite-F}
\end{equation}

\medskip\noindent\emph{Boundary values.}
For fixed $y$, put
\[
 S_k(\varpi,E)=\sum_x|G_\varpi(x,y;E+i/k)|^\vartheta.
\]
These functions are nonnegative and jointly measurable. Integrate
\eqref{eq:infinite-F} over $I$ and apply Fatou's lemma on the
product of probability and Lebesgue measure:
\begin{equation}
 \EE\int_I\liminf_{k\to\infty}S_k(\varpi,E)\dd E
 \le K_y^F|I|<\infty.
 \label{eq:Fatou-boundary}
\end{equation}
Therefore, for almost every realisation and Lebesgue-almost every
$E\in I$, there is a subsequence $k_j\to\infty$ for which
$S_{k_j}(\varpi,E)\le L<\infty$. The subsequence and $L$ may
depend on $\varpi,E,y$.

For $0<\vartheta<2$, any sequence satisfies
\[
 \sum_x|b_x|^2\le\left(\sum_x|b_x|^\vartheta\right)^{2/\vartheta}.
\]
The self-adjoint resolvent identity gives
\begin{equation}
 \Im G_\varpi(y,y;E+i\eta)
 =\eta\sum_x|G_\varpi(x,y;E+i\eta)|^2.
 \label{eq:imaginary}
\end{equation}
Consequently
\begin{equation}
 0\le\Im G_\varpi(y,y;E+i/k_j)
 \le k_j^{-1}L^{2/\vartheta}\longrightarrow0.
 \label{eq:subsequence-zero}
\end{equation}

Let $\nu_{\varpi,y}$ be the spectral measure of $\delta_y$.
Its Stieltjes transform is $G_\varpi(y,y;z)$, and the Poisson
boundary-value formula states
\begin{equation}
 \frac{\mathrm d\nu_{\varpi,y}^{\ac}}{\mathrm dE}(E)
 =\frac1\pi\lim_{\eta\downarrow0}\Im G_\varpi(y,y;E+i\eta)
 \quad\text{for Lebesgue-almost every }E.
 \label{eq:Poisson}
\end{equation}
See \cite[Theorems 3.26--3.27]{Teschl}. At energies where this full
limit exists, it equals the subsequential limit in
\eqref{eq:subsequence-zero}. Hence
$\nu_{\varpi,y}^{\ac}(I)=0$ almost surely. Intersect the
probability-one events over the countably many sites. The projection
$P=\one_I(H_\varpi)P_{\ac}(H_\varpi)$ then satisfies
$\|P\delta_y\|^2=\nu_{\varpi,y}^{\ac}(I)=0$ for every $y$.
The coordinate vectors span a dense subspace, so $P=0$.
Unitary equivalence \eqref{eq:unitary} proves the assertion.
\end{proof}

\begin{remark}
Condition (F) excludes absolutely continuous spectrum. For the uniform
inverse-material law, Section~\ref{sec:localisation} combines the stronger
exponential estimates with spectral averaging to prove spectral and
dynamical localisation.
\end{remark}

\begin{proof}[Completion of the proof of Theorem~\ref{thm:strong-main}]
For $m_n$ uniform on $[1,1+W]$, integration about the midpoint gives
\[
 K_\vartheta=\sup_{a\in\R}\frac1W\int_1^{1+W}|m-a|^{-\vartheta}\dd m
       =\frac{2^\vartheta}{1-\vartheta}W^{-\vartheta}.
\]
Indeed the integral increases up to the midpoint and decreases afterwards,
and its value outside the interval is no larger. Substitution into (C1)
gives exactly \eqref{eq:large-threshold}. Theorem~\ref{thm:FM} gives
\eqref{eq:strong-main-bound}, with
$B_1=(1+W)^\vartheta\kappa/(1-\kappa T_\vartheta)$; Theorem~\ref{thm:noac}
then gives the absence conclusion.
\end{proof}

\section{Uniform estimates near the upper band edge}\label{sec:edge}

We establish fractional-moment estimates near the upper spectral edge for every fixed positive disorder strength, with bounds uniform over finite truncations.

Throughout this section we assume \eqref{eq:reflection} and the
material law \eqref{eq:random-model}.
\subsection{The deficit operator and a Laplacian comparison}
Define
\begin{equation}
 (Jq)_n=(-1)^{n_1+\cdots+n_d}q_n,\qquad
 K=c_M\Id-JCJ,\qquad k(\alpha)=c_M-c(\alpha_M+\alpha).
 \label{eq:K}
\end{equation}
The unitary $J$ shifts quasimomentum by $\alpha_M$. Thus $K\ge0$,
its symbol vanishes only at zero, and its off-diagonal entries have the
same absolute values as those of $C$. Let $\Lap$ be the positive
nearest-neighbour Laplacian, with symbol $\ell$:
\begin{equation}
 (\Lap q)_n=\sum_{j=1}^d(2q_n-q_{n+e_j}-q_{n-e_j}),
 \qquad \ell(\alpha)=2\sum_{j=1}^d(1-\cos\alpha_j).
 \label{eq:Lap}
\end{equation}
By Proposition~\ref{prop:Hessian}, $k(\alpha)\ge a_{\rm L}\ell(\alpha)$
near zero for some $a_{\rm L}>0$. Away from zero both functions are continuous
and $k$ is positive. Decreasing $a_{\rm L}$ gives the global inequality
\begin{equation}
 k(\alpha)\ge a_{\rm L}\ell(\alpha),\qquad K\ge a_{\rm L}\Lap.
 \label{eq:power-bound}
\end{equation}

\subsection{A uniform initial finite-volume estimate}
\label{sec:initial}
Set $\Xi_\varpi=\diag(\xi_n)$ and consider
\begin{equation}
 L_{g,\varpi}=K+g\Xi_\varpi,\qquad
 g\in[g_-,g_+],\quad 0<g_-\le g_+<\infty.
 \label{eq:additive}
\end{equation}
Throughout, the restriction of an operator to a finite set $\Lambda$ always means the compressed operator, e.g., $K_\Lambda=E_\Lambda^*K E_\Lambda$ and
$L_{g,\varpi,\Lambda}=K_\Lambda+g\Xi_{\varpi,\Lambda}$ on $\ell^2(\Lambda)$.
In particular $K_\Lambda\ge0$. This compression retains the interactions of the infinite operator;
the finite capacitance matrix is defined separately in
Section~\ref{sec:finite-crystals}.

\begin{proposition}\label{prop:initial}
Fix $\beta\in(0,1)$ and write
\[
 e_L=L^{-\beta},\quad
 \ell_L=\lfloor L^{\beta/4}\rfloor.
\]
There are $C,c>0$ and $L_0<\infty$ such that, for integers $L\ge L_0$,
\begin{equation}
 p_L:=\sup_{g\in[g_-,g_+]}
 \PP\{\lambda_{\min}(L_{g,\varpi,\Lambda_L})\le e_L\}
 \le C L^d\exp(-cL^{\beta d/4}).
 \label{eq:initial-prob}
\end{equation}
The starting scale $L_0$ may diverge as $g_-\downarrow0$.
\end{proposition}

\begin{proof}[Proof of Proposition~\ref{prop:initial}]
\emph{Laplacian comparison.}
Suppose $\psi$ is a normalised eigenvector for $L_{g,\varpi,\Lambda_L}$ with the eigenvalue at most $e_L$, and extend it by zero to $\Z^d$.
Nonnegativity gives separately
\[
 \langle\psi,K\psi\rangle\le e_L,\qquad
 g\langle\psi,\Xi_\varpi\psi\rangle\le e_L.
\]
The comparison $K\ge a_{\rm L}\Lap$ gives
$\langle\psi,\Lap\psi\rangle\le e_L/a_{\rm L}$. Therefore
\begin{equation}
 \lambda_{\min}(\Lap_{\Lambda_L}^{D}+g\Xi_{\varpi,\Lambda_L})
       \le h_L:=(1+a_{\rm L}^{-1})e_L.
 \label{eq:nearest-comparison}
\end{equation}
Here $\Lap_{\Lambda_L}^{D}=E_{\Lambda_L}^*\Lap E_{\Lambda_L}$
is the Dirichlet compression, or equivalently the operator with zero
extension outside the cube.

\medskip\noindent\emph{Neumann bracketing.}
Partition each coordinate interval $\{-L,\ldots,L\}$ into consecutive subintervals containing between $\ell_L$ and $2\ell_L$ sites. For sufficiently large $L$, this can be done by taking subintervals of $\ell_L$ sites and merging any shorter final subinterval with its predecessor. Let $\mathcal B_L$ denote the resulting partition of $\Lambda_L$ into rectangular blocks, obtained by taking Cartesian products of these coordinate subintervals.
Each edge contributes a nonnegative term to the Laplacian energy. Discarding the terms associated with edges connecting distinct blocks, together with those crossing the outer Dirichlet boundary of $\Lambda_L$, therefore gives 
\begin{equation}
 \Lap_{\Lambda_L}^{D}+g\Xi_{\varpi,\Lambda_L}
 \ge\bigoplus_{B\in\mathcal B_L} (\Lap_B^N+g\Xi_{\varpi,B}),
 \label{eq:bracketing}
\end{equation}
where $\Lap_B^N$ is the Neumann graph Laplacian on $B$, containing only edges whose endpoints both lie in $B$.
The Neumann graph Laplacian $\Lap_B^N$ has a normalised constant
ground state $\phi_B$ and first positive eigenvalue
\[
 \gamma_B=\min_j\{2-2\cos(\pi/N_j)\}\ge \ell_L^{-2},
 \qquad \ell_L\le N_j\le2\ell_L.
\]
The inequality follows from $2-2\cos(\pi/N)\ge4/N^2$.
Put
\begin{equation}
 t_L=\frac1{4\ell_L^2},\qquad
 X_n=\one_{\{\xi_n\ge1/2\}},\qquad
 \overline X_B=|B|^{-1}\sum_{n\in B}X_n.
 \label{eq:bernoulli}
\end{equation}
For large enough $L$, $t_L\le g_-/2$, and thus $g\xi_n\ge t_LX_n$.

We record the needed eigenvalue bound with its proof. If a finite
self-adjoint matrix $T$ has eigenvalues $\lambda_1\le\lambda_2\le\cdots$
and $\lambda_2\ge\gamma$, and a normalised trial vector has mean
$\mu=\langle T\phi,\phi\rangle<\gamma$ and variance
$\sigma^2=\|(T-\mu)\phi\|^2$, then
\begin{equation}
 \lambda_1\ge\mu-\frac{\sigma^2}{\gamma-\mu}.
 \label{eq:Temple}
\end{equation}
Indeed, the spectral expansion gives
$\langle(T-\lambda_1)(T-\gamma)\phi,\phi\rangle\ge0$; rearranging
this inequality proves \eqref{eq:Temple}.

Apply it to $T=\Lap_B^N+t_L\diag(X_n)$ and $\phi=\phi_B$.
The min--max principle gives $\lambda_2(T)\ge\gamma_B\ge4t_L$,
whereas
\[
 \mu=t_L\overline X_B,\qquad
 \sigma^2=t_L^2\overline X_B(1-\overline X_B).
\]
Therefore
\begin{equation}
 \lambda_{\min}(\Lap_B^N+g\Xi_{\varpi,B})
 \ge\frac23t_L\overline X_B.
 \label{eq:block-lower}
\end{equation}
If $\overline X_B\ge1/4$, this is at least $t_L/6$.
On the other hand,
\[
 h_L=O(L^{-\beta}),\qquad
 t_L\asymp L^{-\beta/2},
\]
so $h_L<t_L/6$ for all sufficiently large $L$. Equations
\eqref{eq:nearest-comparison}--\eqref{eq:bracketing} imply that a
low eigenvalue requires $\overline X_B<1/4$ on at least one block.

\medskip\noindent\emph{Probability of a bad block.}
The variables $X_n$ are independent Bernoulli variables of mean $1/2$.
For $N=|B|$, exponential Markov inequality gives
\[
 \PP\{\textstyle\sum_{n\in B}(X_n-1/2)<-N/4\}
 \le e^{-N/4}\bigl(\cosh(1/2)\bigr)^N
 \le e^{-N/8},
\]
where $\cosh x\le e^{x^2/2}$ was used. Every block has at least
$(\ell_L)^d$ sites and there are at most $(2L+1)^d$ blocks. A union
bound yields
\[
 p_L\le(2L+1)^d e^{-(\ell_L)^d/8}.
\]
This proves the stated initial-scale estimate.
\end{proof}

\subsection{Verification of the fractional-moment block criterion}
\label{sec:block}
We verify a finite-volume criterion with constants uniform for
$g\in[g_-,g_+]$.

\subsubsection{Spectral averaging and a finite-volume moment bound}
\begin{lemma}\label{lem:Wegner}
If $\Lambda$ has $N$ sites and $J_0$ is a bounded interval, we have the Wegner estimate
\begin{equation}
 \EE\operatorname{Tr}\one_{J_0}(L_{g,\varpi,\Lambda})
 \le\frac{N|J_0|}{g}.
 \label{eq:Wegner}
\end{equation}
Consequently, for every real $\mathfrak e$, $0<t<1$, and $\eta\ne0$,
\begin{equation}
 \EE\|(L_{g,\varpi,\Lambda}-\mathfrak e-i\eta)^{-1}\|^t
 \le D_t(N,g):=\frac{(2N/g)^t}{1-t}.
 \label{eq:norm-moment}
\end{equation}
\end{lemma}
\begin{proof}
Fix $n\in\Lambda$. For a deterministic parameter $v\in[0,g]$, let $L_{g,\varpi,\Lambda}(v)$ denote the matrix obtained from
$L_{g,\varpi,\Lambda}$ by replacing the diagonal potential $g\xi_n$ at site $n$ with $v$, while retaining all other potentials. Define
\[
F_{n,v}(z) := \left\langle\delta_n, (L_{g,\varpi,\Lambda}(v)-z)^{-1}\delta_n \right\rangle, \qquad
\nu_{n,v}(B) := \left\langle\delta_n, \mathbf1_B(L_{g,\varpi,\Lambda}(v))\delta_n \right\rangle.
\]
Put $D_n=\Lambda\setminus\{n\}$ and $b_n=(K_{mn})_{m\in D_n}$. For $z=E+i\eta$, $\eta>0$, the Schur-complement gives
\[
F_{n,v}(z)=\frac1{v-a(z)}, \qquad a(z)=z-K_{nn} +b_n^*(L_{g,\varpi,D_n}-z)^{-1}b_n
\]
with $\Im a(z)>0$. Thus,
\begin{equation}\label{boundofimFnv}
     \int_0^g\Im F_{n,v}(z)\,\mathrm{d}v= \int_0^g\Im \frac1{v-a(z)}\,\mathrm{d}v  \leq \pi.
\end{equation}

We next derive the corresponding spectral-measure estimate. Write $a=\inf J_0$, $b=\sup J_0$, and fix $\varepsilon>0$. For every fixed configuration of the other variables and every fixed $v$, Stone's formula gives
\begin{equation}\label{stoneformulaFnv}
    \begin{aligned}
 &\lim_{k\to\infty} \frac1\pi\int_{a-\varepsilon}^{b+\varepsilon} \Im F_{n,v}(E+i/k)\,\mathrm{d}E\\
 &\qquad= \nu_{n,v}((a-\varepsilon,b+\varepsilon)) +\frac12\nu_{n,v}(\{a-\varepsilon\})
 +\frac12\nu_{n,v}(\{b+\varepsilon\}) \ge \nu_{n,v}(J_0).
\end{aligned}
\end{equation}

Set
\[
\mathcal F_n :=\sigma(\xi_m:m\in\Lambda\setminus\{n\}),
 \qquad  \mathbb E_n[\cdot] :=\mathbb E[\cdot\mid\mathcal F_n], \qquad V_n:=g\xi_n.
\]
The random variable $V_n$ is independent of $\mathcal F_n$ and is uniformly distributed on $[0,g]$. Fatou's lemma  and \eqref{boundofimFnv}--\eqref{stoneformulaFnv} imply
\[
\begin{aligned}
 \mathbb E_n[\nu_{n,V_n}(J_0)] &= \frac1g\int_0^g\nu_{n,v}(J_0)\,\mathrm{d}v\\
 &\le \liminf_{k\to\infty} \frac1{\pi g} \int_{a-\varepsilon}^{b+\varepsilon} \int_0^g\Im F_{n,v}(E+i/k)\,\mathrm{d}v\mathrm{d}E\\
 &\le \frac{|J_0|+2\varepsilon}{g}.
\end{aligned}
\]
Letting $\varepsilon\downarrow0$ yields $\mathbb E_n[\nu_{n,V_n}(J_0)] \le |J_0|/g$. Using the tower property and summing over $n\in\Lambda$, we obtain
\[
\begin{aligned}
 \mathbb E\operatorname{Tr}\mathbf1_{J_0}(L_{g,\varpi,\Lambda}) &=\sum_{n\in\Lambda}\mathbb E[\nu_{n,V_n}(J_0)] =\sum_{n\in\Lambda}\mathbb E\!\left[\mathbb E_n[\nu_{n,V_n}(J_0)]\right]\\
 &\le \frac{N|J_0|}{g}.
\end{aligned}
\]
This proves \eqref{eq:Wegner}. 
For $r>0$ it implies
\[
 \PP\{\dist(\mathfrak e,\sigma(L_{g,\varpi,\Lambda}))<r\}
 \le\min\{1,2Nr/g\}.
\]
Integrating the tail of the negative $t$-th power of this distance gives
$(2N/g)^t/(1-t)$. Finally the complex resolvent norm is bounded by
the inverse real spectral distance. In particular each fixed real
$\mathfrak e$ is almost surely outside the spectrum of this truncated matrix.
\end{proof}

\subsubsection{An exponential estimate on the good event}
Let $G_{g,\Lambda}(x,y;z)$ denote the integral kernel of the finite-volume
resolvent of $L_{g,\varpi}$. Let $\gamma_0>0$ be the exponential decay
rate of $K$, which may not be the optimal one. As a consequence, the following constant is finite:
\begin{equation}
 T_1=\sum_{r\ne0}|K_{0r}|\,|r|_1e^{\gamma_0|r|_1/2}.
 \label{eq:CT-constants}
\end{equation}
For $0<\nu\le\gamma_0/2$, conjugating by
$e^{\nu|x-y|_1}$ changes the operator by norm at most $\nu T_1$,
by the row and column sum bounds. The diagonal random potential is
unchanged by this conjugation.

On the good event $\{\lambda_{\min}(L_{g,\varpi,\Lambda_L})>e_L\}$, for
$|\mathfrak e|\le e_L/2$ the resolvent norm is at most $2/e_L$,
uniformly in $\eta\ne0$. Choose
\[
 b=\min\{\gamma_0/2,\,[4(T_1+1)]^{-1}\},\qquad \nu=be_L,
\]
and take $L$ large enough that $e_L\le1$. The conjugation error is
at most $e_L/4$, so a Neumann series proves
\begin{equation}
 |G_{g,\Lambda_L}(x,y;\mathfrak e+i\eta)|
 \le\frac4{e_L}e^{-be_L|x-y|_1}
 \quad\text{on the good event}.
 \label{eq:CT}
\end{equation}
This argument is valid for exponentially decaying infinite-range hopping.

Fix $\vartheta=1/8$ and $t=1/2$ for definiteness; any $0<\vartheta<1/4$ and
$\vartheta<t<1$ would work. Let $N_L=(2L+1)^d$. H\"older's inequality on
the bad event $\{\lambda_{\min}(L_{g,\varpi,\Lambda_L})\leq e_L\}$ and Lemma~\ref{lem:Wegner} show that, when
$|u|_\infty\ge L/2$,
\begin{align}
 \EE|G_{g,\Lambda_L}(0,u;\mathfrak e+i\eta)|^\vartheta
 &\le 4^\vartheta e_L^{-\vartheta}e^{-\vartheta b e_LL/2}
 +D_t(N_L,g_-)^{\vartheta/t}p_L^{1-\vartheta/t}\nonumber\\
 &=:\mathcal M_L.
 \label{eq:outer-moment}
\end{align}
This bound is uniform in $g\in[g_-,g_+]$, $|\mathfrak e|\le e_L/2$,
and $\eta\ne0$. Every factor outside the exponentials grows at most
polynomially in $L$, while $e_LL=L^{1-\beta}$ and
\eqref{eq:initial-prob} give stretched-exponential decay.

Put $\tau_r=|K_{0r}|$ for $r\ne0$ and
$\mathfrak X_\vartheta(\Lambda)=
\sum_{u\in\Lambda,\,v\notin\Lambda}\tau_{v-u}^{\vartheta}$.
Exponential hopping gives constants $C_\vartheta',a_\vartheta>0$ such that
\begin{equation}
 \mathfrak X_\vartheta(\Lambda_L)\le C_\vartheta'L^{d-1},\qquad
 \sup_{|u|_\infty<L/2}\sum_{v\notin\Lambda_L}\tau_{v-u}^\vartheta
 \le C_\vartheta'e^{-a_{\vartheta}L}.
 \label{eq:boundary-weights}
\end{equation}
For the first bound, sum the exponential tail according to the integer
distance of $u$ from the complement; each layer has at most
$C_dL^{d-1}$ sites. For the second, every crossing displacement has
length at least $L/2$; polynomial lattice multiplicities can be absorbed
by decreasing the exponential rate.

\subsection{Uniform estimates for truncated matrices} \label{sec:uniformest}
We use \cite[Theorem~3.1]{ASFH}, whose criterion is imposed on
every subset of a testing cube. This form yields bounds uniform
over all truncated matrices, including near their boundaries.

\begin{lemma}\label{lem:all-restrictions-FM}
Fix $[g_-,g_+]\Subset(0,\infty)$. There are $r>0$, $B<\infty$,
$\mu>0$, and $\vartheta=1/8$ such that
\begin{equation}
 \EE\left|\langle\delta_x,
 (K_\Lambda+g\Xi_{\varpi,\Lambda}-\mathfrak e-i\eta)^{-1}
 \delta_y\rangle\right|^\vartheta
 \le Be^{-\mu|x-y|_1}
 \label{eq:all-restrictions-FM}
\end{equation}
for every finite $\Lambda$, $x,y\in\Lambda$, $g\in[g_-,g_+]$,
$|\mathfrak e|\le r$, and $\eta\ne0$. It also holds on the infinite lattice.
\end{lemma}
\begin{proof}
Use Proposition~\ref{prop:initial} with $\beta=1/2$. Thus
$e_L=L^{-1/2}$, $\ell_L=\lfloor L^{1/8}\rfloor$, and the bad-event
probability is at most $CL^d e^{-cL^{d/8}}$. If $V\subset\Lambda_L$,
the matrix $K_V+g\Xi_{\varpi,V}$ is a compression of
$K_{\Lambda_L}+g\Xi_{\varpi,\Lambda_L}$. Its lowest eigenvalue is
therefore at least that of the full cube. A good event for the cube
is consequently good for every $V$ simultaneously. This observation
does not require independence of the events for different subsets.

The proof of \eqref{eq:CT} applies to each such $V$: its hopping row
and column sums are bounded by those on the full lattice. On the bad
event, \eqref{eq:norm-moment} applies with $|V|\le N_L=(2L+1)^d$.
H\"older's inequality, with $t=1/2$ and $\vartheta=1/8$, gives
\begin{equation}
 \sup_{\substack{V\subset\Lambda_L,\ 0,u\in V\\ |u|_\infty\ge L/2}}
 \EE|G_{g,V}(0,u;\mathfrak e+i\eta)|^\vartheta
 \le C L^{b_0}
       \bigl(e^{-cL^{1/2}}+e^{-cL^{d/8}}\bigr)=:\mathcal N_L
 \label{eq:depleted-outer}
\end{equation}
for $|\mathfrak e|\le e_L/2$, uniformly in $g$ and $\eta\ne0$.
Here $b_0$ is a fixed finite exponent. Set a matrix element equal to
zero when either of its indices does not belong to $V$.

Let $\tau_r=|K_{0r}|$ for $r\ne0$ and let
$\mathfrak X_\vartheta(\Lambda_L)$ be as in Section~\ref{sec:block}.
The scalar version of \cite[Theorem~3.1]{ASFH} requires
\begin{equation}
 \left(1+C_{\vartheta}g^{-\vartheta}\mathfrak X_\vartheta(\Lambda_L)\right)
 \sup_{V\subset\Lambda_L}
 \sum_{\substack{u\in\Lambda_L\\v\notin\Lambda_L}}
 \tau_{v-u}^\vartheta\EE|G_{g,V}(0,u;z)|^\vartheta<1.
 \label{eq:all-subsets-criterion}
\end{equation}
The single-site regularity and decoupling assumptions, labelled R1
and R2 in \cite{ASFH}, hold for the uniform law: R1 has exponent
one, and R2 holds for $\vartheta<1/4$ by \cite[Lemma~C.3]{ASFH}.
Exponential summability and translation invariance give the remaining
hypotheses; the deterministic diagonal is absorbed into the energy.
For the inner half-cube use \eqref{eq:norm-moment} and the second
estimate in \eqref{eq:boundary-weights}; for the outer part use
\eqref{eq:depleted-outer}. The left side of
\eqref{eq:all-subsets-criterion} is bounded by
\[
 \left(1+C_{\vartheta}g_-^{-\vartheta}\mathfrak X_\vartheta(\Lambda_L)\right)
 \left[C_\vartheta'N_LD_\vartheta(N_L,g_-)e^{-a_{\vartheta}L}
              +\mathfrak X_\vartheta(\Lambda_L)\mathcal N_L\right].
\]
It tends to zero. Choose one $L_*$ for which it is at most $1/2$.
The criterion gives the conclusion uniformly over all restrictions.
The common strict margin, the common exponential hopping bound, and the
bounded range of $g^{-1}$ give common $B,\mu$. One may take
$r=e_{L_*}/4$. Strong resolvent convergence and Fatou's lemma give
the infinite-volume estimate with the same constants.
\end{proof}

Take $g_-=Wc_M/2$, $g_+=Wc_M$, and choose
$0<\Delta_W<\min\{c_M/2,r\}$ using this lemma. Set
$I_W=(c_M-\Delta_W,c_M)$ and put
\begin{equation}
 \mathscr A_{\varpi,\Lambda}(E)
 :=E M_{\varpi,\Lambda}-J_\Lambda C_\Lambda^tJ_\Lambda
 =K_\Lambda+EW\Xi_{\varpi,\Lambda}-(c_M-E)\Id,
 \label{eq:pencil-localisation}
\end{equation}
where $J_\Lambda$ is the restriction of the staggering operator
\eqref{eq:K}. For every fixed $E\in I_W$, the finite
matrix is invertible almost surely. This follows also from
\eqref{eq:Wegner}, applied with $g=EW$. Taking $\eta\to0$ in
\eqref{eq:all-restrictions-FM} and using Fatou's lemma gives
\begin{equation}
 \EE|\mathscr A_{\varpi,\Lambda}(E)^{-1}(x,y)|^\vartheta
 \le Be^{-\mu|x-y|_1}.
 \label{eq:real-pencil-FM}
\end{equation}
In infinite volume write
$\mathscr A_\varpi(E)=EM_\varpi-JCJ$ and $\check H_\varpi=JH_\varpi J$.
Then
\begin{equation}
 \check H_\varpi-E-i\eta
 =-M_\varpi^{-1/2}
      (\mathscr A_\varpi(E)+i\eta M_\varpi)M_\varpi^{-1/2}.
 \label{eq:complex-pencil}
\end{equation}
The random imaginary part prevents direct identification with the
additive resolvent. Section~\ref{sec:localisation} instead transfers
the fixed-real-energy estimate \eqref{eq:real-pencil-FM} by averaging
the actual material spectral measure.

\section{Spectral and dynamical localisation}\label{sec:localisation}

We convert the real-energy estimates of Section~\ref{sec:edge} into eigenfunction-correlator bounds and deduce spectral and dynamical localisation. The proof combines spectral averaging in one material variable with independent resampling, following the rank-one analysis of \cite{SW} and the fractional-moment approach of  \cite{AM,ASFH,Stolz}.


\subsection{The localisation theorem}

We state the result with correlator bounds uniform over finite truncations, which will also be used in the finite-sample analysis.

\begin{theorem}[Localisation at every positive disorder]
\label{thm:localisation}\label{thm:edge-main}
Assume the geometry of Section~\ref{sec:model}, including
\eqref{eq:reflection}, and the law \eqref{eq:random-model}.
For each prescribed $W>0$, use the interval $I_W$ chosen in
Section~\ref{sec:edge}. There are constants
$A_W<\infty$ and $\nu_W>0$ such that
\begin{equation}
 \EE Q_{\varpi,\Lambda}^t(x,y;I_W)
 \le A_We^{-\nu_W|x-y|_1},\qquad
 I_W=(c_M-\Delta_W,c_M),
 \label{eq:Q-edge}
\end{equation}
uniformly in all finite $\Lambda\subset\Z^d$ and $x,y\in\Lambda$.
The same bound holds for $Q_\varpi$ on the infinite lattice.
Almost surely, the spectrum of $H_\varpi$ in $I_W$ is pure point,
its eigenvalues are simple, and every corresponding eigenfunction
satisfies
\begin{equation}
 |\psi(n)|\le C_{\varpi,\psi,y}
       e^{-\nu|n-y|_1}
 \quad(n\in\Z^d)
 \label{eq:eigenfunction-exponential}
\end{equation}
for every $0<\nu<\nu_W$ and any $y$ with $\psi(y)\ne0$.
The constant in \eqref{eq:eigenfunction-exponential} is allowed to
depend on the eigenfunction. In particular,
\begin{equation}
 \EE\sup_{t\in\R}
 \left|\langle\delta_x,e^{-itH_\varpi}
                  \one_{I_W}(H_\varpi)\delta_y\rangle\right|
 \le A_We^{-\nu_W|x-y|_1}.
 \label{eq:dynamical-edge}
\end{equation}
\end{theorem}

The constants and the interval depend on the fixed value of $W$.
Unitary equivalence \eqref{eq:unitary} gives the same spectral conclusion
for $A_\varpi$ on $\mathcal H_\varpi$. The correlator bound also controls
the reduced wave evolution $\cos(t\sqrt{H_\varpi})\one_{I_W}(H_\varpi)$.

\subsection{Spectral averaging for a material variable}
The following spectral-averaging lemma is independent of translation invariance and of the decay of the capacitance coefficients. It will be used to prove Theorem~\ref{thm:edge-main}.

\begin{lemma}[Material-variable averaging]\label{lem:material-average}
Let $C_*$ be a bounded nonnegative self-adjoint operator on a finite
or countable coordinate space. Fix all inverse material factors except
$m_y=t\in[1,1+W]$, and set
$H_t=M_t^{-1/2}C_*M_t^{-1/2}$. Denote its spectral measure at
$\delta_y$ by $\nu_y^t$. If $I\subset[E_-,\infty)$, $E_->0$, then
for every nonnegative Borel function $F$, independent of $t$,
\begin{equation}
 \frac1W\int_1^{1+W}\int_I F(E)\,\nu_y^t(\dd E)\dd t
 \le \kappa_{W,I}\int_I F(E)\dd E,
 \qquad \kappa_{W,I}:=\frac{1+W}{WE_-}.
 \label{eq:material-average}
\end{equation}
\end{lemma}
\begin{proof}
Put $b=t^{-1}$, $T_t=C_*^{1/2}M_t^{-1/2}$ and
$\phi_y=C_*^{1/2}\delta_y$. Then
\[
 H_t=T_t^*T_t,\qquad
 B_b:=T_tT_t^*=B_0+b|\phi_y\rangle\langle\phi_y|,
\]
where $B_0$ is independent of $b$. Let $\mu_{\phi_y}^b$ be the
spectral measure of $B_b$ at $\phi_y$. Intertwining the functional
calculi of $T_t^*T_t$ and $T_tT_t^*$ gives
\begin{equation}
 b\,\mu_{\phi_y}^b(\dd E)=E\,\nu_y^t(\dd E).
 \label{eq:measure-intertwining}
\end{equation}
Indeed $T_t\delta_y=\sqrt b\,\phi_y$ and
$T_t^*\one_S(B_b)T_t=H_t\one_S(H_t)$ for Borel $S$.
No inverse of $C_*$ is used.

For completeness, rank-one spectral averaging gives
$\int_\R\mu_{\phi_y}^b(\dd E)\dd b=\dd E$ when $\phi_y\ne0$.
If $F_0(z)=\langle\phi_y,(B_0-z)^{-1}\phi_y\rangle$, its resolvent
formula is
\[
 F_b(z)=\frac{F_0(z)}{1+bF_0(z)}
       =\frac1{b+F_0(z)^{-1}}.
\]
For $\Im z>0$, $\Im F_0(z)>0$, and integration of this Poisson
kernel over $b\in\R$ gives $\int_\R\Im F_b(z)\dd b=\pi$.
The spectral-measure boundary formula then proves the averaging identity.
For $\phi_y=0$, the spectral measure of $H_t$ at $y$ is supported at
zero and \eqref{eq:material-average} is immediate.

The law of $b$ has density $(Wb^2)^{-1}$ on $[(1+W)^{-1},1]$.
Using \eqref{eq:measure-intertwining}, the integrand is weighted by
$(WbE)^{-1}\le(1+W)/(WE_-)$. Extending the nonnegative integral
in $b$ to the real line proves the assertion. The argument applies to
unbounded nonnegative $F$ by monotone convergence.
\end{proof}

\subsection{Resampling and finite-volume eigenfunction correlators} \label{sec:6.3}
We work with $\check H_{\varpi,\Lambda}=J_\Lambda
H_{\varpi,\Lambda}^tJ_\Lambda$. Staggering does not change any of
the absolute values in \eqref{eq:Q-definition}. Fix $y$ and replace $m_y$ by an independent copy $m_y'$ with the
same law. A superscript $(y)$ denotes the resulting reference operator.

Suppose $\check H_{\varpi,\Lambda}\psi_j=E_j\psi_j$ and put
$a_j=M_{\varpi,\Lambda}^{-1/2}\psi_j$. Then
\begin{equation}
 \mathscr A_\Lambda^{(y)}(E_j)a_j
 =E_j(m_y'-m_y)a_j(y)\delta_y.
 \label{eq:resampling-identity}
\end{equation}
For fixed other variables and fixed $m_y'$, the reference
spectrum is a finite set independent of $m_y$. Lemma~\ref{lem:material-average}
implies that $\nu_y^{m_y}$ assigns this set zero mass for almost every
$m_y$. Thus, for every eigenvalue with nonzero spectral mass at $y$, the reference
matrix is invertible, except on a probability-zero event. At such an
energy, \eqref{eq:resampling-identity} also proves simplicity: an
eigenvector vanishing at $y$ must be zero. Applying this observation
at every site shows that all positive eigenvalues of the finite random
matrix are almost surely simple.

Let $E_-:=\inf I_W>0$, and set
$d_W:=c_MW\sqrt{1+W}$. Since $E_j\le c_M$ and
$1\le m_n\le1+W$, equation \eqref{eq:resampling-identity} gives
\begin{equation}
 |\psi_j(x)|\le d_W|\psi_j(y)|
       |\mathscr A_\Lambda^{(y)}(E_j)^{-1}(x,y)|.
 \label{eq:resampling-ratio}
\end{equation}
Define
\[
 Q_{\varpi,\Lambda,\vartheta}^t(x,y;I_W)
 :=\sum_{E_j\in I_W}|\psi_j(y)|^{2-\vartheta}|\psi_j(x)|^\vartheta.
\]
The eigenvectors are normalised in ordinary $\ell^2(\Lambda)$.
Raising \eqref{eq:resampling-ratio} to power $\vartheta$ and using
$\nu_y^{m_y}(\{E_j\})=|\psi_j(y)|^2$ yields
\begin{equation}
 Q_{\varpi,\Lambda,\vartheta}^t(x,y;I_W)
 \le d_W^\vartheta\int_{I_W}
 |\mathscr A_\Lambda^{(y)}(E)^{-1}(x,y)|^\vartheta
       \nu_y^{m_y}(\dd E).
 \label{eq:Q-fractional}
\end{equation}
The integrand is independent of the original $m_y$. Average first in
$m_y$ using \eqref{eq:material-average}, and then in the reference
configuration. The latter still has the original iid law, so
\eqref{eq:real-pencil-FM} gives
\begin{equation}
 \EE Q_{\varpi,\Lambda,\vartheta}^t(x,y;I_W)
 \le \kappa_{W,I_W}d_W^\vartheta|I_W|B e^{-\mu|x-y|_1}.
 \label{eq:Q-fractional-bound}
\end{equation}
All exceptional reference energies have zero original spectral mass by
the same averaging lemma. This justifies its use even though the
eigenvalues $E_j$ depend on $m_y$.

For $a,b\ge0$ and $0<\vartheta<1$,
$ab\le a^{\vartheta}b^{2-\vartheta}+a^{2-\vartheta}b^\vartheta$. The finite eigenvalues are simple,
and hence
\begin{align}
 Q_{\varpi,\Lambda}^t(x,y;I_W)
 &=\sum_{E_j\in I_W}|\psi_j(x)\psi_j(y)|\nonumber\\
 &\le Q_{\varpi,\Lambda,\vartheta}^t(x,y;I_W)
       +Q_{\varpi,\Lambda,\vartheta}^t(y,x;I_W).
 \label{eq:Q-from-fractional}
\end{align}
This proves \eqref{eq:Q-edge}, for example with
$A_W=2\kappa_{W,I_W}d_W^\vartheta|I_W|B$ and $\nu_W=\mu$.

\subsection{A range criterion}
The following lemma converts fractional-moment bounds for resolvent columns into almost-sure solvability of the corresponding operator equation.

\begin{lemma}\label{lem:range}
Let $A$ be a bounded self-adjoint random operator on $\ell^2(\Z^d)$.
If, for a fixed $y$ and some $\vartheta\in(0,2)$,
\[
 \sup_{0<\eta\le1}\EE\sum_x
 |\langle\delta_x,(A-i\eta)^{-1}\delta_y\rangle|^\vartheta<\infty,
\]
then $\delta_y\in\operatorname{Ran}A$ almost surely.
\end{lemma}
\begin{proof}
Let $S_k$ be the sum with $\eta=1/k$. Fatou's lemma implies
$\liminf_kS_k<\infty$ almost surely. Choose a subsequence with
$S_{k_j}\le C(\varpi)<\infty$. Since
$\sum_x|b_x|^2\le(\sum_x|b_x|^\vartheta)^{2/\vartheta}$, the vectors
$u_j=(A-i/k_j)^{-1}\delta_y$ are bounded in $\ell^2$. A subsequence
converges weakly to $u$. The identity
$Au_j=\delta_y+(i/k_j)u_j$ and boundedness of $A$ imply
$Au=\delta_y$. No measurable choice of the subsequence or of $u$ is
needed for this pointwise conclusion.
\end{proof}
\subsection{Infinite-volume spectrum and dynamics}

We complete the proof of Theorem~\ref{thm:edge-main} by passing the correlator estimates to infinite volume and establishing almost-sure simplicity of the eigenvalues.

As the cubes \(\Lambda_L\) increase to \(\mathbb Z^d\), the extended matrices
$P_LH_\varpi P_L$ converge strongly to $H_\varpi$ and are uniformly
bounded. Therefore $f(P_LH_\varpi P_L)\to f(H_\varpi)$ strongly
for continuous $f$. If $f$ has compact support in $I_W$, then $f(0)=0$,
and the former operator is the zero extension of $f(H_{\varpi,\Lambda_L}^t)$.
Take successively a maximum over finitely many functions in a countable
dense family in the unit ball of $C_0(I_W)$. Fatou's lemma followed
by monotone convergence proves the infinite-volume form of
\eqref{eq:Q-edge}. Regularity of finite complex spectral measures
identifies this supremum with the total variation on $I_W$, so it also
controls all bounded Borel functions in \eqref{eq:Q-definition}.

For $0<\nu<\nu_W$, the bound implies
\begin{equation}
 \EE\sum_{x\in\Z^d}e^{\nu|x-y|_1}Q_\varpi(x,y;I_W)<\infty.
 \label{eq:weighted-Q}
\end{equation}
Thus the sum is finite almost surely, simultaneously for all $y$ and,
by a countable choice of weights, for every $\nu<\nu_W$.
Since $Q_\varpi\le1$, this gives uniformly vanishing spatial tails
for the orbit
$e^{-itH_\varpi}\one_{I_W}(H_\varpi)\delta_y$, $t\in\R$.
Every such orbit is precompact in $\ell^2$. The spectral theorem
(equivalently, the RAGE criterion) implies that its initial vector
belongs to the pure point subspace; see \cite{Stolz,Teschl}.
The coordinate vectors are total, so the spectrum in $I_W$ is pure point.
Choosing $f(E)=e^{-itE}$ proves \eqref{eq:dynamical-edge}.

We give the additional argument that rules out multiplicities and
therefore proves \eqref{eq:eigenfunction-exponential} for every
eigenfunction. In infinite volume, the weighted version of
\eqref{eq:all-restrictions-FM} gives, for each fixed $E\in I_W$ and $y$,
\[
 \sup_{\eta>0}\EE\sum_x e^{\nu_0|x-y|_1}
 |\langle\delta_x,(\mathscr A_\varpi(E)-i\eta)^{-1}
                          \delta_y\rangle|^\vartheta<\infty
\]
for $0<\nu_0<\mu$. Fatou's lemma, followed by the weak-compactness
argument in Lemma~\ref{lem:range}, produces a solution of
$\mathscr A_\varpi(E)u=\delta_y$ in $\ell^2$. Taking limits of the
coordinates also preserves the finite weighted $\ell^\vartheta$ sum.
By Fubini, for almost every reference configuration this holds for
Lebesgue-almost every $E$. Also exclude the countable set of eigenvalues of the fixed reference
operator $\check H_{\varpi}^{(y)}$, using the resampling notation
of Section~\ref{sec:6.3}. Indeed,
\[
\mathcal A_{\varpi}^{(y)}(E)
=
\bigl(M_{\varpi}^{(y)}\bigr)^{1/2}
\bigl(E\,\mathrm{Id}-\check H_{\varpi}^{(y)}\bigr)
\bigl(M_{\varpi}^{(y)}\bigr)^{1/2}.
\]
Since $M_{\varpi}^{(y)}$ is bounded and boundedly invertible,
$\mathcal A_{\varpi}^{(y)}(E)$ is injective outside this countable
set, and the solution constructed above is unique.
This additional exceptional set depends only on the reference
configuration and is independent of the original material
variable $m_y$.

The sets in this argument are measurable: 
solvability of the reference equation and its solution can be characterised through limits of resolvent columns, while nontriviality of the kernel is detected by spectral atoms.

Fix the outside variables and a reference value $m_y'$ with
these properties. Its exceptional energy set is Lebesgue-null and
independent of the original $m_y$. The infinite-volume version of
Lemma~\ref{lem:material-average} makes its original spectral mass at $y$
zero for almost every $m_y$. At any eigenvalue with nonzero spectral mass at $y$,
\eqref{eq:resampling-identity} holds with an injective reference operator.
An eigenvector with zero $y$ coordinate must therefore vanish. The
evaluation map at $y$ is injective on the eigenspace, whose dimension
is at most one. Fubini and a countable intersection over $y$ prove
simplicity almost surely.

Finally, at a simple eigenvalue $E_j$, its spectral atom satisfies
\[
 |\psi_j(x)\psi_j(y)|
 =|\langle\delta_x,\one_{\{E_j\}}(H_\varpi)\delta_y\rangle|
 \le Q_\varpi(x,y;I_W).
\]
Equation \eqref{eq:weighted-Q} bounds the right side by a finite
random constant times $e^{-\nu|x-y|_1}$. Division by
$|\psi_j(y)|>0$ completes the proof of Theorem~\ref{thm:localisation}.

\begin{corollary}[The large-disorder interval]\label{cor:strong-localisation}
For the uniform inverse-material law, the conclusions of
Theorem~\ref{thm:localisation} hold on the interval $I$ of
Theorem~\ref{thm:strong-main} whenever its large-disorder condition
holds. They also hold under either of the sufficient tests in
Theorem~\ref{thm:FM}, with this same law.
\end{corollary}
\begin{proof}
In finite volume the real inverse operator is related to the boundary
value of the resolvent of $H_{\varpi,\Lambda}^t$ by
\[
 (C_\Lambda^t-EM_{\varpi,\Lambda})^{-1}
 =M_{\varpi,\Lambda}^{-1/2}
 (H_{\varpi,\Lambda}^t-E)^{-1}
 M_{\varpi,\Lambda}^{-1/2}.
\]
For each fixed $E>0$ the inverse exists almost surely by
Lemma~\ref{lem:material-average}. Fatou's lemma and
\eqref{eq:strong-main-bound}, or \eqref{eq:weighted-F}, give the
analogue of \eqref{eq:real-pencil-FM}; the factors $m_n^{-1/2}$
are at most one. The resampling and limiting arguments just proved
apply with $E_-:=\inf I$ and $d_W:=(\sup I)W\sqrt{1+W}$.
Simplicity in infinite volume can alternatively be obtained from the
range argument for the resolvent of $H_\varpi$ and the same averaging lemma.
\end{proof}

\section{An explicit edge window and mobility-edge bounds}
\label{sec:mobility}\label{app:mobility}

We refine the band-edge argument to obtain an explicit localisation window whose energy width is proportional to the disorder strength at weak disorder. This gives an upper bound on the upper endpoint of any absolutely continuous spectral component.

\subsection{Spectral and mean-material edges}
Note that the exact upper spectral edge of the disordered operator is $c_M$. 
We use the unique quadratic maximum established in Section~\ref{sec:deterministic}:


\begin{equation}
 c(\alpha)=c_M\ \Longleftrightarrow\ \alpha=\alpha_M,
 \qquad \mathsf A_M=-\tfrac12D_\alpha^2c(\alpha_M)>0.
 \label{im:eq:app-geometry}
\end{equation}
The equivalence is on the Brillouin torus. Put
\begin{equation}
 E_{\rm top}=c_M,\qquad
 E_a(W)=\frac{c_M}{1+aW},\qquad
 E_{\rm mean}(W)=\frac{c_M}{1+W/2},\qquad 0<a<\tfrac12.
 \label{im:eq:app-edges}
\end{equation}
Here $E_{\rm mean}=c_M/\mu_W$ is the upper edge of $C/\mu_W$. The threshold $E_a(W)$ is a sufficient localisation threshold,
not a definition of a mobility edge. Frequency statements use
$\omega_M$ from \eqref{eq:normalised-frequency}, with $\delta$ and
$\varepsilon$ fixed while varying $W$.
The mean-material edge frequency is $\omega_M/\sqrt{\mu_W}$.

\subsection{The quantitative theorem}

The following theorem makes the localisation interval explicit, while allowing the decay constants to depend on the fixed disorder strength.

\begin{theorem}[A localisation window proportional to the disorder]
\label{im:thm:app-linear-window}
Assume the geometry of Section~\ref{sec:model}, including
\eqref{eq:reflection}, and independent uniform
$\xi_n\in[0,1]$. For each fixed $a\in(0,1/2)$ there is $W_a>0$
such that, for every prescribed $0<W<W_a$, the interval
\begin{equation}
 I_{W,a}=(E_a(W),c_M),\qquad
 \Delta_a(W)=c_M-E_a(W)=\frac{a c_MW}{1+aW}
 \label{im:eq:app-window}
\end{equation}
satisfies
\begin{equation}
 \EE Q_{\varpi,\Lambda}^t(x,y;I_{W,a})
 \le A_{a,W}e^{-\nu_{a,W}|x-y|_1},
 \qquad A_{a,W}<\infty,\quad\nu_{a,W}>0,
 \label{im:eq:app-Q-bound}
\end{equation}
uniformly in all finite $\Lambda$ and $x,y\in\Lambda$. The same
bound holds for $Q_\varpi$. Almost surely the spectral subspace of
$H_\varpi$ in $I_{W,a}$ is pure point and has an orthonormal basis
of exponentially decaying eigenfunctions. Moreover,
\begin{equation}
 \EE\sup_{t\in\R}
 \left|\ip{\delta_x}{e^{-itH_\varpi}
                     \one_{I_{W,a}}(H_\varpi)\delta_y}\right|
 \le A_{a,W}e^{-\nu_{a,W}|x-y|_1}.
 \label{im:eq:app-dynamics}
\end{equation}
In frequency variables the proved interval is
\begin{equation}
 \frac{\omega_M}{\sqrt{1+aW}}<\omega<\omega_M.
 \label{im:eq:app-frequency-window}
\end{equation}
\end{theorem}

The constants depend on the fixed $a$ and $W$; no uniformity as
$a\uparrow1/2$ or $W\downarrow0$ is asserted. The interval is sufficient
for localisation and need not be maximal. 
The proof builds on the spectral-averaging and resolvent estimates of Subsection~\ref{sec:block} and the finite-volume fractional-moment criterion stated in Subsection~\ref{sec:uniformest}.
Its energy width is $\Delta_a(W)\sim a c_MW$, whereas its frequency
width is $\omega_M[1-(1+aW)^{-1/2}]\sim(a/2)\omega_MW$.

\subsection{The deficit comparison}
With the notation of \eqref{eq:K}--\eqref{eq:Lap}, the comparison
\eqref{eq:power-bound} reads
\begin{equation}
 K=c_M\Id-JCJ\ge a_{\rm L}\Lap\ge0.
 \label{im:eq:app-comparison}
\end{equation}
Compression preserves this quadratic-form inequality. The off-diagonal
entries of $K$ have the same absolute values as those of $C$. No sign
assumption on those entries is used below.

\subsection{A uniform low-energy probability estimate at weak coupling}
Consider the additive family in \eqref{eq:additive}, with
\begin{equation}
 L_{g,\varpi}=K+g\Xi_\varpi,
 \qquad g\in[g_0,2g_0],
 \qquad g_0>0.
 \label{im:eq:app-additive}
\end{equation}
For finite $V$, $L_{g,\varpi,V}$ always denotes its
compression. Fix $a<b<1/2$, and put
$\eta_b=(1/2-b)/2>0$. Choose a constant $q_b>0$ such that
\begin{equation}
 \frac{a_{\rm L}}{q_b^2}\ge1+\frac1{4\eta_b},\qquad
 \ell_0=\lfloor q_bg_0^{-1/2}\rfloor,
 \qquad L=\lceil g_0^{-2}\rceil.
 \label{im:eq:app-lengths}
\end{equation}
For sufficiently small $g_0$, $\ell_0\ge2$ and
$L\gg\ell_0$. Partition the coordinate intervals of
$\Lambda_L=[-L,L]^d\cap\Z^d$ into integer intervals of lengths
between $\ell_0$ and $2\ell_0$, merging a short remainder if
necessary. Their products form blocks $B$.

\begin{lemma}[Rare low eigenvalues]\label{im:lem:app-probability}
There are constants $C_b,c_b>0$, independent of sufficiently small
$g_0$, such that
\begin{equation}
 p_{g_0}:=
 \PP\{\lambda_{\min}(L_{g_0,\varpi,\Lambda_L})
                                     \le bg_0\}
 \le C_b L^d\exp(-c_bg_0^{-d/2}).
 \label{im:eq:app-probability}
\end{equation}
On the complementary event, simultaneously for every
$g\in[g_0,2g_0]$ and every nonempty
$V\subset\Lambda_L$, one has
$\lambda_{\min}(L_{g,\varpi,V})>bg_0$.
\end{lemma}
\begin{proof}
Compression of \eqref{im:eq:app-comparison}, followed by deletion of
nonnegative nearest-neighbour edge energies between blocks and at
the outer boundary, gives
\begin{equation}
 L_{g_0,\varpi,\Lambda_L}
 \ge\bigoplus_B(a_{\rm L}\mathsf L_B^N+g_0\Xi_{\varpi,B}).
 \label{im:eq:app-bracketing}
\end{equation}
Here $\mathsf L_B^N$ is the graph Neumann Laplacian. For side
lengths $N_j\in[\ell_0,2\ell_0]$, its first positive eigenvalue is
$\min_j(2-2\cos(\pi/N_j))\ge\ell_0^{-2}$.
Thus $T_B=a_{\rm L}\mathsf L_B^N+g_0\Xi_{\varpi,B}$
has second eigenvalue at least
\[
 \gamma_B=a_{\rm L}\ell_0^{-2}
       \ge(1+(4\eta_b)^{-1})g_0.
\]
In the normalised constant vector its mean and variance satisfy
\[
 m_B=g_0\overline\xi_B\le g_0,
 \qquad v_B=g_0^2
          (\overline{\xi^2}_B-(\overline\xi_B)^2)
                      \le g_0^2/4.
\]
The last inequality holds for every collection of numbers in $[0,1]$.
Temple's inequality, as used in \eqref{eq:Temple}, gives
\begin{equation}
 \lambda_{\min}(T_B)
 \ge m_B-\frac{v_B}{\gamma_B-m_B}
 \ge g_0\overline\xi_B-\eta_bg_0.
 \label{im:eq:app-Temple}
\end{equation}
If $\lambda_{\min}(T_B)\le bg_0$, then
$\overline\xi_B\le b+\eta_b=1/2-\eta_b$. Independence gives
\begin{equation}
 \PP\{\overline\xi_B\le1/2-\eta_b\}
                       \le e^{-2\eta_b^2|B|}.
 \label{im:eq:app-Hoeffding}
\end{equation}
One elementary verification uses
$\EE e^{-t(\xi_n-1/2)}\le\cosh(t/2)\le e^{t^2/8}$;
exponential Markov inequality with $t=4\eta_b$ gives
\eqref{im:eq:app-Hoeffding}. For small $g_0$,
$\ell_0\ge(q_b/2)g_0^{-1/2}$. Hence every block has
at least $(q_b/2)^dg_0^{-d/2}$ sites. There are at
most $(2L+1)^d$ blocks. A union bound in
\eqref{im:eq:app-bracketing} proves \eqref{im:eq:app-probability}.

Finally, increasing $g$ increases the operator, and
compressing a positive lower form bound preserves it. The good
event for the full cube at $g_0$ is therefore good for
all its subsets and all larger couplings simultaneously. No union
bound over the $2^{|\Lambda_L|}$ subsets is taken.
\end{proof}

\subsection{Verification of a fractional-moment criterion}
\label{im:subsec:app-FM}
We give the details needed for uniformity in the coupling and
spectral parameters. Lemma~\ref{lem:Wegner} gives, for any finite
$V$, $N=|V|$, any interval $B_0\subset\R$, and any $g>0$,
\begin{equation}
 \EE\operatorname{Tr}\one_{B_0}(L_{g,\varpi,V})
                  \le N|B_0|/g.
 \label{im:eq:app-Wegner}
\end{equation}
It implies, for $0<p<1$, $u\in\R$, and $\eta\ne0$,
\begin{equation}
 \EE\|(L_{g,\varpi,V}-u-i\eta)^{-1}\|^p
 \le D_p(N,g):=\frac{(2N/g)^p}{1-p}.
 \label{im:eq:app-norm-moment}
\end{equation}
Let $G_{g,V}(x,y;z)$ denote the resolvent kernel, and use the
constant $T_1$ from \eqref{eq:CT-constants}.
For $0<\nu\le\gamma_0/2$, conjugation by $e^{\nu|x-y|_1}$
changes the operator by norm at most $\nu T_1$, by the Schur
test. The diagonal random potential is unchanged. On the good
event of Lemma~\ref{im:lem:app-probability}, the distance to every
real $u\in[0,ag_0]$ is at least
$(b-a)g_0$. Choosing
\[
 \nu=\min\{\gamma_0/2,(b-a)/(2(T_1+1))\}g_0,
 \qquad g_0\le1,
\]
a Neumann series for the conjugated resolvent gives, uniformly
over all $V\subset\Lambda_L$, couplings in
$[g_0,2g_0]$, and $\eta\ne0$,
\begin{equation}
 |G_{g,V}(x,y;u+i\eta)|
 \le\frac{2}{(b-a)g_0}
                 e^{-cg_0|x-y|_1}
 \quad\hbox{on the good event}.
 \label{im:eq:app-CT}
\end{equation}
Take $\vartheta=1/8$ and $p=1/2$. For $|y|_\infty\ge L/2$,
H\"older's inequality on the bad event and
\eqref{im:eq:app-norm-moment} imply
\begin{align}
 \sup_{V\subset\Lambda_L}\EE|G_{g,V}(0,y;u+i\eta)|^\vartheta
 &\le Cg_0^{-\vartheta}e^{-c \vartheta g_0L}
   +D_p(N_L,g_0)^{\vartheta/p}
                          p_{g_0}^{1-\vartheta/p}
   =:\mathcal M(g_0).
 \label{im:eq:app-outer-moment}
\end{align}
A matrix element is defined to be zero if an indicated site is
not in $V$.

Use $\tau_z$ and $\mathfrak X_\vartheta$ from
Subsection~\ref{sec:block}.
For the uniform law, the scalar specialisation of
\cite[Theorem~3.1]{ASFH} states that the strict bound
\begin{equation}
 \left(1+C_\vartheta g^{-\vartheta}\mathfrak X_\vartheta(\Lambda_L)\right)
 \sup_{V\subset\Lambda_L}
 \sum_{x\in\Lambda_L,\,y\notin\Lambda_L}
       \tau_{y-x}^\vartheta\EE|G_{g,V}(0,x;z)|^\vartheta<1
 \label{im:eq:app-criterion}
\end{equation}
implies exponentially weighted fractional-moment bounds uniformly
over truncated matrices. The preparatory estimates and the verification of the regularity and decoupling hypotheses are given in Subsections~\ref{sec:block} and~\ref{sec:uniformest}; the smallness condition in \eqref{im:eq:app-criterion} is verified below.

The constant $C_\vartheta$ is independent
of $g$; the coupling dependence is the factor $g^{-\vartheta}$.

Exponential hopping also gives constants $C,c>0$ such that
\begin{equation}
 \mathfrak X_\vartheta(\Lambda_L)\le C L^{d-1},\qquad
 \sup_{|x|_\infty<L/2}\sum_{y\notin\Lambda_L}\tau_{y-x}^\vartheta
                    \le C e^{-cL}.
 \label{im:eq:app-boundary-sums}
\end{equation}
These are the boundary sums estimated in \eqref{eq:boundary-weights}.
Split \eqref{im:eq:app-criterion} at $|x|_\infty=L/2$. Its left
side is at most
\begin{align}
 &(1+Cg_0^{-\vartheta}L^{d-1})
 \left[C N_L D_\vartheta(N_L,g_0)e^{-cL}
                     +C L^{d-1}\mathcal M(g_0)\right]
 \notag\\
 &\hspace{12mm}\le
 Cg_0^{-P}
       \left(e^{-c/g_0}
                       +e^{-cg_0^{-d/2}}
                       +e^{-cg_0^{-2}}\right)
       \longrightarrow0.
 \label{im:eq:app-contraction-small}
\end{align}
Here $P<\infty$ is fixed, depending on $d,\vartheta,p$, and the constants
can depend on $a,b,a_{\rm L},C$. We used $L=\lceil g_0^{-2}\rceil$,
$g_0 L\ge g_0^{-1}$, and
\eqref{im:eq:app-probability}. All prefactors are polynomial in
$g_0^{-1}$, including the factor $L^d$ in that probability.
This explicitly proves smallness as the coupling tends to zero.

Choose $g_0$ sufficiently small that
\eqref{im:eq:app-contraction-small} is at most $1/2$. For this fixed
$g_0$, the same testing cube and margin work for all
$g\in[g_0,2g_0]$,
$u\in[0,ag_0]$, and both signs of $\eta$.
The contraction kernels in the cited criterion have a common
exponential tail: their finite-cube resolvent factors have the
uniform bounds \eqref{im:eq:app-norm-moment}, and their remaining
factors are hopping tails. Their exponentially weighted norms
therefore tend to their unweighted norms uniformly as the weight
tends to zero. The strict margin $1/2$ gives one common positive
weight and one finite bound. We have proved the following lemma.

\begin{lemma}[Uniform additive estimate]\label{im:lem:app-additive-FM}
For each $0<a<1/2$ and all sufficiently small fixed
$g_0>0$, there are $B_{g_0}<\infty$ and
$\nu_{g_0}>0$ such that
\begin{equation}
 \EE|G_{g,V}(x,y;u+i\eta)|^\vartheta
       \le B_{g_0}e^{-\nu_{g_0}|x-y|_1}
 \label{im:eq:app-additive-FM}
\end{equation}
for all finite $V$, $x,y\in V$,
$g\in[g_0,2g_0]$,
$0\le u\le ag_0$, and $\eta\ne0$.
It also holds in infinite volume. The constants need not stay
bounded or positive uniformly as $g_0\downarrow0$.
\end{lemma}
The infinite-volume assertion follows by strong resolvent convergence
and Fatou's lemma; it is also included in the cited criterion.

\subsection{Transfer to the material-disorder operator}
Fix $0<a<1/2$ and take $W$ sufficiently small. Set
\begin{equation}
 E_-=E_a(W),\qquad
 g_0=WE_-,\qquad g_+=Wc_M,
 \qquad g_+/g_0=1+aW\le2.
 \label{im:eq:app-coupling-range}
\end{equation}
For $E\in I_{W,a}$, define
\begin{equation}
 \mathscr A_{\varpi,\Lambda}(E)
 =K_\Lambda+EW\Xi_{\varpi,\Lambda}-(c_M-E)\Id
 =E M_{\varpi,\Lambda}-J_\Lambda C_\Lambda^tJ_\Lambda.
 \label{im:eq:app-pencil}
\end{equation}
Its parameters satisfy $EW\in[g_0,2g_0]$ and
\begin{equation}
 0<c_M-E<c_M-E_-=aWE_-=ag_0.
 \label{im:eq:app-rectangle}
\end{equation}
This common parameter rectangle gives constants uniform across $I_{W,a}$.
For fixed real $E$, \eqref{im:eq:app-Wegner} implies that the finite
operator is invertible almost surely. Lemma~\ref{im:lem:app-additive-FM}
and Fatou's lemma as $\eta\to0$ yield
\begin{equation}
 \EE|\mathscr A_{\varpi,\Lambda}(E)^{-1}(x,y)|^\vartheta
             \le B_W e^{-\nu_W|x-y|_1},\qquad E\in I_{W,a},
 \label{im:eq:app-real-pencil}
\end{equation}
where dependence on the fixed $a$ is suppressed in $B_W,\nu_W$.
The real-energy congruence \eqref{eq:complex-pencil} does not identify
$H_\varpi$ with a single additive operator. We transfer
\eqref{im:eq:app-real-pencil} by material-variable averaging.

\begin{proof}[Completion of the proof of Theorem~\ref{im:thm:app-linear-window}]
The preceding estimate has the same form as \eqref{eq:real-pencil-FM},
uniformly on $I_{W,a}$ and over all truncated matrices.
Apply the material-variable averaging and resampling argument of
Section~\ref{sec:localisation}, with $E_-=E_a(W)$ and
$d_W=c_MW\sqrt{1+W}$. It gives
\[
 \EE Q_{\varpi,\Lambda}^t(x,y;I_{W,a})
 \le \frac{2(1+W)}{WE_-}\,d_W^{\vartheta}|I_{W,a}|B_W
                  e^{-\nu_W|x-y|_1}.
\]
The limiting argument in that section proves the infinite-volume
correlator bound, pure point spectrum with an exponentially decaying
eigenbasis, and the dynamical bound. Theorem~\ref{thm:finite-main}
then applies on $I_{W,a}$. Finally,
$\omega=\omega_M\sqrt{E/c_M}$ gives
\eqref{im:eq:app-frequency-window}.
\end{proof}

\subsection{A constraint on any upper mobility edge}
Let $\Sigma_{\rm ac}(W)$ be the almost-sure closed support of the
absolutely continuous spectrum, which is deterministic by ergodicity.
When it is nonempty, set $E_{\rm ac}^+(W)=\sup\Sigma_{\rm ac}(W)$.

\begin{corollary}[One-sided mobility-edge constraint]
\label{im:cor:app-mobility}
Under the assumptions of Theorem~\ref{im:thm:app-linear-window}, if
$\Sigma_{\rm ac}(W)\ne\varnothing$, then, for every fixed
$0<a<1/2$ and sufficiently small $W$,
\begin{equation}
 E_{\rm ac}^+(W)\le\frac{c_M}{1+aW}.
 \label{im:eq:app-ac-edge}
\end{equation}
The same bound holds for an upper mobility edge separating an
absolutely continuous region below from a localised region above,
if such a transition exists. Along any sequence $W\downarrow0$
for which $\Sigma_{\rm ac}(W)$ is nonempty,
\begin{equation}
 \liminf_{W\downarrow0}
   \frac{c_M-E_{\rm ac}^+(W)}{W}\ge\frac{c_M}{2}.
 \label{im:eq:app-mobility-liminf}
\end{equation}
Writing $\omega_{\rm ac}^+(W)=\omega_M\sqrt{E_{\rm ac}^+(W)/c_M}$,
the corresponding frequency bound is
\begin{equation}
 \liminf_{W\downarrow0}
  \frac{\omega_M-\omega_{\rm ac}^+(W)}{W}
                       \ge\frac{\omega_M}{4}.
 \label{im:eq:app-frequency-liminf}
\end{equation}
\end{corollary}
\begin{proof}
The absolutely continuous spectral measure vanishes on
$I_{W,a}$ by the theorem, outside $[0,c_M]$ by
Proposition~\ref{prop:spectrum-new}, and at the singleton $c_M$
by absolute continuity. Hence its closed support is contained in
$[0,E_a(W)]$. This proves \eqref{im:eq:app-ac-edge}, including
when the absolutely continuous support has several components.
For each fixed $0<a<1/2$,
\[
 \frac{c_M-E_{\rm ac}^+(W)}{W}
       \ge\frac{a c_M}{1+aW}.
\]
First take the lower limit along the sequence in question, then
let $a\uparrow1/2$. This order of limits does not require a
smallness threshold uniform in $a$. Similarly
$\omega_{\rm ac}^+(W)\le\omega_M(1+aW)^{-1/2}$ and
$[1-(1+aW)^{-1/2}]/W\to a/2$, giving
\eqref{im:eq:app-frequency-liminf}.
\end{proof}

This bound neither establishes nonempty absolutely continuous spectrum
nor gives a matching lower bound on its upper endpoint. A finite matrix
has discrete spectrum regardless of localisation, so a mobility edge
refers here to an infinite-volume transition.

\subsection{Interpretation and the weak-scattering prediction}
\begin{figure}[!htbp]
\centering
\begin{tikzpicture}[x=1cm,y=1cm,>=Latex,font=\small]
 \definecolor{locblue}{RGB}{34,100,151}
 \definecolor{hyporange}{RGB}{173,85,15}
 \definecolor{cohpurple}{RGB}{114,68,145}
 \begin{scope}[xscale=1.65]
  \node[anchor=west,font=\bfseries\small] at (0,6.55)
     {(a) Proved window at small disorder};
  \fill[gray!7] (0.55,0.85) rectangle (6.45,5.4);
  \fill[locblue!16] (0.55,5.4)--(6.45,5.4)--(6.45,3.55)
     .. controls (4.7,3.95) and (2.7,4.62) .. (0.55,5.4)--cycle;
  \draw[->] (0.55,0.8)--(0.55,5.95) node[left] {$E$};
  \draw[->] (0.5,0.85)--(6.65,0.85) node[right] {$W$};
  \node[below left] at (0.55,0.85) {$0$};
  \draw[black,thick] (0.55,5.4)--(6.45,5.4);
  \node[above,fill=white,inner sep=2pt] at (4.0,5.4)
     {$E_{\rm top}=c_M$};
  \draw[locblue,very thick] (0.55,5.4)
     .. controls (2.7,4.62) and (4.7,3.95) .. (6.45,3.55);
  \node[locblue,fill=locblue!8,inner sep=3pt,align=center]
     at (4.3,4.78) {Proved localisation\\$E_a(W)<E<c_M$};
  \draw[black!55,densely dashed,thick] (0.55,5.4)
     .. controls (2.6,4.0) and (4.7,3.02) .. (6.45,2.35);
  \node[black!65,anchor=north east,fill=white,inner sep=3pt]
     at (6.45,2.28) {$E_{\rm mean}=c_M/(1+W/2)$};
  \node[locblue,anchor=north east,fill=white,inner sep=3pt]
     at (6.45,3.53) {$E_a(W)=c_M/(1+aW)$};
  \node[align=center,text=black!75] at (3.1,1.35)
     {Interior spectral type not classified here};
  \node[anchor=west,font=\footnotesize] at (0.55,0.18)
     {Fixed $0<a<1/2$; theorem requires $0<W<W_a$.};
 \end{scope}
 \begin{scope}[yshift=-7.8cm]
  \node[anchor=west,font=\bfseries\small] at (0,6.55)
     {(b) Distinct energy scales at fixed small $W$ in $d=3$};
  \draw[->] (0.7,0.75)--(0.7,5.95) node[left] {$E$};
  \fill[locblue!16] (0.7,3.7) rectangle (6.3,5.45);
  \draw[black,thick] (0.7,5.45)--(6.3,5.45);
  \node[anchor=west,fill=white,inner sep=2pt] at (1.05,5.45)
     {$c_M$: exact spectral edge};
  \draw[locblue,very thick] (0.7,3.7)--(6.3,3.7);
  \node[locblue,anchor=west,fill=white,inner sep=2pt] at (1.05,3.7)
     {$E_a(W)$: sufficient threshold};
  \draw[cohpurple,dashed,thick] (0.7,2.65)--(6.3,2.65);
  \node[cohpurple,anchor=west,fill=white,inner sep=2pt] at (1.05,2.65)
     {$E_{\rm coh}$: approximate coherent edge};
  \draw[hyporange,dash dot,thick] (0.7,2.02)--(6.3,2.02);
  \node[hyporange,anchor=west,fill=white,inner sep=2pt] at (1.05,2.02)
     {$E_{\rm mob}$: conjectured transition};
  \draw[black!55,densely dashed,thick] (0.7,1.12)--(6.3,1.12);
  \node[black!65,anchor=west,fill=white,inner sep=2pt] at (1.05,1.12)
     {$E_{\rm mean}$: mean-material edge};
  \draw[<->,gray!75] (7.0,1.12)--(7.0,5.45);
  \node[anchor=west,align=left,font=\footnotesize] at (7.15,4.85)
     {$O(W)$\\mean shift};
  \draw[<->,cohpurple] (8.9,1.12)--(8.9,2.65);
  \node[cohpurple,anchor=west,align=left,font=\footnotesize] at (9.05,1.48)
     {$O(W^2)$\\coherent correction};
  \draw[<->,hyporange] (7.65,2.02)--(7.65,2.65);
  \node[hyporange,anchor=west,align=left,font=\footnotesize] at (7.8,3.05)
     {$O(W^4)$\\predicted separation};
  \draw[hyporange,thin] (7.9,2.82)--(7.65,2.36);
 \end{scope}
\end{tikzpicture}
\caption{The proved localisation window and the distinct mobility-edge
question. Panel (a) shows the sufficient threshold $E_a(W)$ and the
mean-material edge; the region below $E_a(W)$ is not classified by
Theorem~\ref{im:thm:app-linear-window}. Panel (b) separates the exact
spectral edge from the sufficient threshold and the three-dimensional
heuristic levels. The displayed ordering of the latter is illustrative,
not an additional theorem. Both panels are schematic and not to scale.}
\label{im:fig:mobility-scales}
\end{figure}
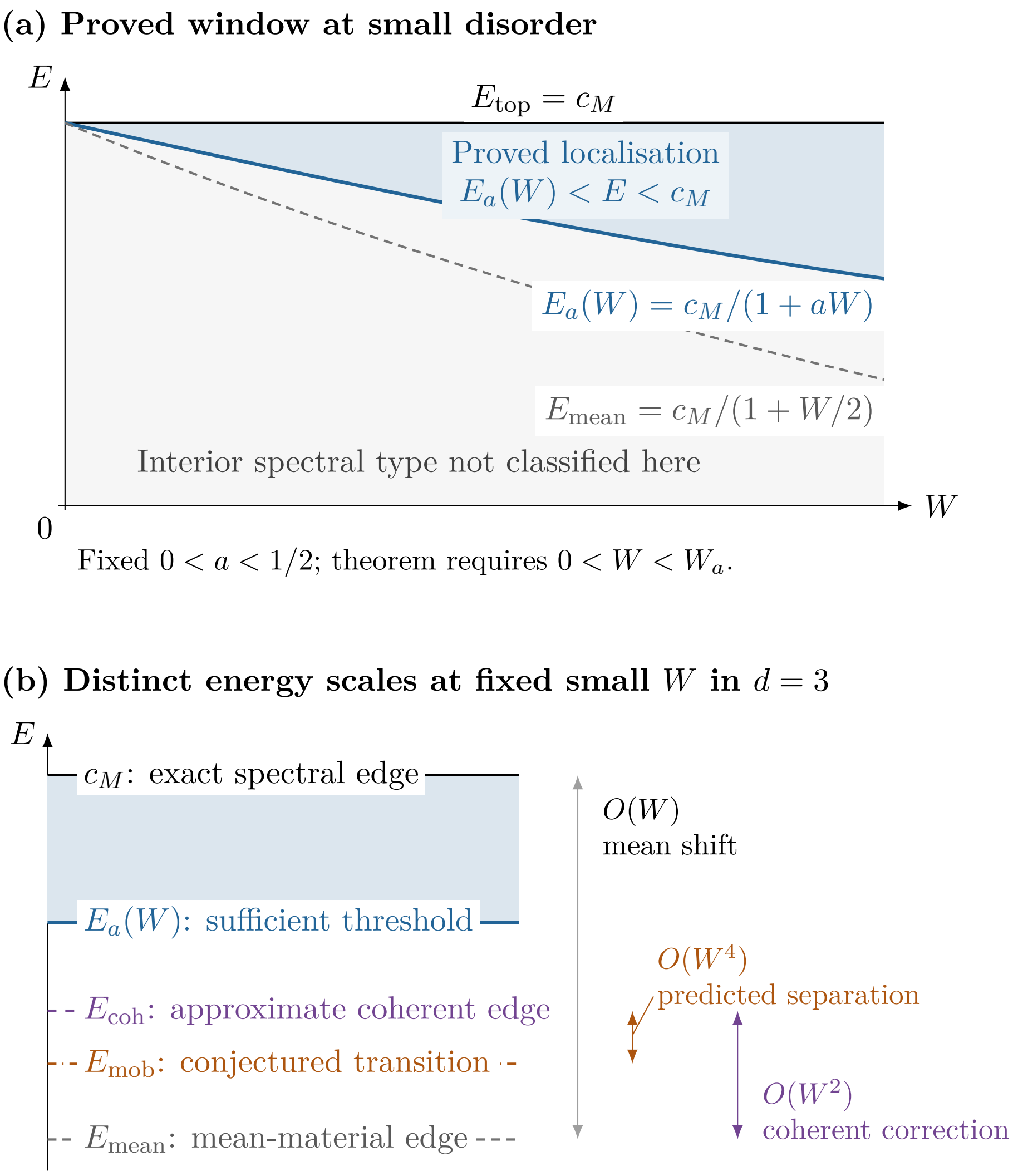

In dimension three a weak-scattering calculation predicts a separation
of order $W^4$ between a mobility edge and an approximate coherent
edge. The latter differs from $E_{\rm mean}(W)$ at order $W^2$;
both differ from the exact spectral edge at order $W$ for the present
uncentred inverse-material law. Appendix~\ref{app:heuristic} gives
the calculation and the rigorous unperturbed density-of-states
asymptotic on which it is based. The $W^4$ prediction is heuristic
and is not an error estimate for the bound in
Corollary~\ref{im:cor:app-mobility}. Figure~\ref{im:fig:mobility-scales}
separates the proved and predicted scales.

\FloatBarrier

\section{Eigenfunction extent at regular interior energies}
\label{sec:interior}

The estimates in this section hold for every material configuration in
the prescribed range. We use the operators and Fourier convention of
Section~\ref{sec:model}, with
\begin{equation}
 c_r=c_{-r}\in\R,\qquad |c_r|\le A_0e^{-\gamma_0|r|_1}.
 \label{im:eq:C}
\end{equation}
We use the normalised measure $\dd\bar\alpha=(2\pi)^{-d}\dd\alpha$ on $Y^*$.

\subsection{Centred material variables}
With $\mu_W=\EE m_n$ as above, set
\begin{equation}
 \zeta_n=\xi_n-\frac12,\qquad
 Z_\varpi=\diag(\zeta_n),\qquad \lambda=\frac W{\mu_W}.
 \label{im:eq:centering}
\end{equation}
Then $\EE\zeta_n=0$, $\EE\zeta_n\zeta_m=\sigma^2\delta_{nm}$,
$\sigma^2=1/12$, where $\delta_{nm}$ is the Kronecker delta, and
$\|Z_\varpi\|\le1/2$. There are two useful
exact identities:
\begin{align}
 (C-s)q&=EW Z_\varpi q,\qquad s=\mu_WE,
 \label{im:eq:pencil}\\
 \widetilde H_\varpi:=\mu_WH_\varpi
   &=(\Id+\lambda Z_\varpi)^{-1/2}C
                    (\Id+\lambda Z_\varpi)^{-1/2}.
 \label{im:eq:tildeH}
\end{align}
Multiplication of $H_\varpi$ by the scalar $\mu_W$ does not change
its eigenvectors, but rescales every spectral value. In particular,
\begin{equation}
 E=\frac{s}{\mu_W},\qquad
 \omega=\frac{v_r\sqrt\delta}{\varepsilon}
                   \sqrt{\frac{s}{\mu_W}},\qquad
 \frac\omega{\omega_M}=\sqrt{\frac{s}{\mu_Wc_M}}.
 \label{eq:centred-frequency-conversion}
\end{equation}
Almost surely, the upper edge of $\widetilde H_\varpi$ is $\mu_Wc_M$;
the reference symbol $c$ has upper edge $c_M$.
The random potential in \eqref{im:eq:pencil} depends
on $E$; it must not be treated as one fixed additive operator when
studying all spectral energies at once.

\subsection{Regular-energy quasimodes}
Let $I\Subset(0,c_M)$ be a compact interval and assume there exist
$a>0$ and $b_*>0$ such that
\begin{equation}
 |\nabla c(\alpha)|\ge b_*
 \quad\hbox{whenever }\dist(c(\alpha),I)\le a.
 \label{im:eq:regular}
\end{equation}
In particular, $I$ avoids all critical values of $c$. Regularity at
one selected carrier is insufficient: disorder scatters to the whole
energy surface. Coordinate reflection symmetry does not remove the
critical energies associated with vertices other than $M$.
Here $I$ is an interval of centred energy $s$, so the corresponding
eigenvalues of $H_\varpi$ lie in
$I/\mu_W:=\{s/\mu_W:s\in I\}$. This convention also applies to
the finite-sample interior estimates below.

For a normalised $q\in\ell^2(\Z^d)$ set
\begin{equation}
 \operatorname{IPR}(q)=\sum_n|q(n)|^4,\qquad
 P(q)=\operatorname{IPR}(q)^{-1},\qquad
 R_2(q;x_0)=\left(\sum_n|n-x_0|^2|q(n)|^2\right)^{1/2}.
 \label{im:eq:measures}
\end{equation}
The last quantity is allowed to be infinite. It is a radius measured
by the second moment, not the inverse of an exponential decay rate.

\begin{lemma}[A regular-energy resolvent integral]\label{im:lem:coarea}
Under \eqref{im:eq:regular}, there are $K_I<\infty$ and $h_*>0$ such that
\begin{equation}
 \sup_{e\in I}\int_{Y^*}\frac{\dd\bar\alpha}
                 {h^2+(c(\alpha)-e)^2}\le\frac{K_I}{h},
 \qquad 0<h\le h_*.
 \label{im:eq:coarea}
\end{equation}
\end{lemma}
\begin{proof}
On the compact inverse image of a slightly enlarged interval, the
coarea formula and \eqref{im:eq:regular} give a uniformly bounded
density
\[
 \rho_c(t)=\int_{\{c=t\}}
          \frac{\dd S(\alpha)}{(2\pi)^d|\nabla c(\alpha)|},
\]
where $\dd S$ is the surface measure. Boundedness follows, for example,
from a finite collection of submersion charts. For $|c-e|<a/2$, integration in $t=c(\alpha)$
is bounded by
$\|\rho_c\|_\infty\int_\R(h^2+t^2)^{-1}\dd t
=\pi\|\rho_c\|_\infty/h$. On the complement the denominator is
at least $a^2/4$, and $\int_{Y^*}\dd\bar\alpha=1$. Increasing the
constant proves \eqref{im:eq:coarea} for any fixed sufficiently small
$h_*$. No assumption on critical values outside the enlarged interval
is needed.
\end{proof}

\begin{proposition}[Extent of a regular-energy quasimode]\label{im:prop:quasimode}
Let $\|q\|_2=1$, $e\in I$, and
$r_q=\|(C-e)q\|_2$. For $0<r_q\le r_*$, with $r_*>0$ sufficiently
small depending only on $c$ and $I$,
\begin{align}
 \|q\|_\infty^2&\le2K_Ir_q,\qquad
 P(q)\ge(2K_Ir_q)^{-1},\label{im:eq:point-bound}\\
 R_2(q;x_0)&\ge\frac{b_*^2}{4b_1r_q}-\frac{b_2}{2b_1},
       \qquad x_0\in\R^d,\label{im:eq:radius-bound}
\end{align}
where $b_1=\sup_{Y^*}|\nabla c|>0$ and
$b_2=\sup_{Y^*}|\Delta_\alpha c|$. For any $\eta_0\in(0,1]$,
a set containing at least the fraction $\eta_0$ of the squared norm
has at least $\eta_0/(2K_Ir_q)$ lattice sites.
\end{proposition}
\begin{proof}
By Fourier inversion and Cauchy--Schwarz, for each $n$ and $h>0$,
\begin{align*}
 |q(n)|^2
 &\le\left(\int\frac{\dd\bar\alpha}{h^2+(c-e)^2}\right)
       \left(\int[h^2+(c-e)^2]|\widehat q|^2\dd\bar\alpha\right)\\
 &\le\frac{K_I}{h}(h^2+r_q^2).
\end{align*}
Taking $h=r_q$ proves the first inequality. The participation and
cardinality bounds follow from
$\sum|q|^4\le\|q\|_\infty^2\|q\|_2^2$ and
$\sum_{n\in S}|q(n)|^2\le |S|\|q\|_\infty^2$.

For the radius estimate, the case $R_2=\infty$ is immediate. Otherwise
$\widehat q\in H^1(Y^*)$. Let $X_jq(n)=n_jq(n)$ and let $V_j$
be the Fourier multiplier $\partial_jc$. On this domain introduce
the symmetric operator
\[
 \mathcal A_{x_0}=\tfrac12\sum_{j=1}^d
       \bigl[V_j(X_j-x_{0j})+(X_j-x_{0j})V_j\bigr].
\]
In our Fourier convention $X_j=-i\partial_{\alpha_j}$, hence
\begin{equation}
 i[\mathcal A_{x_0},C]=|\nabla c|^2,\qquad
 \|\mathcal A_{x_0}q\|_2
                    \le b_1R_2(q;x_0)+\tfrac12b_2.
 \label{im:eq:commutator}
\end{equation}
These identities hold first on smooth Fourier functions and then on
$H^1$ by density.

The spectral weight where $|c-e|\ge a$ is at most $r_q^2/a^2$.
Thus \eqref{im:eq:regular} gives
\[
 \ip q{i[\mathcal A_{x_0},C]q}
   =\int|\nabla c|^2|\widehat q|^2\dd\bar\alpha
   \ge b_*^2(1-r_q^2/a^2).
\]
Writing $Cq=e q+(C-e)q$, the same commutator expectation has absolute
value at most $2r_q\|\mathcal A_{x_0}q\|_2$. Choose
$r_*\le a/\sqrt2$ and $r_*\le h_*$. Then
$b_*^2/2\le2r_q(b_1R_2+b_2/2)$, which is
\eqref{im:eq:radius-bound}. If $r_q=0$, the first estimate with $h\downarrow0$
would imply $q=0$, so there is no nonzero normalised vector of that kind.
\end{proof}

\begin{theorem}[Weak-disorder lower bounds on eigenfunction extent]\label{im:thm:eigenfunctions}
Assume \eqref{im:eq:C} and \eqref{im:eq:regular}. There are constants
$W_0,C_I,c_I,C_I'>0$ such that the following holds for every
configuration $0\le\xi_n\le1$ and $0<W\le W_0$. If
\begin{equation}
 Cq=E M_\varpi q,\qquad \|q\|_2=1,\qquad \mu_WE\in I,
 \label{im:eq:eigen-assumptions}
\end{equation}
then
\begin{equation}
 \|q\|_\infty^2\le C_IW,\qquad
 P(q)\ge\frac{c_I}{W},\qquad
 R_2(q;x_0)\ge\frac{c_I}{W}-C_I'
                   \quad\hbox{for every }x_0\in\R^d.
 \label{im:eq:eigen-bounds}
\end{equation}
The conclusions, with adjusted uniform constants, hold for the
ordinary $\ell^2$-normalised eigenvector of $H_\varpi$.
\end{theorem}
\begin{proof}
Identity \eqref{im:eq:pencil} gives
$\|(C-\mu_WE)q\|_2\le EW/2$. Since $\mu_WE\in I$, $E$ is
bounded above uniformly. For sufficiently small $W$, this residual
is at most $r_*$. Apply Proposition~\ref{im:prop:quasimode}; replacing
the residual by its upper bound weakens the estimates in the required
direction. For a normalised eigenvector $\psi$ of $H_\varpi$, put
$q=M_\varpi^{-1/2}\psi$. Its norm is between $(1+W)^{-1/2}$ and
one. Normalise $q$ and use $1\le m_n\le1+W$. Since $M_\varpi$
commutes with position, both pointwise and second-moment estimates
transfer with factors uniformly bounded for $W\le W_0$.
\end{proof}

The bounds are compatible with exponentially localised eigenfunctions
whose extent diverges as $W\downarrow0$. An exponential upper bound
with an uncontrolled prefactor does not yield a decay-rate estimate
from \eqref{im:eq:eigen-bounds} alone. The participation number measures
the number of occupied sites, whereas $R_2$ also measures their separation.

\section{Verification for crystals of spheres}\label{sec:spheres}
Let $D=B_r(0)$ with $0<r<1/2$. In dimension three this is a simple cubic crystal of
spheres; in dimension two it is the corresponding square crystal of disks.

\subsection{Explicit geometric constants}
Choose $r<R<1/2$ and an $H^1$ radial cutoff equal to one in $B_r$, harmonic
in the annulus $B_R\setminus \overline{B_r}$, and zero outside $B_R$. Its normalised energy is
\begin{equation}
 E_\chi=\frac1{|D|}
 \begin{cases}
 2\pi/\log(R/r),&d=2,\\
 4\pi/(r^{-1}-R^{-1}),&d=3.
 \end{cases}
 \label{eq:annular}
\end{equation}
The three-dimensional formula follows by differentiating
$(|x|_2^{-1}-R^{-1})/(r^{-1}-R^{-1})$ and integrating its squared
gradient; in two dimensions use $\log(R/|x|_2)/\log(R/r)$.
The constants $C_P,\gamma_0,A_0$ in
\eqref{eq:CP} and \eqref{eq:explicit-A-alpha} are now explicit.

At $\alpha_M$, all quasiperiodic Fourier wavevectors have squared
length at least $d\pi^2$. Since an admissible function equals one
in $D$,
\begin{equation}
 c_*:=d\pi^2\le c_M\le E_\chi.
 \label{eq:sphere-edge-bounds}
\end{equation}
These are bounds for the normalised symbol. Without volume normalisation
they read $d\pi^2|D|\le\operatorname{Cap}_{\alpha_M,D}\le|D|E_\chi$.
For $d=3$, $r=1/5$, and $R=2/5$, they give
$3\pi^2\le c_M\le150$.

\subsection{The fixed-interval hypothesis}
For any $\vartheta\in(0,1)$ and $0<\mu<\vartheta\gamma_0$, the geometric series gives
\begin{equation}
 T_\vartheta(\mu)\le A_0^\vartheta
 \left[\coth\left(\frac{\vartheta\gamma_0-\mu}{2}\right)^d-1\right].
 \label{eq:Ts-explicit}
\end{equation}
In particular put $\vartheta=1/2$, $\mu=\gamma_0/4$, and
\begin{equation}
 \overline T=\sqrt{A_0}\,[\coth(\gamma_0/8)^d-1],\qquad
 I_*=(c_*/2,3c_*/4),\qquad W_{\rm sph}=64\overline T^{\,2}/c_*.
 \label{eq:sphere-choice}
\end{equation}
For the inverse-material law of width $W=W_{\rm sph}$,
\[
 q_1\le\frac{4\overline T}{\sqrt{Wc_*}}=\frac12.
\]
Thus Theorem~\ref{thm:strong-main} and
Corollary~\ref{cor:strong-localisation} apply on the fixed interval $I_*$;
they also apply for any larger width. These conservative sufficient
bounds do not estimate the onset of localisation. Table~\ref{tab:spheres} gives rounded numerical evaluations
for $r=1/5$, $R=2/5$.

\begin{table}[htbp]\centering\small
\begin{tabular}{@{}lrr@{}}\toprule
Quantity & Disks ($d=2$) & Spheres ($d=3$)\\\midrule
$|D|$ & $0.1256637$ & $0.0335103$\\
$C_P$ & $0.9526854$ & $1.7677394$\\
$E_\chi$ & $72.1347520$ & $150.0000000$\\
$\gamma_0$ & $0.0927781$ & $0.0408255$\\
$A_0$ & $260.5264575$ & $508.6315987$\\
$\overline T$ & $1.200039\times10^{5}$ & $1.697030\times10^{8}$\\
$c_*$ & $19.7392088$ & $29.6088132$\\
$W_{\rm sph}$ & $4.669185\times10^{10}$ & $6.224978\times10^{16}$\\
\bottomrule\end{tabular}
\vspace{2mm}
\caption{Explicit normalised geometric bounds. The very large sufficient
widths reflect the rough analytic upper bound on the interaction sum.}
\label{tab:spheres}\end{table}

\subsection{The band-edge hypotheses}
Proposition~\ref{prop:exp}, Theorem~\ref{thm:gradient}, and
Proposition~\ref{prop:Hessian} give exponential decay of the coefficients
and a unique nondegenerate upper maximum. They imply
\eqref{eq:power-bound}, so Theorem~\ref{thm:localisation} applies for
every fixed $W>0$ and $0<r<1/2$. The truncated and finite capacitance matrices satisfy Theorems~\ref{thm:finite-main} and
\ref{thm:cluster-main}, respectively. For fixed $a\in(0,1/2)$ and
sufficiently small $W$, Section~\ref{sec:mobility} gives the window
of width $a c_MW/(1+aW)$; its smallness threshold is not evaluated
numerically here. No large-disorder threshold from
Table~\ref{tab:spheres} is required for these edge results.

Coordinate-permutation symmetry makes the edge Hessian a scalar
multiple of the identity, as used in Appendix~\ref{app:heuristic}.
The interior estimates of Section~\ref{sec:interior} apply to intervals
avoiding all critical values, including those at the other reflection
vertices. For the direct material law, Proposition~\ref{prop:other-laws}
gives absence of absolutely continuous spectrum near the rescaled edge.

\FloatBarrier
\part{Truncated capacitance matrices}\label{part:truncated}
The estimates in this part are the intermediate step from the infinite
crystal to the finite capacitance matrix. Truncated matrices retain
the periodic interaction and enjoy bounds uniform up to their boundary.
We also prove a spacing estimate that applies to both finite constructions.

\section{Uniform localisation for truncated matrices} \label{sec:principal}
We derive quantitative concentration bounds for eigenvectors of truncated matrices and verify that their constants remain uniform for the contrast-dependent capacitance family at small contrast.

\subsection{Simultaneous concentration}

We formulate the simultaneous eigenvector bounds both on the lattice and in physical coordinates, using the following description of the sample.

Let $\Omega\subset\R^d$ be a nonempty bounded Lipschitz domain and set
\begin{equation}
 I_\varepsilon=\{n\in\Z^d:\varepsilon(n+\overline Y)\subset\Omega\},
 \qquad N_\varepsilon=|I_\varepsilon|\asymp\varepsilon^{-d}.
 \label{eq:sample-indices}
\end{equation}
Define the isometry $J_\varepsilon:\ell^2(I_\varepsilon)\to L^2(\Omega)$
by $(J_\varepsilon q)(x)=\varepsilon^{-d/2}q_n$ for
$x\in\varepsilon(n+Y)$, and zero elsewhere.

\begin{theorem}[Simultaneous localisation of finite-volume edge modes]
\label{thm:finite-main}\label{cor:simultaneous-concentration}
Assume the geometry above, including \eqref{eq:reflection}, and the
material law \eqref{eq:random-model}. For every prescribed $W>0$
there are $\Delta_W\in(0,c_M/2)$, $A_W<\infty$, and $\nu_W>0$
such that, with $I_W=(c_M-\Delta_W,c_M)$,
\begin{equation}
 \EE Q_{\varpi,\Lambda}^t(x,y;I_W)
          \le A_We^{-\nu_W|x-y|_1}
 \label{eq:finite-main-Q}
\end{equation}
for every finite $\Lambda\subset\Z^d$ and $x,y\in\Lambda$.
For each $p>0$, there are $R_{W,p},C_{W,p}<\infty$, independent
of $\Lambda$, such that, with probability at least
$1-C_{W,p}N^{-p}$, $N=|\Lambda|\ge2$, every normalised eigenvector
$\psi_j$ with eigenvalue $E_j\in I_W$ has a centre $n_j\in\Lambda$
for which
\begin{align}
 |\psi_j(n)|&\le N^{p+5/2}e^{-\nu_W|n-n_j|_1/2},
                                      \quad n\in\Lambda,
                       \label{eq:finite-main-exponential}\\
 \sum_{|n-n_j|_1>R_{W,p}\log(2+N)}|\psi_j(n)|^2
                 &\le N^{-p}.
                       \label{eq:simultaneous-concentration}
\end{align}
Both estimates hold simultaneously for all modes in $I_W$.
For $\Lambda=I_\varepsilon$, put $x_{j,\varepsilon}=\varepsilon n_j$
and $r_\varepsilon=\varepsilon R_{W,p}\log(2+N_\varepsilon)
+\sqrt d\,\varepsilon/2$. Then, on the same event,
\begin{equation}
 \int_{\Omega\setminus B(x_{j,\varepsilon},r_\varepsilon)}
     |J_\varepsilon\psi_j(x)|^2\dd x\le N_\varepsilon^{-p},
 \qquad r_\varepsilon=O_{W,p}(\varepsilon\log(1/\varepsilon)).
 \label{eq:physical-concentration}
\end{equation}
For $q_j=M_{\varpi,I_\varepsilon}^{-1/2}\psi_j$, the normalised field
$J_\varepsilon q_j/\|q_j\|_2$ satisfies the same concentration estimate
with right-hand side $(1+W)N_\varepsilon^{-p}$.
\end{theorem}

The constants are uniform in sample size for fixed $W$. Along
$\varepsilon=k^{-1}$, the concentration holds eventually almost surely
if $dp>1$, by Borel--Cantelli. The same conclusion holds on the
large-disorder intervals of Corollary~\ref{cor:strong-localisation}.

\subsection{Proof of the finite-volume theorem}
\begin{proof}[Proof of Theorem~\ref{thm:finite-main}]
The correlator estimate is \eqref{eq:Q-edge}. Write $N=|\Lambda|\ge2$.
Markov's inequality and a union bound show that
\[
 Q_{\varpi,\Lambda}^t(x,y;I_W)
       \le N^{p+2}e^{-\nu_W|x-y|_1/2}
       \quad(x,y\in\Lambda)
\]
fails with probability at most
\[
 A_WN^{-p-2}\sum_{x,y\in\Lambda}e^{-\nu_W|x-y|_1/2}
                  \le C_WN^{-p-1}.
\]
On this common event choose $n_j$ where $|\psi_j|$ is maximal.
Then $|\psi_j(n_j)|\ge N^{-1/2}$. The finite eigenvalues are simple
almost surely by the resampling argument, and therefore
$|\psi_j(n)\psi_j(n_j)|\le Q_{\varpi,\Lambda}^t(n,n_j;I_W)$.
This proves \eqref{eq:finite-main-exponential} for all modes at once.
Summing its square outside lattice radius $R$ gives
\[
 \sum_{|n-n_j|_1>R}|\psi_j(n)|^2
             \le C_WN^{2p+5}e^{-\nu_WR/2}.
\]
Choose $R=R_{W,p}\log(2+N)$ with $R_{W,p}$ sufficiently large
to obtain \eqref{eq:simultaneous-concentration}.

Every point of $\varepsilon(n+Y)$ lies at Euclidean distance at most
$\varepsilon|n-n_j|_1+\sqrt d\,\varepsilon/2$ from $\varepsilon n_j$.
The isometry $J_\varepsilon$ consequently gives
\eqref{eq:physical-concentration}. Finally,
$(1+W)^{-1/2}\le\|M^{-1/2}\psi_j\|_2\le1$ and
$|M^{-1/2}\psi_j(n)|\le|\psi_j(n)|$. Normalising the amplitudes
costs at most the factor $1+W$ in the squared tail.
Corollary~\ref{cor:strong-localisation} proves the same assertions
on its large-disorder interval.
\end{proof}

\begin{remark}[Number of edge modes]
The statement is not restricted to a fixed number of modes. By
Theorem~\ref{thm:IDS}, a fixed nonempty open subinterval of $I_W$
contains a positive density of eigenvalues in truncated matrices on growing cubes:
choose a nonnegative continuous function supported in that interval
with positive density-of-states integral. Thus the simultaneous
estimate controls a number of modes proportional to the sample size
as the cubes grow.
\end{remark}

\subsection{Uniformity in the contrast parameter}

The simultaneous period and contrast limit requires constants that remain uniform as \(\delta\to0 \). We obtain this uniformity for the family \(C_\delta \), allowing the choice \(\delta=\varepsilon^2 \).

\begin{proposition}\label{prop:uniform-localisation}
Assume \eqref{eq:reflection} and the material law
\eqref{eq:random-model}. For the family in
Proposition~\ref{prop:uniform-family} and every fixed $W>0$, there are $\Delta_W>0$, $A_W<\infty$, and $\nu_W>0$,
independent of sufficiently small $\delta$, such that
\begin{equation}
 \EE Q_{\delta,\varpi,\Lambda}^t(x,y;I_{W,\delta})
       \le A_We^{-\nu_W|x-y|_1},\qquad
 I_{W,\delta}=(c_{M,\delta}-\Delta_W,c_{M,\delta}),
 \label{eq:uniform-Q}
\end{equation}
for all finite $\Lambda$ and $x,y\in\Lambda$. Here $Q_\delta^t$ is
\eqref{eq:Q-definition} with $C$ replaced by $C_\delta$.
For every $p>0$, every deterministic sufficiently small $\delta\ge0$,
and every finite $\Lambda$ with $N=|\Lambda|\ge2$, the simultaneous
conclusions of Theorem~\ref{thm:finite-main}  hold with probability at least
$1-C_{W,p}N^{-p}$, with constants $C_{W,p}$ and $R_{W,p}$
independent of $\delta$ and $\Lambda$.
The event may depend on $\delta$ and $\Lambda$; simultaneity
concerns all eigenvectors with eigenvalues in $I_{W,\delta}$.

On a prescribed bounded interval $I=(E_0,E_1)$ with $E_0>0$, the
same conclusions hold, without reflection symmetry, if
\begin{equation}
 W>\frac1{E_0}
 \left(\frac{2^\vartheta T_\vartheta^*(\mu)}{1-\vartheta}\right)^{1/\vartheta},
 \qquad
 T_\vartheta^*(\mu)=
 \sup_{0\le\delta\le\delta_0}
 \sum_{r\ne0}|c_{\delta,r}|^\vartheta e^{\mu|r|_1},
 \quad 0<\mu<\vartheta\gamma_* .
 \label{eq:uniform-large}
\end{equation}
For each fixed $0<a<1/2$, the quantitative window
$(c_{M,\delta}/(1+aW),c_{M,\delta})$ likewise has uniform constants
for sufficiently small fixed $W$. In particular, these uniform probability bounds apply for each
$\varepsilon$ under the deterministic choice $\delta=\varepsilon^2$.
\end{proposition}
\begin{proof}
Let $J$ be the staggering operator in \eqref{eq:K} and put
$K_\delta=c_{M,\delta}\Id-JC_\delta J$.
Proposition~\ref{prop:uniform-family} gives
$K_\delta\ge a_*\Lap$, a common exponential coefficient bound,
and $c_*\le c_{M,\delta}\le c^*$.
In Proposition~\ref{prop:initial} and
Lemma~\ref{lem:all-restrictions-FM}, take the common coupling range
$[Wc_*/2,Wc^*]$. The Neumann-block gap, the Bernoulli probability,
the resolvent conjugation constant, and the hopping boundary sums
then have constants independent of $\delta$. One testing cube and
one strict contraction margin work for the whole family.
Choose $\Delta_W<\min(c_*/2,r)$, with $r$ from that criterion.
The real-operator estimate is uniform; material averaging uses the
common lower energy $c_*/2$ and resampling uses the common upper
energy $c^*$. This proves \eqref{eq:uniform-Q}. The proof of
Theorem~\ref{thm:finite-main} uses only this bound.

For \eqref{eq:uniform-large}, the one-site recursion of
Theorem~\ref{thm:FM} has the common contraction factor
$2^\vartheta T_\vartheta^*(\mu)/((1-\vartheta)(WE_0)^\vartheta)<1$.
The same averaging argument applies. For the quantitative window,
repeat Section~\ref{sec:mobility} with the common deficit comparison
and decay constants. Its small-coupling estimates are uniform for
$g_0=Wc_{M,\delta}/(1+aW)$, since $c_{M,\delta}$
stays in $[c_*,c^*]$. This proves the final assertion.
\end{proof}

\section{Eigenvalue spacing under material disorder}\label{sec:spacing}
Matching individual modes in Section~\ref{sec:cluster-localisation} requires a quantitative bound on the probability of small eigenvalue gaps.
The following consequence of Lemma~\ref{lem:material-average} applies
to any deterministic nonnegative finite matrix. In particular, it does
not require exponential decay of the entries of the finite capacitance matrix.
\begin{proposition}[A material-variable Minami estimate]\label{fc:prop:Minami}
Let $C_*$ be a deterministic nonnegative $N\times N$ matrix and
$\mathcal N_J=\Tr\one_J(M^{-1/2}C_*M^{-1/2})$, with the law
\eqref{eq:random-model} and $W>0$. For an interval $J\subset[E_0,\infty)$, $E_0>0$,
\begin{equation}
 \EE[\mathcal N_J]\le\kappa N|J|,\qquad
 \EE[\mathcal N_J(\mathcal N_J-1)]\le\kappa^2N^2|J|^2,
 \label{fc:eq:Minami}
\end{equation}
where $\kappa=(1+W)/(WE_0)$. On a fixed bounded positive-energy region, the probability that two eigenvalues have separation less than $h$ is at most $C_WN^2h$ for small $h>0$. Positive eigenvalues are almost surely simple.
\end{proposition}

\begin{proof}
The first estimate follows by summing \eqref{eq:material-average} over coordinates. For the second, independently resample $m_y$, and write $\mathcal N^{(y)}_J$ for the reference count. The finite matrices $H$ and $B=C_*^{1/2}M^{-1}C_*^{1/2}$ have identical spectra, including multiplicities. Changing one material variable changes $B$ by a sign-definite rank-one matrix. Rank-one interlacing consequently gives, for any original and reference values,
\begin{equation}
 \mathcal N_J-1\le\mathcal N^{(y)}_J.
 \label{fc:eq:interlace}
\end{equation}
For example, the eigenvalue counting functions of two sign-definite rank-one perturbations differ at each energy by either zero or one with a common sign; taking their difference at the two ends of $J$ proves \eqref{fc:eq:interlace}.

Since $\mathcal N_J=\sum_y\nu_y(J)$, one has
\[
 \mathcal N_J(\mathcal N_J-1)
 \le\sum_y\nu_y(J)\mathcal N^{(y)}_J.
\]
Condition on all other variables and on the resampled one. The reference count is independent of the original $m_y$, so \eqref{eq:material-average} bounds its contribution by
$\kappa|J|\mathcal N^{(y)}_J$. Averaging again and applying the first estimate to the reference operator gives
$\kappa^2N^2|J|^2$. This proves \eqref{fc:eq:Minami}. This is the material-disorder analogue of the quadratic counting estimate in \cite{Minami,CGK2009}.

Cover the fixed energy region by $O(h^{-1})$ overlapping intervals of length $3h$, so that every pair of energies at distance less than $h$ lies in one interval. Since $\PP\{\mathcal N_J\ge2\}\le\frac12\EE[\mathcal N_J(\mathcal N_J-1)]$, a union bound gives $C_WN^2h$. Let \(h\downarrow0\) for each fixed \(N\), and apply the argument on the intervals \([1/k,k] \), \(k=2,3,\ldots\), to obtain simplicity at every positive energy.
\end{proof}

\section{Interior estimates for truncated matrices}\label{sec:principal-interior}
We extend the interior eigenfunction bounds of Section~\ref{sec:interior} to truncated matrices by controlling the residual created at the boundary and bounding the number of exceptional modes.

\subsection{Boundary residuals and eigenfunction extent}\label{im:sec:finite}
Use the projection $P_\Lambda$ and the truncated matrix $C_\Lambda^t$
from \eqref{eq:finite}. The zero extension of $C_\Lambda^t$ is
$P_\Lambda C P_\Lambda$.
Extend a normalised eigenvector amplitude $q$ by zero to
$\widetilde q\in\ell^2(\Z^d)$. If
$C_\Lambda^tq=EM_{\varpi,\Lambda}q$ and $e=\mu_WE$, then
\begin{equation}
 (C-e)\widetilde q=EW Z_\varpi\widetilde q+b_\Lambda,
 \qquad b_\Lambda=(1-P_\Lambda)C\widetilde q.
 \label{im:eq:boundary-residual}
\end{equation}
The two terms on the right have disjoint supports. Hence
\begin{equation}
 \|(C-e)\widetilde q\|_2^2
   \le E^2W^2/4+\|b_\Lambda\|_2^2.
 \label{im:eq:boundary-error}
\end{equation}
Proposition~\ref{im:prop:quasimode} applies with this full-lattice residual.

\begin{lemma}[Exponential boundary leakage estimate]\label{im:lem:leakage}
Let $d_\Lambda(n)=\dist_{\ell^\infty}(n,\Z^d\setminus\Lambda)$ for
$n\in\Lambda$. For every $a_0\in(0,\gamma_0)$ there is a constant
$K_{a_0}$, independent of $\Lambda$, such that
\begin{equation}
 \|(1-P_\Lambda)C\widetilde q\|_2
 \le K_{a_0}\|e^{-a_0d_\Lambda}q\|_2
 \le K_{a_0}\left(\|\one_{\{d_\Lambda<\ell\}}q\|_2
                                      +e^{-a_0\ell}\|q\|_2\right).
 \label{im:eq:leakage}
\end{equation}
\end{lemma}
\begin{proof}
For $x\notin\Lambda$ and $y\in\Lambda$,
$d_\Lambda(y)\le|x-y|_1$. The matrix with entries
$c_{x-y}e^{a_0d_\Lambda(y)}$ is therefore bounded in absolute value
by $A_0e^{-(\gamma_0-a_0)|x-y|_1}$. Its row and column sums are
bounded by $A_0\sum_{r\in\Z^d}e^{-(\gamma_0-a_0)|r|_1}$.
The Schur test proves the first inequality after factoring out
$e^{-a_0d_\Lambda(y)}q(y)$. Splitting at $d_\Lambda=\ell$ proves
the second.
\end{proof}

\begin{corollary}[Interior finite-sample modes]\label{im:cor:finite}
Let $I\Subset(0,c_M)$ be a compact interval such that
\eqref{im:eq:regular} holds. Suppose that $\mu_WE\in I$, $\|q\|_2=1$, and an eigenvector of the truncated matrix satisfies
\begin{equation}
 \|\one_{\{d_\Lambda<\ell\}}q\|_2\le C_0W,\qquad
 e^{-a_0\ell}\le W.
 \label{im:eq:bulk-mode}
\end{equation}
For sufficiently small $W$, the bounds \eqref{im:eq:eigen-bounds}
hold uniformly in $\Lambda$, with constants also depending on $C_0$.
\end{corollary}
\begin{proof}
Equations \eqref{im:eq:boundary-error} and \eqref{im:eq:leakage} bound
the full-lattice residual by $C W$. Apply
Proposition~\ref{im:prop:quasimode}.
\end{proof}

This hypothesis need not hold for every finite eigenvector. It excludes
substantial boundary concentration, which should not be mistaken for
bulk Anderson localisation. Nevertheless, there is a uniform count
of the exceptions.

\begin{theorem}[A bound on the number of boundary-affected modes]\label{im:thm:most}
Let $I\Subset(0,c_M)$ be a compact interval such that
\eqref{im:eq:regular} holds. Fix $a_0\in(0,\gamma_0)$ and $0<W\le W_0<1$, and set
$\ell_W=\lceil a_0^{-1}\log(1/W)\rceil+1$. Write
$S_{\Lambda,W}=\{n\in\Lambda:d_\Lambda(n)<\ell_W\}$.
Let $(\psi_j)_{j=1}^{|\Lambda|}$ be any orthonormal eigenbasis of
$H_{\varpi,\Lambda}^t=M_{\varpi,\Lambda}^{-1/2}
C_\Lambda^tM_{\varpi,\Lambda}^{-1/2}$. At most
\begin{equation}
 \frac{|S_{\Lambda,W}|}{W^2}
 \label{im:eq:bad-count}
\end{equation}
of these eigenvectors can have squared mass in $S_{\Lambda,W}$
greater than $W^2$. Every remaining eigenvector whose energy satisfies
$\mu_WE_j\in I$ obeys the pointwise, participation, and radius
bounds \eqref{im:eq:eigen-bounds}, with uniform constants after
normalisation in the chosen variables.

For cubes $\Lambda_L=\{-L,\ldots,L\}^d$ with $L\ge2\ell_W$, the
fraction of possible exceptions among the entire eigenbasis is at most
\begin{equation}
 C\frac{\log(1/W)+1}{L W^2}.
 \label{im:eq:bad-fraction}
\end{equation}
Thus for each fixed sufficiently small $W>0$ this fraction tends to
zero as the sample grows. The assertion is valid for every material
configuration in the bounded range, not only in expectation.
\end{theorem}
\begin{proof}
Completeness of the orthonormal basis gives
\[
 \sum_{j=1}^{|\Lambda|}\|\one_{S_{\Lambda,W}}\psi_j\|_2^2
                =\operatorname{Tr}\one_{S_{\Lambda,W}}=|S_{\Lambda,W}|.
\]
Each exceptional vector contributes more than $W^2$, proving
\eqref{im:eq:bad-count}. For a nonexceptional vector, put
$q_j=M^{-1/2}\psi_j/\|M^{-1/2}\psi_j\|_2$. Then
\[
 \|\one_{S_{\Lambda,W}}q_j\|_2^2
       \le(1+W)\|\one_{S_{\Lambda,W}}\psi_j\|_2^2
       \le(1+W_0)W^2.
\]
Also $e^{-a_0\ell_W}\le W$. Corollary~\ref{im:cor:finite} applies,
and bounded multiplication by $M^{1/2}$ transfers the estimates back
to $\psi_j$. For cubes, elementary layer counting gives
$|S_{\Lambda_L,W}|\le C_d\ell_W(2L+1)^{d-1}$, proving
\eqref{im:eq:bad-fraction}.
\end{proof}

The denominator in \eqref{im:eq:bad-fraction} counts all modes. To conclude
that a vanishing fraction of the modes with $\mu_WE_j\in I$ are exceptions, one also
needs their count to be a positive fraction of $|\Lambda_L|$; for
random growing samples this can be provided by an appropriate
density-of-states limit with positive mass in $I/\mu_W$. No
eigenvalue simplicity assumption is needed for Theorem~\ref{im:thm:most};
the count holds for any fixed orthonormal eigenbasis.

For finite capacitance matrices, the residual acquires
$(C_\Lambda^t-C_\Lambda^f)q$. Proposition~\ref{fc:prop:boundary}
bounds this additional term by an exponentially weighted boundary norm;
see Corollary~\ref{cor:finite-interior-extent}. Frequency-dependent
radiating reductions \cite{ALLSU} require a different argument, since their modes
need not form an orthonormal basis of a self-adjoint matrix.

\subsection{Physical scales and the interior spectral question}
Set $R_{2,\rm phys}=\varepsilon R_2$ and allow the disorder to
depend on the cell size, $W=W_\varepsilon$. The individual bound becomes
\begin{equation}
 R_{2,\rm phys}\ge c_I\frac{\varepsilon}{W_\varepsilon}
                            -C_I'\varepsilon.
 \label{im:eq:physical-radius}
\end{equation}
For fixed $W>0$, this lower bound tends to zero as the cells shrink.
At $W_\varepsilon\asymp\sqrt\varepsilon$ it is of order
$\sqrt\varepsilon$, and the exceptional fraction in
\eqref{im:eq:bad-fraction} need not be small. Thus these estimates
do not establish macroscopic delocalisation in either scaling.

For every fixed sufficiently small $W$, the boundary-exception fraction
does vanish in growing cubes. Since the density-of-states measure has
full support by Theorem~\ref{thm:IDS}, every fixed regular interval
$I$ with nonempty interior gives a positive density of modes with
$E_j\in I/\mu_W$. The exceptional
fraction also vanishes relative to those modes, by testing the
density-of-states limit with a nonnegative continuous function supported
inside $I/\mu_W$. Exponential localisation on a large lattice
length remains compatible with all these lower bounds.

\FloatBarrier
\part{Finite resonator clusters}\label{part:finite}

\section{Finite capacitance matrices}\label{sec:finite-crystals}

We now define the capacitance matrix of a finite cluster and quantify the boundary correction caused by removing the exterior resonators.
We retain the same material realisation in the two finite constructions.
The static model is treated first; all boundary constants below are
uniform for the family in \eqref{eq:uniform-cell}.

\FiniteFigure

\subsection{Variational definition}
The truncated matrices $C_\Lambda^t$ and $H_{\varpi,\Lambda}^t$ were
defined in \eqref{eq:finite}. To define the finite capacitance matrix, use
the energy spaces
\[
 \mathscr E_d=
 \begin{cases}
 \{u\in H^1_{\rm loc}(\R^2):\nabla u\in L^2(\R^2)\},&d=2,\\
 \{u\in H^1_{\rm loc}(\R^3)\cap L^6(\R^3):
                         \nabla u\in L^2(\R^3)\},&d=3.
 \end{cases}
\]
For a nonempty finite set $\Lambda$ define
\begin{equation}
 q^*C_\Lambda^fq=\frac1{|D|}
  \min_{U\in\mathscr E_d,\ U=q_n\ {\rm in}\ D+n\ (n\in\Lambda)}
                 \int_{\R^d}|\nabla U|^2,
 \qquad
 H_{\varpi,\Lambda}^f=B_{\varpi,\Lambda}^{1/2}
                C_\Lambda^fB_{\varpi,\Lambda}^{1/2}.
 \label{eq:physical-finite-cap}
\end{equation}
This is the static limit of the variational construction in
\cite[Section~3.1]{EM}. In dimension three, it is the usual free-space
capacitance matrix divided by $|D|$. In dimension two, constants are
admissible and $C_\Lambda^f\one=0$, where $\one$ denotes the constant
vector with entries one.
The direct method gives the minimiser. In dimension three use the Sobolev
inequality in the homogeneous energy space; in dimension two an imposed
value on one resonator fixes the additive constant, giving local $H^1$
compactness. The constraints are closed and the energy is strictly convex
on functions with zero constrained traces.

For $C_\Lambda^t$, every omitted resonator remains present with prescribed
potential zero. In \eqref{eq:physical-finite-cap} it is absent and no
constraint is imposed in its former location. Hence
\begin{equation}
 0\le C_\Lambda^f\le C_\Lambda^t\le c_M\Id,
 \qquad 0\le H_{\varpi,\Lambda}^f
       \le H_{\varpi,\Lambda}^t\le c_M\Id
 \label{eq:finite-ordering}
\end{equation}
for the main law, since $0<B_\varpi\le\Id$. The first comparison follows
by inserting the globally constrained minimiser as a competitor for the
finite-cluster problem.

For $\delta>0$ the corresponding contrast-dependent matrix is
\begin{equation}
 q^*C_{\delta,\Lambda}^fq
 =\frac1{|D|}
 \min_{\substack{U\in\mathscr E_d\\ \fint_{D+n}U=q_n\ (n\in\Lambda)}}
 \int_{\R^d}a_{\delta,\Lambda}|\nabla U|^2,\qquad
 a_{\delta,\Lambda}=\delta^{-1}\one_{\cup_{n\in\Lambda}(D+n)}
                       +\one_{\R^d\setminus\cup_{n\in\Lambda}(D+n)} .
 \label{eq:finite-delta}
\end{equation}
Set $C_{0,\Lambda}^f=C_\Lambda^f$ and, for $a\in\{t,f\}$,
$H_{\delta,\varpi,\Lambda}^a
=M_{\varpi,\Lambda}^{-1/2}C_{\delta,\Lambda}^a
 M_{\varpi,\Lambda}^{-1/2}$.
The same variational comparison holds for every sufficiently small
$\delta$, with a common upper bound. In dimension two the constant
vector lies in the kernel of $C_{\delta,\Lambda}^f$; all localisation
intervals considered below are separated from zero.

\subsection{The boundary comparison}\label{fc:sec:boundary}

With the depth $d_\Lambda$ from Lemma~\ref{im:lem:leakage}, set
\begin{equation}
 \partial_r\Lambda=\{n\in\Lambda:d_\Lambda(n)\le r\}.
 \label{fc:eq:depth}
\end{equation}
The variational framework of \cite{EM} gives an exponentially
weighted boundary comparison.

\begin{proposition}[Boundary correction]\label{fc:prop:boundary}
There are $K,\gamma>0$, independent of finite $\Lambda$ and $0\le\delta\le\delta_0$, such that
\begin{equation}
 0\le R_{\delta,\Lambda}:=C^t_{\delta,\Lambda}-C^f_{\delta,\Lambda}
 \le K\diag\bigl(e^{-\gamma d_\Lambda(n)}\bigr)
 \quad\text{as quadratic forms}.                       \label{fc:eq:boundary-form}
\end{equation}
Also $0\le C^f_{\delta,\Lambda}\le C^t_{\delta,\Lambda}\le K\Id$.
\end{proposition}

\begin{proof}
This is Lemma 5.1 of \cite{EM}. Its proof, based on Lemma 4.4 there, is uniform in the cell contrast and applies to arbitrary finite cell sets. We recall the argument, writing $R=R_{\delta,\Lambda}$.

The full-lattice minimiser with zero-extended means is admissible for the finite problem, and the finite coefficient is no larger than the full periodic coefficient. This proves positivity of $R$. Local cutoff lifts equal to one near $D_n$, with disjoint cell supports and gradient zero in the resonators, give a common upper bound $K\Id$.

Suppose the prescribed means $q$ are supported at depth greater than $m$, and let $U$ minimise the finite energy. The prescribed means vanish in the first \(m\) layers of retained cells. No mean constraint is imposed outside the retained array. Choose layer cutoffs that are constant in each resonator; their gradients are then supported where the coefficient is one. Testing the Euler equation with $\chi^2U$ and using the cell Poincar\'e inequality for zero-mean functions gives
\[
 \mathcal E(r)\le\frac{C}{k^2}\mathcal E(r+k),\qquad r+k\le m,
\]
Here $\mathcal E(r)$ is the weighted energy in the exterior together with the first $r$ layers, and $\mathcal E(\infty)$ is the total weighted energy of $U$. The cell Poincar\'e constant and cutoff bounds depend on $D\Subset Y$, not on $\delta$ or the sample. Choosing a fixed sufficiently large $k$ and iterating yields
\begin{equation}
 \mathcal E(r)\le C\rho^{m-r}\mathcal E(\infty),\qquad 0<\rho<1.
 \label{fc:eq:layer-decay}
\end{equation}
Cut off $U$ between depths $m/3$ and $2m/3$ to make a full-lattice competitor with means $E_\Lambda q$. The removed part has zero means on all retained resonators, hence is energy-orthogonal to $U$. Equation~\eqref{fc:eq:layer-decay} then gives
\begin{equation}
 \langle R_{\delta,\Lambda}q,q\rangle
       \le C\rho^{m/3}\|q\|^2
       \quad\text{if }\supp q\subset\{d_\Lambda>m\}.
 \label{fc:eq:deep-form}
\end{equation}
For arbitrary $q$, write $q=\sum_{m\ge0}Q_mq$, where $Q_m$ projects onto the sites with $m<d_\Lambda(n)\le m+1$. Positivity allows use of $R^{1/2}$. The triangle inequality followed by weighted Cauchy--Schwarz and \eqref{fc:eq:deep-form} gives
\[
 \|R^{1/2}q\|^2
 \le C\left(\sum_{m\ge0}\rho^{m/6}\|Q_mq\|\right)^2
 \le C\sum_{m\ge0}\rho^{m/6}\|Q_mq\|^2.
\]
This is \eqref{fc:eq:boundary-form} after changing the constants. The construction of the cutoffs uses the fixed positive gap between $D$ and the cell faces; harmless fixed shifts in layer indices do not change the estimate. Passing to the static variational limit gives the same statement at $\delta=0$.
\end{proof}

The bound is a positive quadratic-form estimate; it does not assert exponential decay of every entry of $C^f$ in the separation between its indices. In particular, interactions between sites close to the exterior boundary are not controlled by their mutual distance in \eqref{fc:eq:boundary-form}.

\begin{corollary}[Random boundary correction and surface rates]\label{fc:cor:boundary}
Write $S=M_{\varpi,\Lambda}^{-1/2}$ and suppress $\delta,\varpi,\Lambda$
within this corollary. For \eqref{eq:random-model}, put $T=H^t-H^f=SRS$. Then, pathwise,
\begin{equation}
 0\le T\le K\diag(e^{-\gamma d_\Lambda(n)}).            \label{fc:eq:random-boundary}
\end{equation}
If $P_r$ projects onto $\ell^2(\partial_r\Lambda)$ and $Q_r=\Id-P_r$, then
\begin{equation}
 \|TQ_r\|\le K_1e^{-\gamma r/2},\qquad
 \|T-P_rTP_r\|\le2K_1e^{-\gamma r/2},\qquad
 \rank(P_rTP_r)\le|\partial_r\Lambda|.                 \label{fc:eq:lowrank}
\end{equation}
Let $E_{1,L}^a\le\cdots\le E_{N_L,L}^a$ be the eigenvalues of
$H_{\delta,\varpi,\Lambda_L}^a$, counted with multiplicity, for
$a\in\{t,f\}$. Uniformly in $\delta$ and the realisation,
\begin{align}
 \frac1{N_L}\sum_{j=1}^{N_L}(E_{j,L}^t-E_{j,L}^f)
       &=\frac1{N_L}\Tr T=O(L^{-1}),                  \label{fc:eq:trace-rate}\\
 \frac1{N_L}\sum_{j=1}^{N_L}|E_{j,L}^t-E_{j,L}^f|^2
       &=O(L^{-1}).                                    \label{fc:eq:square-rate}
\end{align}
For complete cells in a bounded Lipschitz domain at scale $\varepsilon$, the corresponding right sides are $O(\varepsilon)$.
\end{corollary}

\begin{proof}
Diagonal multiplication commutes with the boundary weight, and $\|S\|\le1$, so \eqref{fc:eq:random-boundary} follows by congruence. For any vector $u$,
\[
 \|TQ_ru\|^2\le\|T\|\langle Q_ru,TQ_ru\rangle
                \le K^2e^{-\gamma r}\|u\|^2.
\]
Moreover $T-P_rTP_r=TQ_r+Q_rTP_r$, proving \eqref{fc:eq:lowrank}. Ordered eigenvalues satisfy $E_j^f\le E_j^t$, and their sum difference is $\Tr T$. The weighted layer sum is $O(L^{d-1})$ for cubes, proving \eqref{fc:eq:trace-rate}. Finally
$\|T\|_{\mathrm{HS}}^2=\Tr T^2\le\|T\|\Tr T$ and \eqref{fc:eq:HW} prove \eqref{fc:eq:square-rate}. For a Lipschitz sample, the cell boundary-layer count is $O(\varepsilon^{1-d}r)$ for $1\le r\ll\varepsilon^{-1}$; summation with the exponential weights proves the asserted rate.
\end{proof}

In particular, the Wasserstein distance of order one between the two empirical eigenvalue measures is $O(\varepsilon)$. This rate compares two models at the same random realisation. It is not a rate of convergence of either empirical measure to the infinite-volume density of states.

\FloatBarrier
\section{Random asymptotic equivalence and spectral convergence}\label{fc:sec:equivalence}

The deterministic comparison in \cite[Lemma 3.1]{ADHconv} is boundedness in operator norm together with convergence in the \emph{normalised Frobenius norm}. It is not convergence of the difference in operator norm. The following extension is purely deterministic and hence holds for every realisation.

\begin{theorem}[Pathwise weighted asymptotic equivalence]\label{fc:thm:equivalence}
Let $C_N^t,C_N^f$ be Hermitian $N\times N$ matrices satisfying
\[
 \sup_N(\|C_N^t\|+\|C_N^f\|)<\infty,\qquad
 a_N:=N^{-1/2}\|C_N^t-C_N^f\|_{\mathrm{HS}}\longrightarrow0.
\]
For any diagonal matrices $M_N$ with
$0<m_-\le(M_N)_{nn}\le m_+<\infty$, let $H_N^a=M_N^{-1/2}C_N^aM_N^{-1/2}$. Then
\begin{equation}
 N^{-1/2}\|H_N^t-H_N^f\|_{\mathrm{HS}}\le m_-^{-1}a_N.
 \label{fc:eq:HS}
\end{equation}
Their ordered eigenvalues $E_{j,N}^a$, $a\in\{t,f\}$, satisfy
\begin{equation}
 \frac1N\sum_{j=1}^N|E_{j,N}^t-E_{j,N}^f|^2
                \le m_-^{-2}a_N^2.                   \label{fc:eq:HW}
\end{equation}
Consequently their empirical eigenvalue measures have the same weak limits. No independence or ergodicity is required for this comparison.
\end{theorem}

\begin{proof}
Put $S_N=M_N^{-1/2}$. Then
$H_N^t-H_N^f=S_N(C_N^t-C_N^f)S_N$, and the Hilbert--Schmidt ideal inequality gives \eqref{fc:eq:HS}. The Hermitian eigenvalue inequality of Hoffman--Wielandt gives \eqref{fc:eq:HW}. Equivalently, it follows by diagonalising both matrices, using that the squared moduli of the change-of-basis entries form a doubly stochastic matrix, and applying the rearrangement inequality to the ordered eigenvalues. For a Lipschitz function $F$,
\[
 \left|\frac1N\Tr F(H_N^t)-\frac1N\Tr F(H_N^f)\right|
 \le\Lip(F)\left(\frac1N\sum_j|E_{j,N}^t-E_{j,N}^f|^2\right)^{1/2}.
\]
The spectra stay in a common compact interval. Uniform approximation of continuous functions by Lipschitz functions completes the proof.
\end{proof}

For \eqref{eq:random-model}, $m_-=1$, so \eqref{fc:eq:HS} is
independent of $W$. The two matrices with the same material realisation
have the common deterministic density-of-states measure of the random model of Theorem~\ref{thm:IDS};
that measure generally depends on $W$. The direct-factor law of
Appendix~\ref{sec:other-laws} gives the same comparison with the
constant $1+W$.

\subsection{Strong and spectral-set convergence}
We fix $W>0$ and write $d_H$ for the Hausdorff distance between
nonempty compact subsets of $\R$.
\begin{theorem}\label{thm:finite-limit}
For every fixed realisation and $a\in\{t,f\}$,
\begin{equation}
 E_{\Lambda_L}H_{\varpi,\Lambda_L}^aE_{\Lambda_L}^*
       \longrightarrow H_\varpi\quad\hbox{strongly},
 \qquad \Lambda_L=[-L,L]^d\cap\Z^d.
 \label{eq:strong-both}
\end{equation}
Resolvents converge strongly off the real axis and local spectral
measures converge weakly. Almost surely,
\begin{equation}
 d_H(\sigma(H_{\varpi,\Lambda_L}^a),[0,c_M])\longrightarrow0 .
 \label{eq:Hausdorff}
\end{equation}
The same statements hold at each fixed sufficiently small $\delta$,
with $H_{\delta,\varpi}$ and $[0,c_{M,\delta}]$.
If $\delta=\delta_L\to0$, the limits are the static ones.
\end{theorem}
\begin{proof}
The zero-extended truncated matrices are $P_LH_\varpi P_L$,
where $P_L\to\Id$ strongly. They converge strongly since $H_\varpi$
is bounded. By \eqref{fc:eq:random-boundary} and
$\|Tq\|^2\le\|T\|\langle Tq,q\rangle$, the difference of the two
extended finite matrices tends to zero on every finitely supported
vector: its support has depth tending to infinity. Uniform operator
bounds extend this convergence to all of $\ell^2$. This proves
\eqref{eq:strong-both}. The resolvent identity and polynomial
approximation give the asserted resolvent and local-measure limits.

The almost-sure spectrum is $[0,c_M]$ by
Proposition~\ref{prop:spectrum-new}. Every energy in this interval
has finitely supported approximate eigenvectors. Strong convergence
applied to each such fixed vector puts finite eigenvalues arbitrarily
close to that energy. All finite spectra lie in $[0,c_M]$ by
\eqref{eq:finite-ordering}. A finite net proves Hausdorff convergence.
For fixed $\delta$, the same spectrum proof applies: $c_\delta$
is continuous, nonnegative, vanishes at zero, and has maximum
$c_{M,\delta}$, while favourable material boxes approximate
$C_\delta$. For $\delta_L\to0$ use
$\|C_{\delta_L}-C\|=\|c_{\delta_L}-c\|_\infty\to0$ and the
uniform boundary comparison. The finite spectra are contained in
$[0,c_{M,\delta_L}]$, whose endpoints converge. No general
absence of spectral pollution in gaps is being inferred from strong
convergence.
\end{proof}

\subsection{The common density of states} \label{sec:14.2}
For fixed $W>0$ and $\delta$, define the probability measure
$\nu_{W,\delta}$ by the identity below for bounded Borel functions $F$:
\begin{equation}
 \int F\,\dd\nu_{W,\delta}
  =\EE\langle\delta_0,F(H_{\delta,\varpi})\delta_0\rangle .
 \label{eq:DOS}
\end{equation}
Its cumulative distribution is the integrated density of states.
Write $\delta_E^{\rm D}$ for the unit point mass at energy $E$.

\begin{theorem}\label{thm:IDS}
Along growing cubes, almost surely,
\[
 \frac1{|\Lambda_L|}\sum_{j=1}^{|\Lambda_L|}\delta_{E_{j,L}^a}^{\rm D}
       \rightharpoonup\nu_{W,\delta},\qquad a\in\{t,f\}.
\]
For $\delta_L\to0$ both limits are $\nu_{W,0}$ on one
probability-one event for the prescribed sequence. The support of
$\nu_{W,\delta}$ is $[0,c_{M,\delta}]$. It has a locally bounded
density on $(0,\infty)$ and no atom at zero. Consequently the
normalised eigenvalue-counting functions converge at every energy.
\end{theorem}
\begin{proof}
For truncated matrices, use stationarity and ergodicity of the infinite family.
The ergodic theorem applies to the diagonal entries of each polynomial
in $H_{\delta,\varpi}$; see \cite{Kirsch}. Cut off hopping beyond
distance $R$. The norm error is bounded uniformly in the realisation
by the summable coefficient tail. For a polynomial of degree $k$ in
this finite-range operator, finite and infinite diagonal entries agree
outside a boundary strip of width $kR$. Its relative size vanishes.
Let $L\to\infty$, then $R\to\infty$, and approximate continuous
functions by polynomials on the common compact spectral interval.
A countable dense family of polynomials gives one probability-one
event for weak convergence. The trace estimate
\eqref{fc:eq:trace-rate} transfers it to finite capacitance matrices.

For moving contrast,
$\sup_{\varpi,\Lambda}\|H_{\delta,\varpi,\Lambda}^t
-H_{0,\varpi,\Lambda}^t\|\le\|c_\delta-c\|_\infty\to0$.
Apply the static result and the uniform trace comparison.
If an open interval has zero density-of-states mass, positivity and
stationarity make its spectral projection vanish at every coordinate
almost surely, hence vanish as an operator. This identifies the
support with the almost-sure spectrum.

Lemma~\ref{lem:material-average} gives
$\nu_{W,\delta}(J)\le(1+W)|J|/(WE_0)$ for
$J\subset[E_0,\infty)$, $E_0>0$.
The variational principle shows that $c_\delta$ vanishes only at
zero: zero energy forces the admissible function to be constant
and every Bloch phase to be one. Thus $\ker C_\delta=\{0\}$
as a multiplier on $L^2(Y^*)$. Bounded invertibility of
$M_\varpi^{1/2}$ implies $\ker H_{\delta,\varpi}=\{0\}$.
There is no atom at zero. The distribution function is continuous,
which proves the final assertion.
\end{proof}

The same proof applies to tempered F{\o}lner sequences, by the
pointwise ergodic theorem of \cite{Lindenstrauss}. For example, the sets
$I_{1/k}$ associated with a fixed bounded
Lipschitz domain form a tempered F{\o}lner sequence:
$|I_{1/k}|\asymp k^d$, fixed-width boundary fractions vanish,
and $\bigcup_{j<k}(I_{1/k}-I_{1/j})$ lies in a cube of $O(k^d)$
sites. The same density-of-states conclusion therefore holds
along these shrinking-cell samples, whether or not they are nested. 
The boundary comparison applies to each bounded Lipschitz sample separately.
At $\delta=\varepsilon^2$, the frequency distribution is the
push-forward of the reduced energy distribution under
$E\mapsto v_r\sqrt E$. This map is independent of $\varepsilon$,
so the moving-contrast limit is the push-forward of $\nu_{W,0}$.
More generally, for fixed positive $\varepsilon,\delta$, if
$\rho_{W,\delta}$ is the density of $\nu_{W,\delta}$ in energy,
the corresponding frequency density at $\omega>0$ is
\begin{equation}
 \frac{2\varepsilon^2\omega}{\delta v_r^2}\,
 \rho_{W,\delta}\!\left(\frac{\varepsilon^2\omega^2}
                              {\delta v_r^2}\right).
 \label{eq:frequency-DOS}
\end{equation}
The factor is the Jacobian $\dd E/\dd\omega$; an energy density
cannot be read directly as a frequency density.

A density-of-states measure can have a density in a localisation
regime: averaging atomic random spectral measures need not produce
a singular measure. Neither spectral-set convergence nor
asymptotic equivalence proves the absence of absolutely continuous
spectrum. That conclusion follows from Part~\ref{part:infinite}.

\section{Localisation of finite-cluster modes}\label{sec:cluster-localisation}\label{fc:sec:transfer}

We use the uniform estimates for truncated matrices from
Theorem~\ref{thm:localisation}, Corollary~\ref{cor:strong-localisation},
and Proposition~\ref{prop:uniform-localisation}. Let $I=(E_-,E_+)$ be bounded, $E_->0$, and suppose that for the truncated matrices
\begin{equation}
 \EE Q^t_{\delta,\varpi,\Lambda}(x,y;I)
       \le A e^{-\nu|x-y|_1}                           \label{fc:eq:principal-Q}
\end{equation}
uniformly in $\Lambda$ and the parameters considered. Here $Q^t$
is the correlator in \eqref{eq:Q-definition}, with $C_\delta$ in place
of $C$. The estimate holds near the upper edge for every fixed $W>0$
and on positive intervals satisfying \eqref{eq:uniform-large}.

The approximation of interior localised modes is related to the defect-mode comparison in \cite{ADHdef}. Here one cannot assume a fixed isolation gap: the finite random eigenvalues become dense. Proposition~\ref{fc:prop:Minami} provides a sufficiently small gap scale with high probability.

\begin{theorem}[Localisation outside an exceptional family]\label{fc:thm:transfer}\label{thm:cluster-transfer}
Assume \eqref{fc:eq:random-boundary}, \eqref{eq:random-model}, and
\eqref{fc:eq:principal-Q}, and let $N=|\Lambda|\ge2$.
For every $p>0$, there are $C_p,R_p<\infty$, independent of the finite sample, such that with probability at least $1-C_pN^{-p}$ the following holds. There is an excluded set $\mathcal B_{\delta,\varpi,\Lambda,p}(I)$
of eigenvalue indices in $I$, with
\begin{equation}
 |\mathcal B_{\delta,\varpi,\Lambda,p}(I)|
 \le C_p\bigl|\partial_{R_p\log(2+N)}\Lambda\bigr| \label{fc:eq:exceptions}
\end{equation}
such that every normalised eigenvector $\psi_j^f$ with $E_j^f\in I$
and $j\notin\mathcal B_{\delta,\varpi,\Lambda,p}(I)$ has a centre
$n_j\in\Lambda$ satisfying
\begin{equation}
 \sum_{|n-n_j|_1>R_p\log(2+N)}|\psi_j^f(n)|^2\le N^{-p}.
 \label{fc:eq:finite-local}
\end{equation}
Each retained eigenvector is, up to phase, within $C_pN^{-(p+5)}$
in $\ell^2$ of a localised eigenvector of the truncated matrix. The excluded set
consists of the indices not obtained by the injective matching in the proof. The constants may depend on $W$, the positive-energy bounds, and the constants in \eqref{fc:eq:principal-Q}. They are uniform for the contrast-dependent family as $\delta\to0$. Intervals may vary with $\delta$ if their endpoints stay in a common bounded subset of $(0,\infty)$ and \eqref{fc:eq:principal-Q} remains uniform.
\end{theorem}

\begin{proof}
Fix $p>0$. In this proof write $H^{t,f}=H^{t,f}_{\delta,\varpi,\Lambda}$
and $T=H^t-H^f$; let $K$ be the common norm bound in
\eqref{fc:eq:random-boundary}. We use a spectral gap scale $g_N=N^{-b}$ with $b=p+3$, and an error scale $\eta_N=N^{-q}$ with $q=2p+8$. These satisfy
\begin{equation}
 N^2g_N\le N^{-p-1},\qquad
 N\eta_N\le N^{-p-1},\qquad
 \eta_N/g_N=N^{-(p+5)}.                                \label{fc:eq:scales}
\end{equation}
Enlarging fixed constants handles bounded values of $N$.

\emph{Step 1: one event for all eigenvectors of the truncated matrix.}
From \eqref{fc:eq:principal-Q}, Markov's inequality and a union bound over pairs of sites give, with probability $1-O(N^{-p})$, an exponential bound with a polynomial prefactor for every eigenvector of $H^t$ in $I$. Choosing the constant in a logarithmic radius sufficiently large, one obtains simultaneously centres $n_j$ and
\begin{equation}
 \|\one_{\{|n-n_j|_1>\rho_N\}}\psi_j^t\|_2
       \le\frac{\eta_N}{8(K+1)},\qquad
 \rho_N=a_p\log(2+N).                                 \label{fc:eq:principal-tail}
\end{equation}
Indeed, simplicity gives
$|\psi_j^t(x)\psi_j^t(y)|\le Q^t(x,y;I)$; at a maximum site $|\psi_j^t(n_j)|\ge N^{-1/2}$. A common estimate
$Q^t(x,y;I)\le N^{p+3}e^{-\nu|x-y|_1/2}$ has failure probability $O(N^{-p-2})$. Summing its resulting squared eigenvector bound outside $a_p\log(2+N)$ gives \eqref{fc:eq:principal-tail} for any prescribed polynomial error after increasing $a_p$.

\emph{Step 2: the boundary perturbation is almost finite rank.}
Choose $r_N=a'_p\log(2+N)$ large enough that, with $P=P_{r_N}$,
\begin{equation}
 \|T(\Id-P)\|\le\eta_N/8,\qquad
 \|T-PTP\|\le\eta_N/4.                               \label{fc:eq:small-deep}
\end{equation}
This uses \eqref{fc:eq:lowrank} and holds deterministically. Set
$r=\rank P=|\partial_{r_N}\Lambda|$. Then
\[
 H^f=H^t-PTP+\mathcal E_N,\qquad \|\mathcal E_N\|\le\eta_N/4.
\]
The large part of the perturbation has rank at most $r$.

\emph{Step 3: eigenvalue separation and interval endpoints.}
Apply Proposition~\ref{fc:prop:Minami} to $C_*=C^f_{\delta,\Lambda}$. In a fixed positive-energy neighbourhood of $\overline I$, all distinct eigenvalues of $H^f$ are at distance at least $g_N$, except with probability $O(N^2g_N)=O(N^{-p-1})$. The first estimate in \eqref{fc:eq:Minami}, applied to $H^t$, also shows that there is no eigenvalue of $H^t$ within $2\eta_N$ of either endpoint of $I$, except with probability $O(N\eta_N)$. Both estimates are uniform in the deterministic capacitance matrix. Take the intersection with the event of Step 1.

\emph{Step 4: match each interior-centred eigenvector of the truncated matrix.}
Consider an eigenvector from \eqref{fc:eq:principal-tail} with
\begin{equation}
 d_\Lambda(n_j)>r_N+\rho_N+1.                          \label{fc:eq:deep-centre}
\end{equation}
Its localisation ball lies in the range of $\Id-P$. By \eqref{fc:eq:principal-tail} and \eqref{fc:eq:small-deep},
\begin{equation}
 \|(H^f-E_j^t)\psi_j^t\|=\|T\psi_j^t\|
       \le\eta_N/8+K\eta_N/[8(K+1)]<\eta_N/4.         \label{fc:eq:residual}
\end{equation}
There is an eigenvalue $E_k^f$ within $\eta_N/4$ of $E_j^t$, and the endpoint event ensures it belongs to $I$. The gap event makes it the unique such eigenvalue. Expanding $\psi_j^t$ in an eigenbasis of $H^f$ gives
\[
 \|(\Id-P_k^f)\psi_j^t\|\le\frac{\eta_N}{g_N-\eta_N},
\]
where $P_k^f$ is its one-dimensional projection. Thus, for a suitable phase,
\begin{equation}
 \|\psi_k^f-e^{i\varphi_j}\psi_j^t\|_2
       \le3\eta_N/g_N.                                \label{fc:eq:matching}
\end{equation}
Two orthogonal eigenvectors of $H^t$ cannot be matched to the same $P_k^f$: each would have squared projection larger than $1/2$ onto a one-dimensional space, contradicting Bessel's inequality. The matching is injective. Equations~\eqref{fc:eq:principal-tail}, \eqref{fc:eq:matching}, and \eqref{fc:eq:scales} prove \eqref{fc:eq:finite-local} for each matched eigenvector, with room in the exponents to absorb fixed constants.

\emph{Step 5: count the unmatched finite modes.}
Eigenvectors of $H^t$ failing \eqref{fc:eq:deep-centre} have all but $O(\eta_N^2)$ of their mass in
$\partial_{r_N+2\rho_N+2}\Lambda$. Orthogonality therefore bounds their number by
\begin{equation}
 2\bigl|\partial_{r_N+2\rho_N+2}\Lambda\bigr|.          \label{fc:eq:bad-centres}
\end{equation}
This bounds the number of boundary-centred modes of the truncated matrix without claiming that the eventual exceptional finite modes themselves must be supported in that layer.

A rank-$r$ perturbation changes a cumulative eigenvalue count by at most $r$. The additional norm error in Step 2 can move a counting threshold by at most $\eta_N/4$. Since there are no eigenvalues of $H^t$ within $2\eta_N$ of the endpoints, the eigenvalue counts for $H^f$ and $H^t$ on $I$ differ by at most $2r$. Consequently, the number of finite eigenvectors in $I$ not reached by the injective matching is at most
\[
 2r+2\bigl|\partial_{r_N+2\rho_N+2}\Lambda\bigr|
 \le4\bigl|\partial_{r_N+2\rho_N+2}\Lambda\bigr|.
\]
Define $\mathcal B_{\delta,\varpi,\Lambda,p}(I)$ to be precisely
these unmatched indices. This proves \eqref{fc:eq:exceptions}. More explicitly, if $n_A(E)=\Tr\one_{(-\infty,E]}(A)$, rank perturbation and norm perturbation give
\[
 n_{H^t}(E-\eta_N/4)-r
 \le n_{H^f}(E)
 \le n_{H^t}(E+\eta_N/4)+r.
\]
Apply this at both endpoints. Positivity of $PTP$ actually improves the interval bound to $r$ on the endpoint event, but $2r$ suffices. Finally add the three failure probabilities from Steps 1 and 3. This proves the theorem.
\end{proof}

\begin{remark}[The excluded family and the decay statement]
We call the unmatched indices in the theorem \emph{boundary-affected}
only in the sense of this construction. The bound on their number
comes from a boundary-supported perturbation and the modes of the truncated matrix
centred near the boundary. It does not show that each excluded
finite-cluster eigenvector is itself supported in a boundary layer.
The excluded set can depend on $p$, the interval, and the realisation.

The matched vectors satisfy an exponential bound with a polynomial prefactor inherited from $\psi_j^t$, with an additional uniformly small error from \eqref{fc:eq:matching}. Thus the result proves concentration and closeness to an exponentially localised profile. It does not prove a pure exponential bound at every separation for $\psi_k^f$ itself; the small residual tail must not be omitted.
\end{remark}

\subsection{Localisation as the cells shrink}\label{fc:sec:epsilon}

Use the cell set $I_\varepsilon$ and its cardinality $N_\varepsilon$
from \eqref{eq:sample-indices}, and take $\delta=\varepsilon^2$.
For $1\le r\ll\varepsilon^{-1}$, its boundary count obeys
\begin{equation}
 |\partial_r I_\varepsilon|\le C_\Omega\varepsilon^{1-d}(r+1).
 \label{fc:eq:surface-count}
\end{equation}
Indeed a retained cell within $r$ lattice steps of an omitted cell lies in a physical tubular neighbourhood of $\partial\Omega$ of thickness $C\varepsilon(r+1)$. Its volume is $O(\varepsilon(r+1))$ for a bounded Lipschitz domain; divide by the cell volume.

\begin{theorem}[Localisation in finite resonator clusters]\label{thm:cluster-main}\label{fc:cor:physical}
Assume the geometry of Section~\ref{sec:model}, including
\eqref{eq:reflection}, and the law \eqref{eq:random-model}.
For every fixed $W>0$, let
\begin{equation}
 I_{W,\delta}=(c_{M,\delta}-\Delta_W,c_{M,\delta})
 \label{fc:eq:edge-window}
\end{equation}
be the interval furnished by Proposition~\ref{prop:uniform-localisation}.
For every $p>0$ and all sufficiently small $\varepsilon$, with probability at least $1-C_{W,p}N_\varepsilon^{-p}$, there is an excluded set $\mathcal B_{\varepsilon,p}$ with
\begin{equation}
 |\mathcal B_{\varepsilon,p}|
 \le C_{W,p}\varepsilon^{1-d}\log(1/\varepsilon)
 \label{fc:eq:physical-exceptions}
\end{equation}
such that every $\ell^2$-normalised eigenvector of the random finite matrix
$H^f_{\varepsilon^2,\varpi,I_\varepsilon}$ with eigenvalue in
$I_{W,\varepsilon^2}$ and index outside $\mathcal B_{\varepsilon,p}$
satisfies \eqref{fc:eq:finite-local}. Here $\mathcal B_{\varepsilon,p}$ abbreviates
$\mathcal B_{\varepsilon^2,\varpi,I_\varepsilon,p}(I_{W,\varepsilon^2})$
from Theorem~\ref{thm:cluster-transfer}. The cell interpolations of
the retained eigenvectors concentrate in physical balls of radius
\begin{equation}
 r_\varepsilon=C_{W,p}\varepsilon\log(1/\varepsilon),
 \label{fc:eq:physical-radius}
\end{equation}
with squared tail at most $N_\varepsilon^{-p}$. For the original amplitudes $q=M^{-1/2}\psi$, normalising in $\ell^2$ costs at most a factor $1+W$ in that squared tail.

The same conclusion holds on a prescribed bounded positive interval
satisfying \eqref{eq:uniform-large}, without reflection symmetry,
and on the quantitative weak-disorder window of
Proposition~\ref{prop:uniform-localisation}.
It also holds for the static finite matrices $H^f_{\varpi,I_\varepsilon}$,
with the corresponding static intervals.
For the family $C_{\varepsilon^2}$ at $\delta=\varepsilon^2$, the
edge interval in frequency is
\begin{equation}
 v_r\sqrt{c_{M,\varepsilon^2}-\Delta_W}<\omega
           <v_r\sqrt{c_{M,\varepsilon^2}}.
 \label{eq:finite-frequency-window}
\end{equation}
The eigenvectors and their spatial estimates are unchanged by this
reparametrisation. The endpoints converge to
$v_r\sqrt{c_M-\Delta_W}$ and $v_r\sqrt{c_M}$.
\end{theorem}

\begin{proof}
Use Theorem~\ref{fc:thm:transfer}, the uniform correlator bound for truncated matrices,
and \eqref{fc:eq:surface-count}. The isometry $J_\varepsilon$ defined
in Section~\ref{sec:principal}
converts lattice radii $R\log(2+N_\varepsilon)$ to Euclidean radii at most
$\varepsilon R\log(2+N_\varepsilon)+\sqrt d\varepsilon/2$.
Also $(1+W)^{-1/2}\le\|M^{-1/2}\psi\|_2\le1$, which proves the final amplitude statement. The other intervals follow from Proposition~\ref{prop:uniform-localisation}
and the same transfer theorem.
\end{proof}

The exceptional fraction among all $N_\varepsilon$ modes is
\[
 O_{W,p}(\varepsilon\log(1/\varepsilon))\longrightarrow0.
\]
When the chosen interval has positive limiting density of states, it is also a vanishing fraction of the modes \emph{within that interval}. For the fixed-disorder edge interval, this is the case: the almost-sure spectrum of the infinite model is $[0,c_M]$, the support of its density of states is the same interval, and the limiting window $(c_M-\Delta_W,c_M)$ is nonempty. The moving-contrast eigenvalue-distribution result provides the corresponding positive asymptotic mode count along the sequences of growing samples considered in Section~\ref{sec:14.2}.

Along any sequence with $\sum_kN_{\varepsilon_k}^{-p}<\infty$, the concentration statement holds eventually almost surely. For example, $\varepsilon_k=1/k$ and $p>1/d$ suffice. The constants can deteriorate as $W\downarrow0$; arbitrary fixed positive disorder is distinct from a disorder strength tending to zero with $\varepsilon$.

\LocalisationFigure

\subsection{Interior extent and the scope of the finite-cluster result}

The boundary comparison also transfers the interior eigenfunction bounds to finite clusters, with an explicit bound on the number of excluded modes.

\begin{corollary}\label{cor:finite-interior-extent}
For the static finite capacitance matrix, the interior eigenfunction bounds
of Theorem~\ref{im:thm:eigenfunctions} hold outside a family of at
most $C|\partial_{\ell_W}\Lambda|/W^2$ modes, where
$\ell_W=C(1+\log(1/W))$ and $0<W\le W_0$.
The energy condition remains $\mu_WE_j\in I$, with the regular
interval $I$ of \eqref{im:eq:regular}. For each fixed small $W$,
the exceptional proportion vanishes along growing regular samples.
\end{corollary}
\begin{proof}
For a normalised finite-cluster amplitude $q$, its zero extension obeys
\[
 (C-\mu_WE)\widetilde q
 =EWZ_\varpi\widetilde q
   +E_\Lambda(C_\Lambda^t-C_\Lambda^f)q
   +(1-P_\Lambda)C\widetilde q .
\]
Lemma~\ref{im:lem:leakage} controls the last term, and
\eqref{fc:eq:boundary-form} together with
$\|Rq\|^2\le\|R\|\langle Rq,q\rangle$ controls the middle term.
Their sum is $O(W)$ whenever the squared $\ell^2$ mass in
$\partial_{\ell_W}\Lambda$ is $O(W^2)$ and $\ell_W$ is chosen
as above. Proposition~\ref{im:prop:quasimode} applies.
For an orthonormal eigenbasis in the self-adjoint variables,
the sum of squared boundary masses is $|\partial_{\ell_W}\Lambda|$.
The counting and bounded-change-of-variable argument in
Theorem~\ref{im:thm:most} proves the assertion.
\end{proof}

The matching theorem controls individual finite-cluster eigenvectors;
it requires the spacing estimate as well as the boundary comparison.
It leaves the unmatched family unclassified. The variational matrix
also differs from the outgoing acoustic problem, whose resonances can
be complex. Transfer to that problem requires a reduction and field
reconstruction estimate uniform in sample size and compatible with the
shrinking-cell, fixed-disorder limit.

\FloatBarrier
\section{Numerical illustrations}\label{sec:numerics}

We illustrate the static finite-cluster problem with square arrays of
disks. The calculations use the finite capacitance matrix of
\eqref{eq:physical-finite-cap} and retain all pair interactions.
They use the static limit $C_0=C$ of the reduced model; physical
frequencies are assigned through \eqref{eq:frequency-conversion}
with positive contrast.

\subsection{Geometry, discretisation, and observables}
Let $D$ be the disk of radius $0.2$ centred at the origin, and set
\[
 \mathcal S_L=\{0,\ldots,L-1\}^2,\qquad
 N=|\mathcal S_L|=L^2,\qquad L\in\{36,51,70\}.
\]
Here $L$ counts cells along one side; $\mathcal S_L$ is distinguished
from the centred boxes $\Lambda_L$ used above. The lattice spacing is
one. After rescaling by $\varepsilon=L^{-1}$ and translating the
sample, its cells fill the unit square.

We discretise the logarithmic single-layer potential on each disk by
a Fourier--Galerkin method, retaining the constant, sine, and cosine
modes through order two. The potential at infinity is an unknown
constant, and the total charge is constrained to vanish, consistently
with the two-dimensional energy convention in
\eqref{eq:physical-finite-cap}. Eliminating the nonconstant harmonics
and dividing the resulting potential-to-charge map by $|D|$ gives the numerical
matrix $C^{f,(2)}_{\mathcal S_L}$. We form
\begin{equation}
 H^{f,(2)}_{\varpi,\mathcal S_L}
 =M_{\varpi,\mathcal S_L}^{-1/2}C^{f,(2)}_{\mathcal S_L}
                  M_{\varpi,\mathcal S_L}^{-1/2},
 \qquad m_n=1+W\xi_n,
 \label{num:eq:matrix}
\end{equation}
with the law \eqref{eq:random-model}. We use $500$ independent
realisations for each $W\in\{0.1,0.3,1,4\}$ and each $L$. The $W=0$ control is deterministic and is computed once.

Write $(E_j,\psi_j)$ for the eigenpairs of \eqref{num:eq:matrix},
with $\|\psi_j\|_2=1$ and positive eigenvalues in increasing order,
$j=1,\ldots,N-1$. The zero eigenvalue, whose eigenvector is
proportional to $M_{\varpi,\mathcal S_L}^{1/2}\one$, is omitted.
As in \eqref{eq:unitary}, $q_j=M_{\varpi,\mathcal S_L}^{-1/2}\psi_j$;
thus the plotted mass $|\psi_j(n)|^2$ equals $m_n|q_j(n)|^2$.
The reference periodic edge is computed independently from the
grounded-cell problem at $\alpha_M=(\pi,\pi)$, giving
$c_M\simeq50.4013$. All normalised energies in the figures use this
numerical reference. For an individual mode,
$E_j/c_M=(\omega_j/\omega_M)^2$; thus a point at $E/c_M=0.45$
corresponds to $\omega/\omega_M=\sqrt{0.45}\simeq0.671$.

For a normalised mode $\psi$, use the participation number $P(\psi)$
from \eqref{im:eq:measures}. Let $n_*$ be a site where its mass is
maximal and define
\begin{equation}
 R_{90}(\psi)=\min\left\{r\ge0:
       \sum_{|n-n_*|_2\le r}|\psi(n)|^2\ge0.9\right\}.
 \label{num:eq:radius}
\end{equation}
Then $P/N$ measures the fraction of sites effectively occupied by the
mode, and $R_{90}/L$ is its physical concentration radius. This
radius is centred at a maximum site and is distinct from the
second-moment radius used in Section~\ref{sec:interior}.

For the spatial profiles only, we select modes satisfying
\begin{equation}
 \beta_2(\psi):=\sum_{n\in\partial_2\mathcal S_L}|\psi(n)|^2
                  \le0.2,
 \label{num:eq:boundary-filter}
\end{equation}
where $\partial_2\mathcal S_L$ consists of the two outermost layers,
as in \eqref{fc:eq:depth}. The upper profile uses the largest eigenvalue
among these modes. The interior profile uses the eigenvalue closest
to $0.45c_M$ among them. Every realisation supplies one mode to each
average. Condition~\eqref{num:eq:boundary-filter}
is a numerical selection rule. It does not identify the exceptional
family of Theorem~\ref{thm:cluster-transfer}.

\subsection{Ensemble averages}
Let $S=500$, and let $\widetilde\psi^{(s)}$ be the zero extension of
the selected mode in realisation $s$. The spatial plots display
\begin{equation}
 \overline\rho_L(k/L)=\frac{N}{S}\sum_{s=1}^S
  \left|\widetilde\psi^{(s)}(n_*^{(s)}+k)\right|^2,
 \qquad k\in\Z^2.
 \label{num:eq:mean-profile}
\end{equation}
Thus intensities are translated to their maximum sites before
averaging, with no rotation or periodic wrapping. Since localisation
centres depend on the realisation, this centred mean differs from
the intensity averaged at fixed physical sites. Its normalisation is
$L^{-2}\sum_k\overline\rho_L(k/L)=1$. The quantities displayed in
the panel titles are the means of the individual values of $P$ and
$R_{90}/L$, not the corresponding statistics of the mean profile.
Angle brackets in the figures denote these empirical means.

For spectral participation, all positive modes are retained, including
those failing \eqref{num:eq:boundary-filter}. At each ordered rank $j$
we plot
\begin{equation}
 \left(\frac{\overline E_j}{c_M},\frac{\overline P_j}{N}\right),
 \qquad
 \overline E_j=\frac1S\sum_{s=1}^SE_j^{(s)},\qquad
 \overline P_j=\frac1S\sum_{s=1}^SP(\psi_j^{(s)}).
 \label{num:eq:rank-average}
\end{equation}
These are averages at fixed rank, rather than conditional averages
at fixed energy. In frequency terms,
$\overline E_j/c_M=\overline{(\omega_j/\omega_M)^2}$;
its square root is a root-mean-square frequency ratio, not the mean
frequency ratio. The ribbons give $\pm1.96$ sample standard errors
of the vertical mean. 

\subsection{Concentration near the top of the spectrum}
\begin{figure}[p]
 \centering
 \includegraphics[width=.92\textwidth]{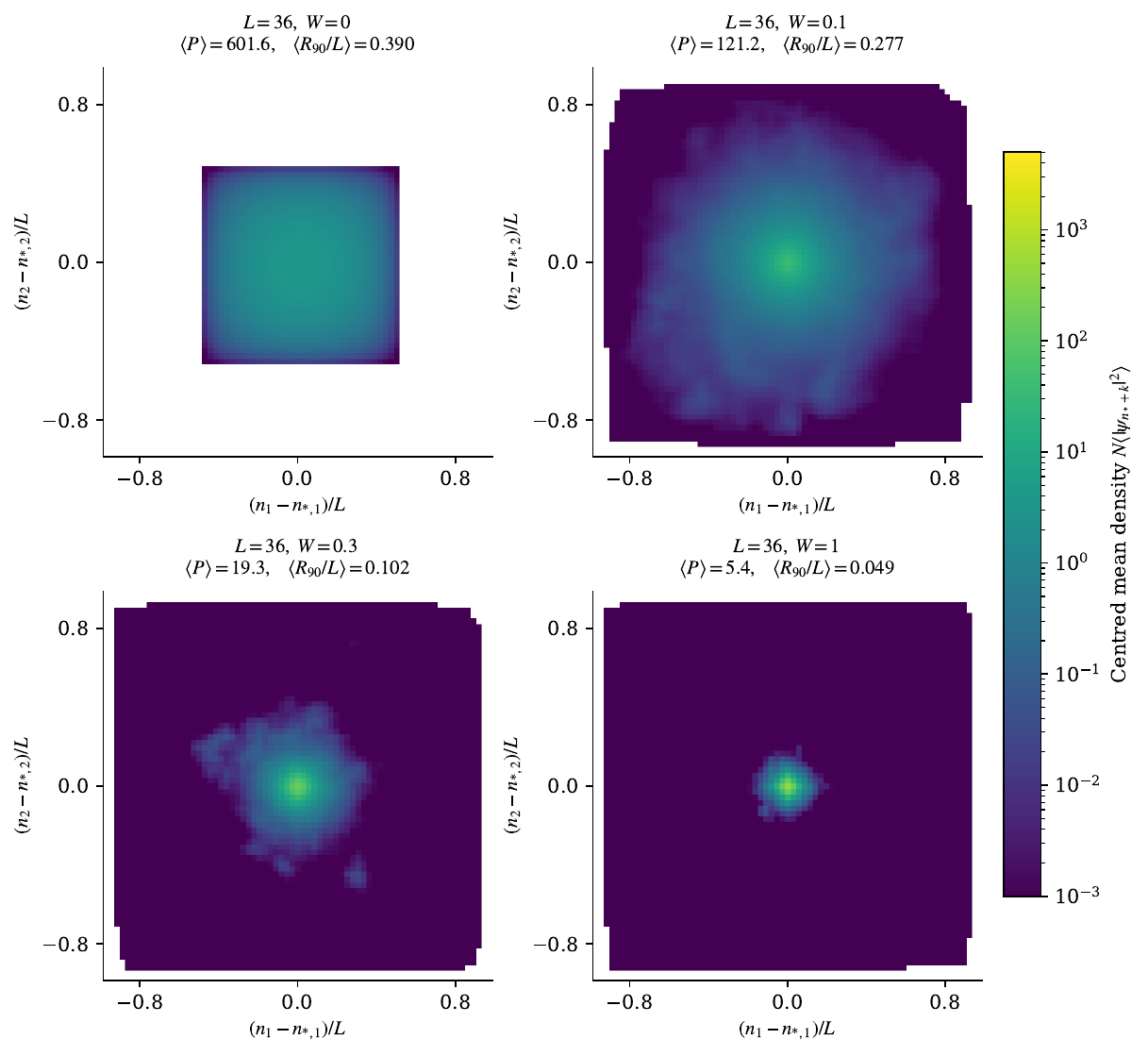}
 \caption{Selected high-energy profiles for $L=36$ ($N=1296$).
 Each panel uses the highest-energy mode satisfying
 \eqref{num:eq:boundary-filter}. Colours show the centred density
 \eqref{num:eq:mean-profile}; titles give the means of $P$ and
 $R_{90}/L$. Each positive-disorder panel averages all $500$
 realisations. The $W=0$ panel is the deterministic control.}
 \label{fig:num-upper36}
\end{figure}

\begin{figure}[p]
 \centering
 \includegraphics[width=.92\textwidth]{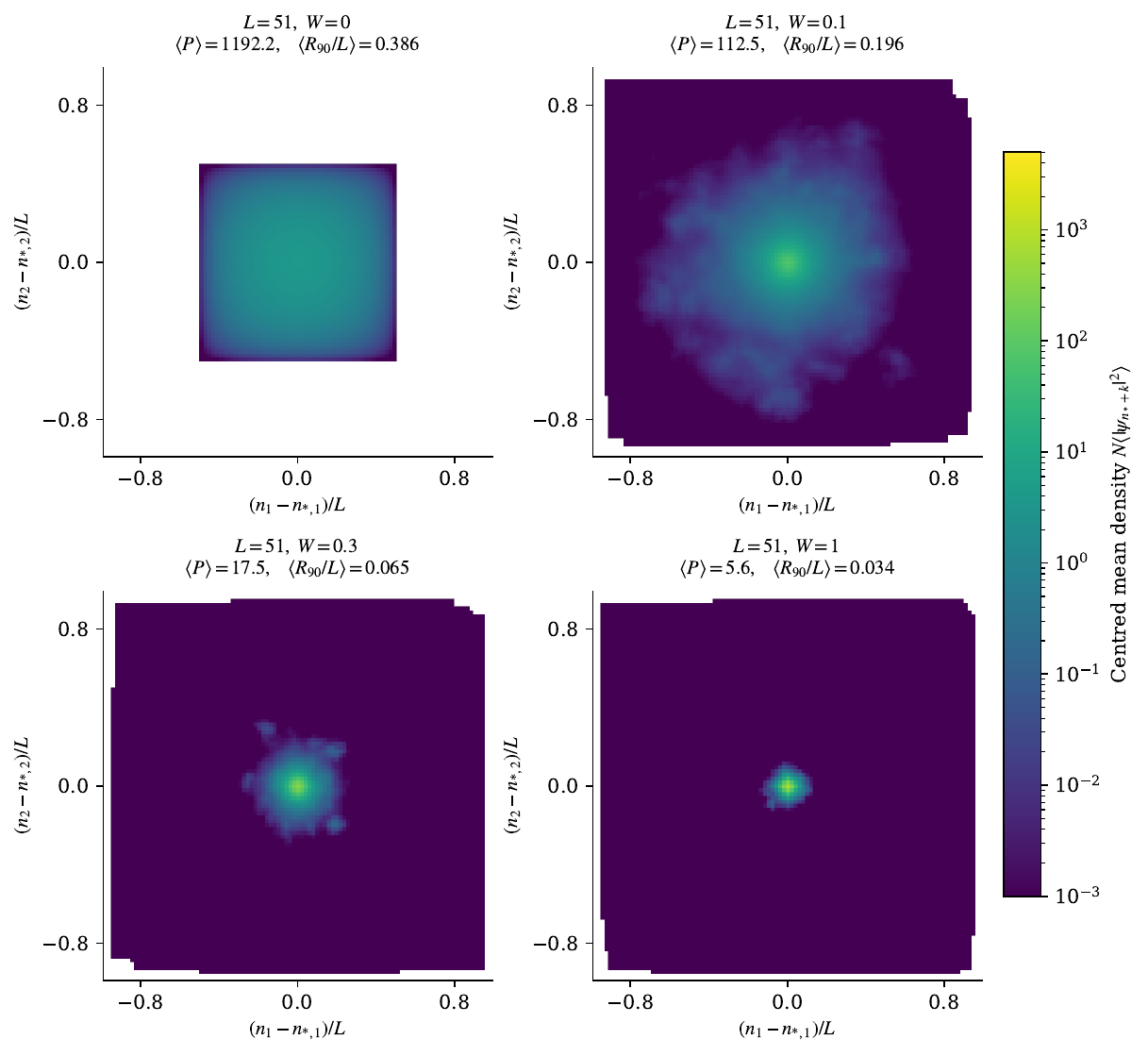}
 \caption{Selected high-energy profiles for $L=51$ ($N=2601$),
 with the same selection, normalisation, and averaging as in
 Figure~\ref{fig:num-upper36}. All $500$ realisations enter each
 positive-disorder panel.}
 \label{fig:num-upper51}
\end{figure}

\begin{figure}[p]
 \centering
 \includegraphics[width=.92\textwidth]{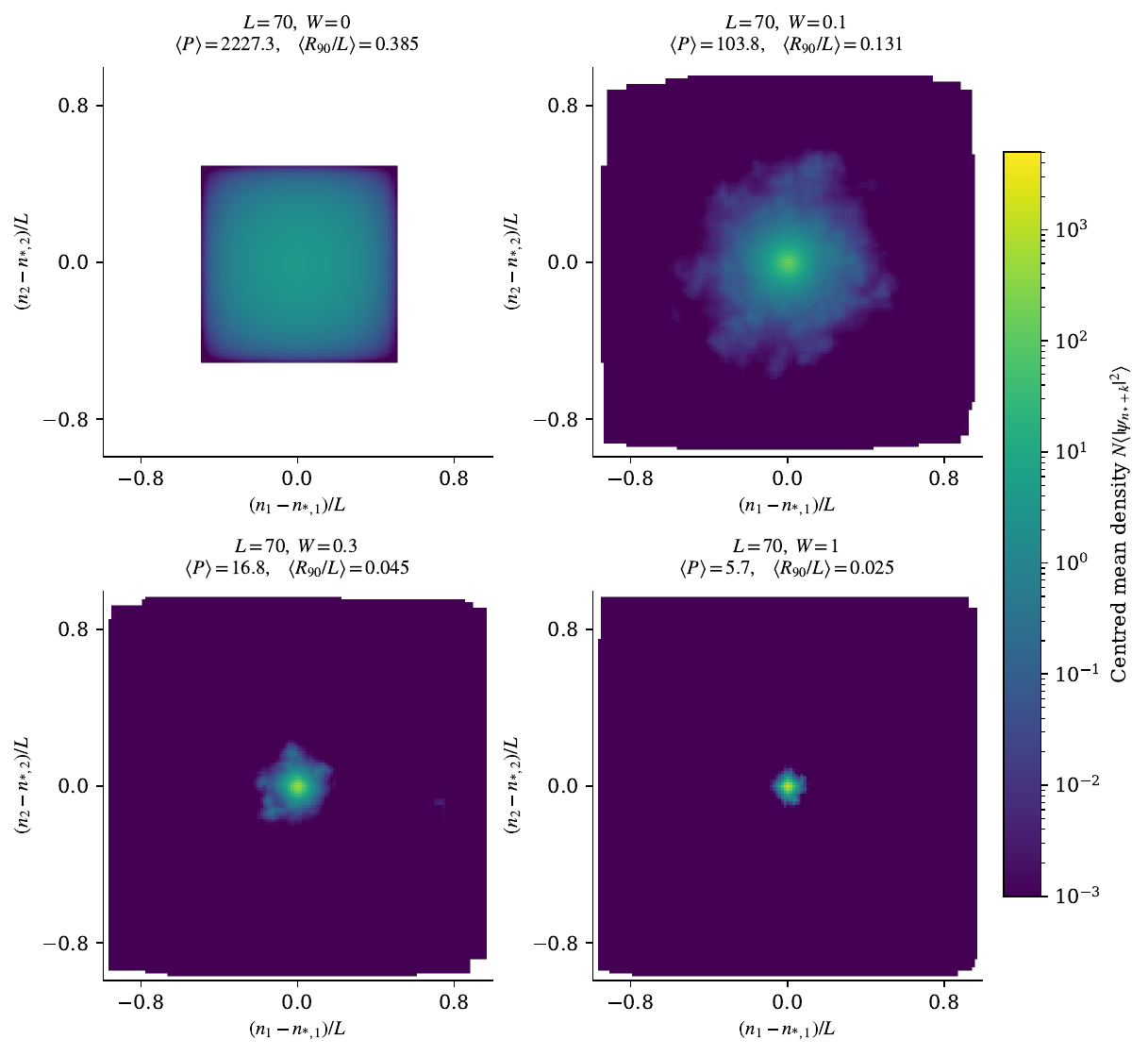}
 \caption{Selected high-energy profiles for $L=70$ ($N=4900$),
 with the same selection, normalisation, and averaging as in
 Figure~\ref{fig:num-upper36}. All $500$ realisations enter each
 positive-disorder panel. The physical displacement is $(n-n_*)/L$.}
 \label{fig:num-upper70}
\end{figure}

Figures~\ref{fig:num-upper36}--\ref{fig:num-upper70} show the upper
profiles for the three sizes. Table~\ref{tab:num-sizes} records the
associated statistics. Without disorder, the participation number is
approximately $0.46N$, and the physical radius remains close to
$0.39$. At fixed positive disorder, the participation numbers grow
much more slowly than $N$, and the physical radii decrease with $L$.
For example, at $W=0.1$, the mean participation number changes from
$121.2$ to $103.8$ between $L=36$ and $L=70$, while the mean physical
radius decreases from $0.277$ to $0.131$. At $W=1$, the selected
modes have participation numbers between $5.4$ and $5.7$ and physical
radii decreasing from $0.049$ to $0.025$. The centred profiles display
the same increasing concentration.

\begin{table}[htbp]
 \centering\small
 \setlength{\tabcolsep}{5pt}
 \begin{tabular}{@{}lcrrrrrr@{}}
 \toprule
 Selection & $W$ & \multicolumn{3}{c}{$\overline P$}
                    & \multicolumn{3}{c}{$\overline{R_{90}/L}$}\\
 \cmidrule(lr){3-5}\cmidrule(l){6-8}
 & & $L=36$ & $L=51$ & $L=70$ & $L=36$ & $L=51$ & $L=70$\\
 \midrule
 Upper & $0$   & $601.60$ & $1192.15$ & $2227.30$ & $0.390$ & $0.386$ & $0.385$\\
 Upper & $0.1$ & $121.23$ & $112.54$  & $103.76$  & $0.277$ & $0.196$ & $0.131$\\
 Upper & $0.3$ & $19.33$  & $17.50$   & $16.78$   & $0.102$ & $0.065$ & $0.045$\\
 Upper & $1$   & $5.40$   & $5.55$    & $5.70$    & $0.049$ & $0.034$ & $0.025$\\
 Interior & $4$ & $3.42$  & $3.43$    & $3.53$    & $0.061$ & $0.043$ & $0.032$\\
 \bottomrule
 \end{tabular}
 \vspace{2mm}
 \caption{Mean selected-mode statistics for the finite capacitance matrices. Each positive-disorder entry averages all $500$
 realisations; the first row is deterministic. The selection and
 observables are defined in \eqref{im:eq:measures},
 \eqref{num:eq:radius}, and \eqref{num:eq:boundary-filter}.}
 \label{tab:num-sizes}
\end{table}

\begin{figure}[p]
 \centering
 \includegraphics[width=\textwidth]{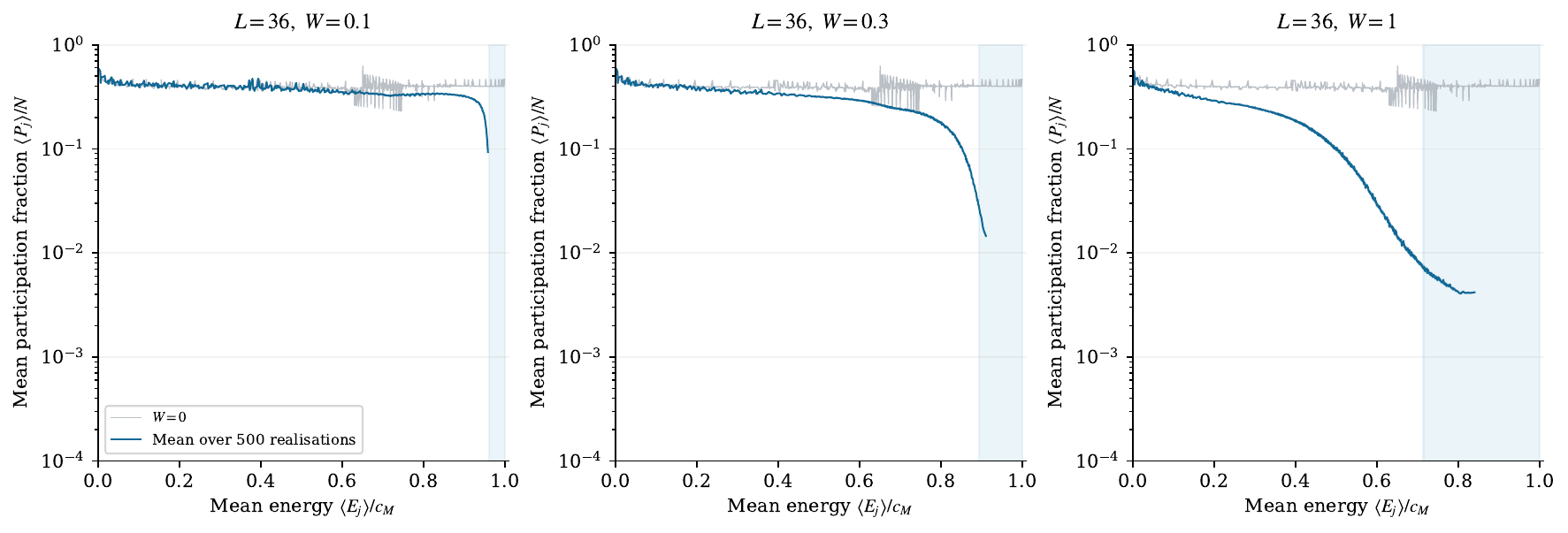}\par\medskip
 \includegraphics[width=\textwidth]{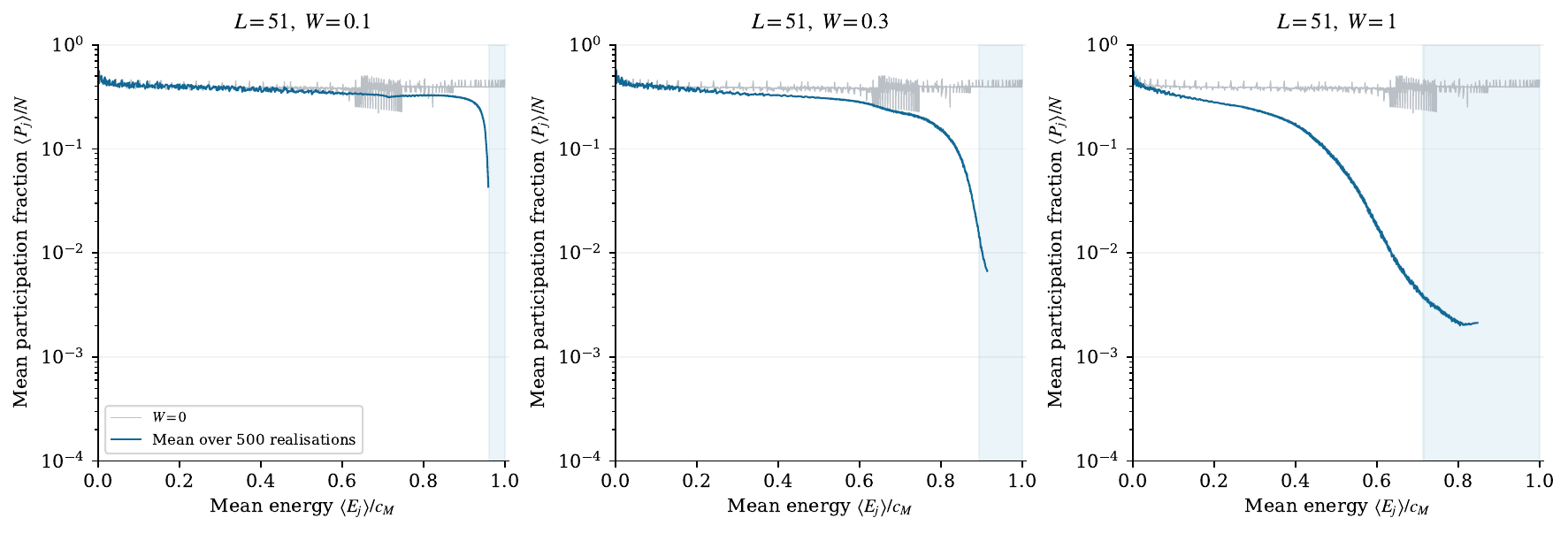}\par\medskip
 \includegraphics[width=\textwidth]{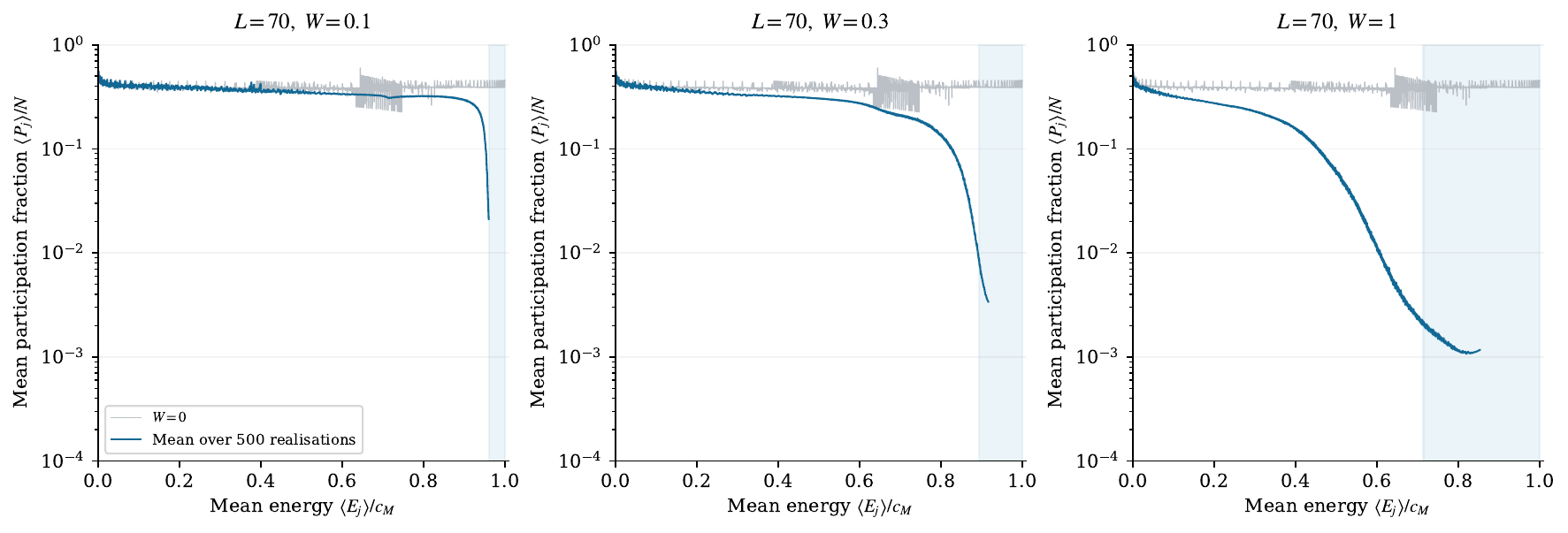}
 \caption{Spectral participation for $L=36,51,70$ (top to bottom)
 and $W=0.1,0.3,1$ (left to right). Blue curves are the rank averages
 \eqref{num:eq:rank-average} over all $500$ realisations, with
 $\pm1.96$ standard-error ribbons for the vertical mean. All positive
 modes are included. Grey curves give the periodic controls. Shading
 marks $((1+0.4W)^{-1},1)$ in $E/c_M$, a comparison window rather
 than a certified localisation interval. The corresponding interval
 for individual frequencies is $(1+0.4W)^{-1/2}<\omega/\omega_M<1$.}
 \label{fig:num-spectra}
\end{figure}

Figure~\ref{fig:num-spectra} complements the selected profiles by
showing spectral participation for all positive modes. The reduction
in participation is most pronounced in the upper part of the spectrum
and extends to lower normalised energies as $W$ increases. The blue
shading marks the comparison interval
$((1+0.4W)^{-1},1)$ on the $E/c_M$ axis. Its form is motivated by
Theorem~\ref{im:thm:app-linear-window}, but its use in the plot does not
verify that theorem's small-disorder threshold. In particular, the
selected upper modes need not lie in this interval. At $W=0.1$, only
$14$, $34$, and $58$ of the $500$ samples, respectively, contain a
positive eigenvalue in it. Thus these sample sizes resolve
concentration of modes near the top of the spectrum more readily than the sparsely
populated region closest to the infinite-volume edge.

\subsection{Interior modes at stronger disorder}
\begin{figure}[p]
 \centering
 \includegraphics[width=.97\textwidth]{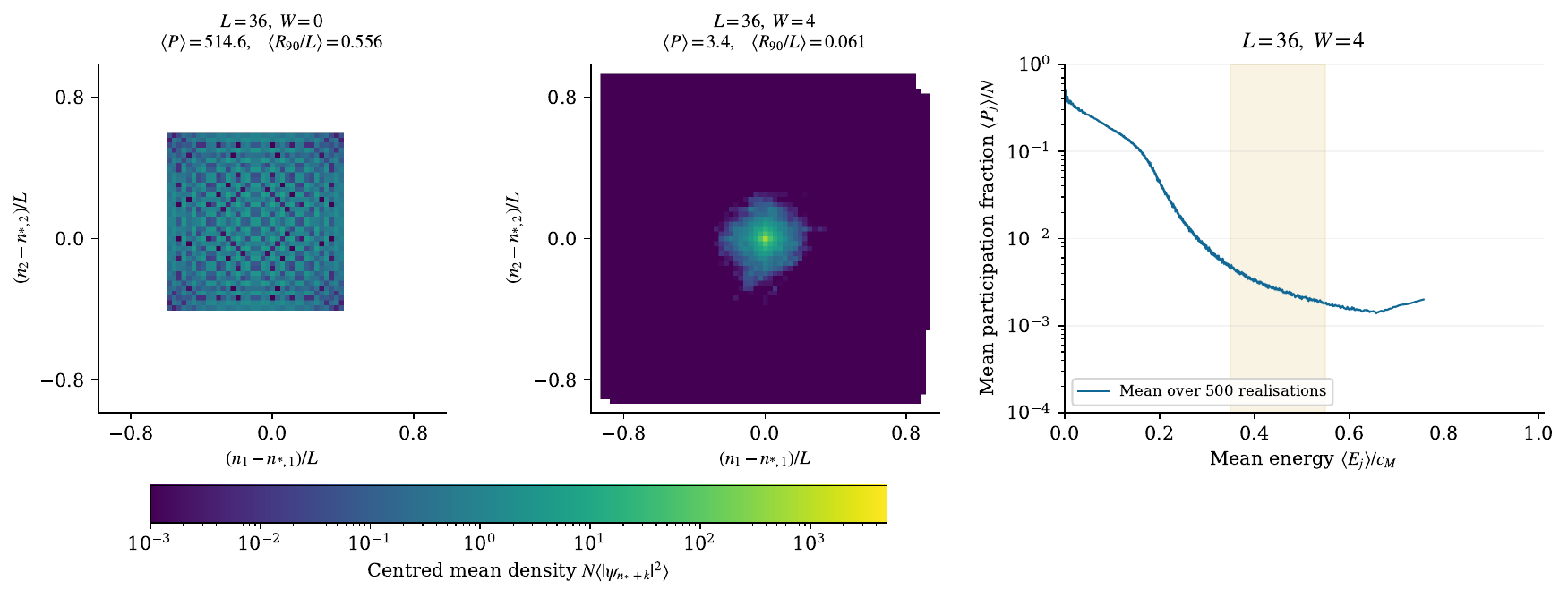}\par\smallskip
 \includegraphics[width=.97\textwidth]{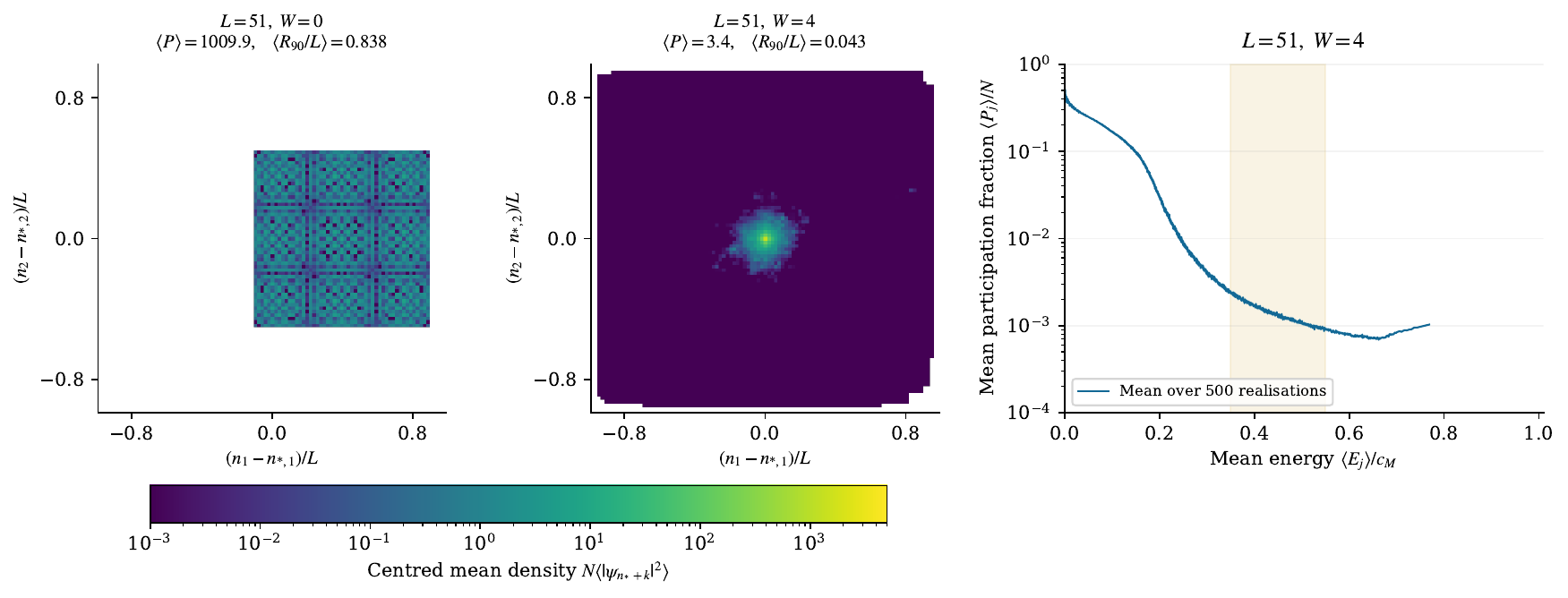}\par\smallskip
 \includegraphics[width=.97\textwidth]{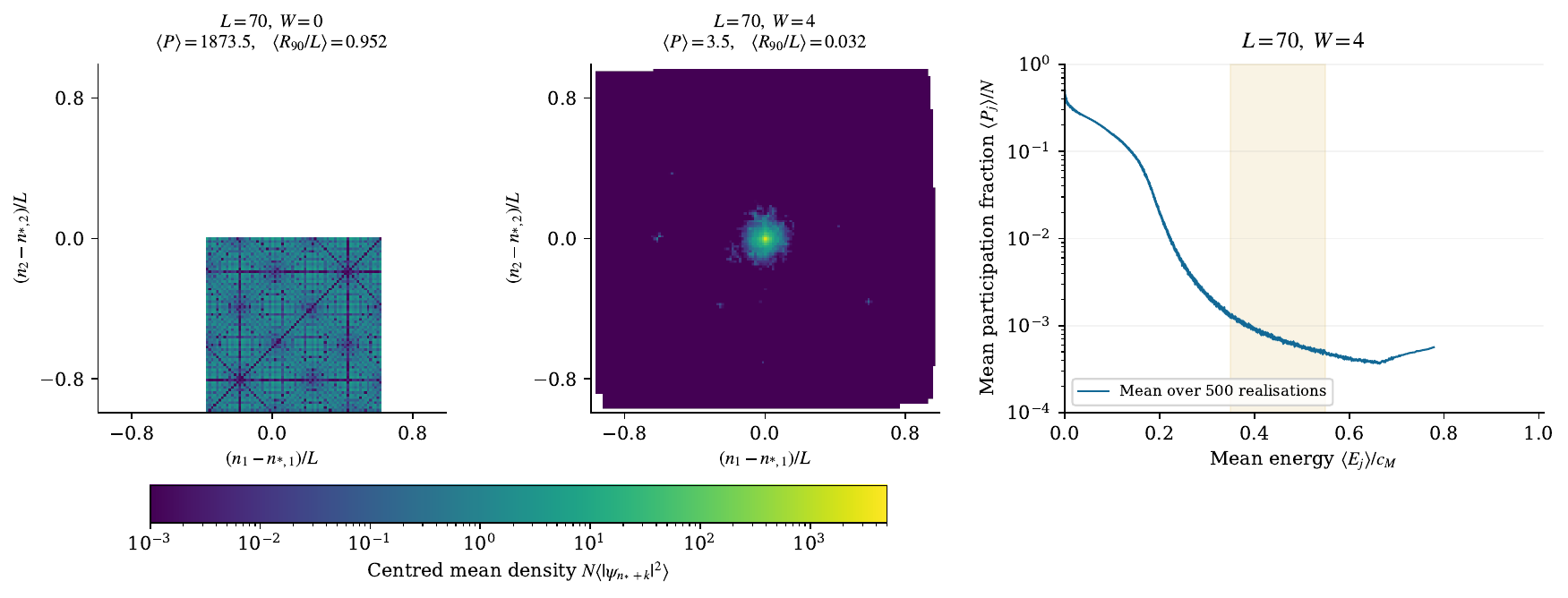}
 \caption{Interior modes for $L=36,51,70$ (top to bottom).
 The first two columns show the periodic control and the centred
 mean at $W=4$, selecting the mode closest to $0.45c_M$ subject to
 \eqref{num:eq:boundary-filter}. Each random profile averages all
 $500$ realisations. The right column gives rank-averaged
 participation for every positive mode at $W=4$, with
 $\pm1.96$ standard-error ribbons. Shading marks
 $0.35<E/c_M<0.55$. The profile normalisation is
 \eqref{num:eq:mean-profile}.}
 \label{fig:num-interior}
\end{figure}

Figure~\ref{fig:num-interior} compares the periodic interior profiles
with those at $W=4$, using the target energy $0.45c_M$. At this
disorder strength, the mean participation number remains between
$3.4$ and $3.6$ for all three sizes, while the mean physical radius
decreases from $0.061$ to $0.032$; see Table~\ref{tab:num-sizes}.
The spectral panels show small participation fractions throughout
the displayed comparison interval $0.35<E/c_M<0.55$.
These observations illustrate concentration in the interior of the
unperturbed band at stronger disorder, consistently with the
fixed-interval localisation mechanism. They do not evaluate the
sufficient disorder threshold in \eqref{eq:uniform-large}.
Here interior refers to the energy band $(0,c_M)$ of $C$. At $W=4$,
the target $E=0.45c_M$ has centred energy $s=1.35c_M$ and is
outside the regular-centred-energy regime of
Theorem~\ref{im:thm:eigenfunctions}.

Taken together, the figures support the finite-cluster concentration
described by Theorem~\ref{thm:cluster-main}. They do not test its
simultaneous exceptional-mode count or establish the asymptotic
$\varepsilon\log(1/\varepsilon)$ radius from three sizes. Nor do
participation numbers and $90\%$ radii alone certify exponential
tails. Large participation in the interior at weak disorder remains
compatible with localisation on scales exceeding the samples.
The calculations use a fixed Fourier discretisation of the static
matrix; they do not measure the error of the full acoustic reduction.

\FloatBarrier
\section{Concluding remarks}\label{sec:conclusions}
We have established spectral and dynamical Anderson localisation for
a random capacitance operator with long-range interactions. The proofs
retain the full exponentially decaying capacitance interaction.
Localisation holds near the upper band edge at every fixed positive
disorder strength and on prescribed positive intervals at sufficiently
large disorder. At weak disorder, the explicit edge window has energy
width asymptotic to $a c_MW$, and frequency width asymptotic to
$(a/2)\omega_MW$, for each fixed $a\in(0,1/2)$. It gives an upper bound on the upper endpoint of any absolutely continuous spectral component, if present.

For finite resonator clusters, uniform estimates for truncated matrices combine with the variational boundary comparison and
material-variable eigenvalue spacing. With probability tending to one as the sample size increases, the resulting matching proves concentration of individual modes of the finite matrix outside an exceptional
family of size bounded by a constant times the number of sites in a logarithmic boundary layer. Their physical
radii are $O(\varepsilon\log(1/\varepsilon))$, and the exceptional
fraction vanishes in every localised window of positive limiting
density of states. The estimates are uniform for the contrast-dependent
capacitance family at small contrast. Weighted asymptotic equivalence
also gives a common limiting eigenvalue distribution for the two
finite constructions.

The computations for $36^2$, $51^2$, and $70^2$ disks, with $500$
realisations at each positive disorder strength, show decreasing
participation fractions and physical concentration radii for selected
high-energy modes. At stronger disorder, similar concentration occurs
in an interior comparison interval. The periodic controls remain spread
through the sample. These observations illustrate the finite-cluster
results without determining asymptotic decay rates or a mobility edge.

The localisation properties of the unmatched modes, the interior spectral type at weak disorder, and the passage to the full transmission problem remain open in this analysis. The interior extent bounds quantify spatial scales but do
not prove delocalisation. A uniform capacitance reduction is needed to
transfer the finite-matrix conclusions to acoustic fields.

The frequency-dependent capacitance formulation of
\cite{ALLSU} suggests an extension beyond the subwavelength regime.
Let $\lambda_*>0$ be a simple Neumann eigenvalue of $-\Delta$ in
the reference resonator $D$. The corresponding reference frequency
is $\omega_*=v_r\sqrt{\lambda_*}$ at unit period, or
$v_r\sqrt{\lambda_*}/\varepsilon$ at physical period $\varepsilon$.
Near such frequencies, the regular quasiperiodic reduction in
\cite[Section~4]{ALLSU} describes the bands generated by the
interior Neumann mode. We expect localisation near an isolated
edge of these bands for every fixed positive disorder strength in
the reduced modal problem. A proof would require uniform exterior
non-resonance, exponential decay of the modal interactions, an
isolated nondegenerate edge, suitable spectral averaging, and
uniform control of the random reduction.

At unit period, the relevant spectral parameter is the detuning
$(\omega-\omega_*)/\delta$. The disorder scale must also preserve
the modal separation. For
the inverse material factors used here, one possible scaling is
$W=\delta W_{\rm mod}$, with any fixed $W_{\rm mod}>0$, so that
$m_n=1+\delta W_{\rm mod}\xi_n$. The decoupled frequencies
$\omega_*/\sqrt{m_n}$ then differ from $\omega_*$ by $O(\delta)$,
the scale of the regular modal band splitting. Arbitrary strength
in this prospective statement refers to $W_{\rm mod}$; unrestricted
unscaled material variations can mix distinct modal bands.
Localisation beyond the subwavelength regime therefore remains
a conditional prospect. For finite outgoing arrays, radiation losses
and comparison of the resulting resonance modes require additional
estimates.

\medskip
\noindent\textbf{Funding declaration.}
The work of W.\,J.\ is partially supported by the NSFC Grant
No.\,12571220 and by the New Cornerstone Investigator Program 100001127.

\medskip
\noindent\textbf{Conflict of interest.}
The authors declare no competing financial or non-financial interests
relevant to this work.

\medskip
\noindent\textbf{Declaration on the use of AI.}
During the preparation of this manuscript, the authors used ChatGPT
(OpenAI, GPT-Astra 6) as an interactive writing, editing, and analytical assistant.
The system was employed to explore mathematical ideas, assist in the
development and checking of proof arguments, improve the organisation
and exposition, and assist in preparing the LaTeX source, numerical
code, and illustrations. All mathematical results presented in this article originate either from the cited sources or from the authors' own work. The authors independently verified the mathematical arguments, references, numerical illustrations, and final formulations presented here. The authors assume full responsibility for the content and originality of this manuscript.

\appendix

\section{The three-dimensional weak-scattering scale}
\label{app:heuristic}\label{im:subsec:app-heuristic}
The mobility-edge predictions in this appendix are heuristic.
They do not enter the proof of Theorem~\ref{im:thm:app-linear-window}
or Corollary~\ref{im:cor:app-mobility}. We use the centred variables
$\mu_W=1+W/2$, $\lambda=W/\mu_W$, $s=\mu_WE$, and
$\sigma^2=\EE(\xi_n-1/2)^2=1/12$ from
Section~\ref{sec:interior}. The centred operator is
$\widetilde H_\varpi=\mu_WH_\varpi$ as in
\eqref{im:eq:tildeH}.

In dimension three, assume
for the explicit calculation that
$\mathsf A_M=a_M\Id$, $a_M>0$, as for a cubic geometry with
coordinate-permutation symmetry. Near the maximum,
\begin{equation}
 c(\alpha_M+\beta)=c_M-a_M|\beta|^2+O(|\beta|^4).
 \label{im:eq:app-parabola}
\end{equation}
Reciprocity and $\alpha_M=-\alpha_M$ modulo $2\pi\Z^d$ remove
the odd Taylor terms. Coarea in these local coordinates gives
the rigorous unperturbed asymptotic
\begin{equation}
 \rho_c(c_M-\Delta)
  =\frac{\sqrt\Delta}{4\pi^2a_M^{3/2}}(1+O(\Delta)),
 \qquad\Delta\downarrow0.
 \label{im:eq:app-edge-DOS}
\end{equation}
Indeed the quadratic level set has area $4\pi|\beta|^2$, gradient
modulus $2a_M|\beta|$, and $|\beta|=\sqrt{\Delta/a_M}$; division
by $(2\pi)^3$ gives the coefficient. Smooth even corrections
give the stated relative remainder. For an anisotropic positive
$\mathsf A_M$, the coefficient is instead
$(4\pi^2\sqrt{\det\mathsf A_M})^{-1}$.

The simplest weak-scattering description replaces the true tail
of the averaged spectrum by a coherent edge. In the centred
spectral variable it predicts
\begin{equation}
 s_{\rm coh}(\lambda)=c_M+O(\lambda^2),\qquad
 E_{\rm coh}(W)=\frac{s_{\rm coh}(\lambda)}{\mu_W}
                    =E_{\rm mean}(W)+O(W^2).
 \label{im:eq:app-coherent-shift}
\end{equation}
These formulas describe an approximation, not a proven asymptotic
for a spectral support endpoint. Almost surely, the upper edge is $c_M$ in
energy $E$, or $\mu_Wc_M$ in centred energy $s$, by
Proposition~\ref{prop:spectrum-new}.

To specify the Born coefficient, expansion of \eqref{im:eq:tildeH}
gives, uniformly in the material configuration,
\begin{equation}
 \widetilde H_\varpi=C+\lambda V_1+O_{\ell^2\to\ell^2}(\lambda^2),
 \qquad V_1=-\tfrac12(Z_\varpi C+CZ_\varpi).
 \label{eq:heuristic-first-order}
\end{equation}
On a periodic lattice with $N$ sites, let $C$ denote its periodisation
and let $\phi_\alpha(n)=N^{-1/2}e^{in\cdot\alpha}$ for the allowed
quasimomenta. Then $C\phi_\alpha=c(\alpha)\phi_\alpha$ and
\begin{equation}
 \EE\bigl|\ip{\phi_{\alpha'}}{V_1\phi_\alpha}\bigr|^2
 =\frac{\sigma^2}{4N}\bigl(c(\alpha)+c(\alpha')\bigr)^2.
 \label{eq:heuristic-variance}
\end{equation}
Indeed the matrix element of $Z_\varpi$ is
$N^{-1}\sum_n(\xi_n-1/2)e^{in\cdot(\alpha-\alpha')}$, whose
variance is $\sigma^2/N$ by independence. On the energy surface
$c(\alpha)=c(\alpha')=s$, the coefficient in
\eqref{eq:heuristic-variance} is $\sigma^2s^2/N$.
For the auxiliary evolution $e^{-it\widetilde H_\varpi}$, whose time
is conjugate to $s$, using the distributional identity
$t^{-1}|\int_0^t e^{iu\tau}\dd\tau|^2\to2\pi\delta_{\rm D}(u)$,
where $\delta_{\rm D}$ is the Dirac distribution, in the first
Duhamel term gives the golden-rule prediction
\begin{equation}
 \Gamma_{\rm Born}=2\pi\lambda^2\sigma^2s^2\rho_c(s),
 \qquad
 \rho_c(s)=\int_{\{c=s\}}\frac{\dd S(\alpha)}
                                  {(2\pi)^3|\nabla c(\alpha)|}.
 \label{eq:heuristic-golden-rule}
\end{equation}
Here $s$ is a regular energy, and the continuum of quasimomenta is
taken before the large-time limit. The fixed-time perturbation
expansion does not justify a microscopic kinetic limit.

Replacing the unperturbed density in \eqref{eq:heuristic-golden-rule}
by the parabolic density measured from the coherent edge, with
$\Delta_{\rm coh}=s_{\rm coh}-s>0$, gives the further approximation
\begin{equation}
 \Gamma_{\rm Born}
 \simeq 2\pi\lambda^2\sigma^2c_M^2
            \frac{\sqrt{\Delta_{\rm coh}}}{4\pi^2a_M^{3/2}},\qquad
 |\nabla c|\simeq2\sqrt{a_M\Delta_{\rm coh}}.
 \label{im:eq:app-Born-rate}
\end{equation}
Consequently
\begin{equation}
 \ell_{\rm mf}\simeq\frac{|\nabla c|}{\Gamma_{\rm Born}}
       =\frac{4\pi a_M^2}{\lambda^2\sigma^2c_M^2}.
 \label{im:eq:app-mean-free-path}
\end{equation}
This is a mean free path in cell units, not an eigenfunction decay
length. To convert this calculation to positive-frequency evolution,
use \eqref{eq:centred-frequency-conversion} and put
\[
 F(s)=\frac{v_r\sqrt\delta}{\varepsilon}\sqrt{\frac{s}{\mu_W}},
 \qquad F'(s)=\frac{v_r\sqrt\delta}
                    {2\varepsilon\sqrt{\mu_Ws}}.
\]
On a regular positive energy surface, the group velocity in cell
units is $F'(s)\nabla c$, and the first-order on-shell perturbation
matrix elements acquire the factor $F'(s)$. In the golden rule,
$\delta_{\rm D}(F(s')-F(s))=\delta_{\rm D}(s'-s)/F'(s)$,
so the collision rate is $F'(s)\Gamma_{\rm Born}$. The factor
cancels in their ratio. The physical mean free path is therefore
$\varepsilon\ell_{\rm mf}$. This change of variables does not
justify the kinetic approximation.
Extrapolating the heuristic Ioffe--Regel rule
$|\beta|\ell_{\rm mf}=\chi_{\rm IR}$, where $\chi_{\rm IR}>0$ is
an undetermined constant of order one, gives
\begin{equation}
 \Delta_{\rm IR}
 \simeq\frac{\chi_{\rm IR}^2\sigma^4c_M^4}
                  {16\pi^2a_M^3}\lambda^4,
 \qquad\sigma^4=(\sigma^2)^2.
 \label{im:eq:app-IR}
\end{equation}
At this scale the weak-scattering approximation is itself at its
limit of validity. Thus neither the coefficient nor the mobility
transition is proved by the extrapolation. The same distinction
between a quadratic coherent-edge shift and a fourth-order
mobility-edge separation appears in the self-consistent treatment
of a different three-dimensional Anderson model in
\cite[Section~IV, equation~(35)]{PZZO}.

Writing $s_{\rm mob}=\mu_WE_{\rm mob}^{\rm heur}$ for the predicted
transition in centred energy, these approximations give
\begin{equation}
 s_{\rm coh}-s_{\rm mob}\asymp\lambda^4,
 \qquad
 E_{\rm mob}^{\rm heur}(W)
       =\frac{c_M}{1+W/2}+O(W^2),
 \qquad
 c_M-E_{\rm mob}^{\rm heur}(W)
       =\tfrac12c_MW+O(W^2).
 \label{im:eq:app-mobility-heuristic}
\end{equation}
Equivalently,
$\omega_{\rm mob}^{\rm heur}(W)
=\omega_M(1-W/4+O(W^2))$. An $O(W^2)$ uncertainty in the
coherent edge is larger than the predicted $O(W^4)$ separation;
one cannot obtain a fourth-order expansion for the mobility edge
by retaining only the leading coherent-edge correction.

These formulas are specific to the inverse-material law
$M=\operatorname{diag}(1+W\xi_n)$ used here. Replacing it by a
centred law or by a direct material multiplier changes the mean
shift and the exact spectral endpoints. In two dimensions the
usual scalar, time-reversal-invariant scaling picture instead
predicts eventual localisation even at weak disorder, with very
large lengths \cite{AALR}. For the nearest-neighbour Anderson model with Gaussian disorder, time-averaged quantum diffusion on suitable weak-disorder scales and lower bounds on localisation length are established in \cite{BDH}.
These results do not predict a conducting bulk separated by a mobility edge of the form above. None of these physical expectations establishes the spectral type of our operator in the remaining interior band.

\section{Other material distributions}\label{sec:other-laws}
The main finite-volume theorem is stated for the uniform inverse-material
law. The following absence statement also covers direct uniform material
perturbations. The explicit coefficient $1/2$ in the quantitative
weak-disorder theorem is specific to the mean of the main material law.

\begin{proposition}\label{prop:other-laws}
Assume the geometry of Section \ref{sec:model}, including \eqref{eq:reflection}.
For $m_n=1+W\xi_n$ and every fixed $W>0$, absence of absolutely
continuous spectrum on a nonempty upper-edge interval remains valid if
the independent variables
$\xi_n$ have a common bounded density supported in $[0,1]$, zero
belongs to its support, and, for some $a_0>0$,
$q_0:=\PP\{\xi_n\ge a_0\}>0$.
All constants may depend on that law. Consequently, if instead the
direct factors $b_n$ are independent and uniform on $[1,1+w]$,
$w>0$, the operator $B_\varpi^{1/2} C B_\varpi^{1/2}$ has no
absolutely continuous spectrum on some nonempty interval
\[
 ((1+w)c_M-\widetilde\Delta_w,(1+w)c_M)
\]
almost surely, and its spectrum is $[0,(1+w)c_M]$ almost surely.
\end{proposition}
\begin{proof}
For the first assertion, modify the block estimate by using $X_n=\one_{\{\xi_n\ge a_0\}}$, and take the starting
scale so that $t_L\le g_-a_0$. The bound
\eqref{eq:block-lower} is unchanged. A low eigenvalue therefore
requires $\overline X_B<q_0/2$, once $h_L<t_Lq_0/3$.
For $N=|B|$ and $t_0=1/2$, the exponential Markov bound gives
\[
 \PP\{\textstyle\sum_{n\in B}X_n<q_0N/2\}
 \le e^{t_0 q_0N/2}(1-q_0+q_0e^{-t_0})^N
 \le e^{-q_0N/8}.
\]
Indeed $1-e^{-1/2}>3/8$ and $1-x\le e^{-x}$. Thus the same
stretched-exponential initial estimate holds with changed constants.
If the density is bounded by $R$, spectral averaging gives $NR|J_0|/g$
in \eqref{eq:Wegner} and $(2NR/g)^t/(1-t)$ in
\eqref{eq:norm-moment}. A bounded density with bounded support
satisfies the regularity assumptions of the same fractional-moment
criterion, with constants depending on the law. The all-subsets block argument gives the analogue of
\eqref{eq:all-restrictions-FM}. At fixed $E$ in a sufficiently small
upper-edge interval, Lemma~\ref{lem:range} gives
$\delta_y\in\operatorname{Ran}(EM_\varpi-JCJ)$ almost surely.
Since
\[
 JH_\varpi J-E
 =-M_\varpi^{-1/2}(EM_\varpi-JCJ)M_\varpi^{-1/2},
\]
also $\delta_y\in\operatorname{Ran}(JH_\varpi J-E)$.
For a bounded self-adjoint $A$, if $\delta_y=(A-E)u$, then
$\|(A-E-i\eta)^{-1}\delta_y\|\le\|u\|$ and
$\operatorname{Im}\langle\delta_y,(A-E-i\eta)^{-1}\delta_y\rangle
\le\eta\|u\|^2\to0$. Solvability of \((A-E)u=\delta_y\) in \(\ell^2(\mathbb Z^d)\) is a measurable condition, since it is equivalent to
\[
\sup_{k\geq1}\bigl\|(A-E-i/k)^{-1}\delta_y\bigr\|_2<\infty.
\]
Fubini and the Poisson boundary formula, as in
Theorem~\ref{thm:noac}, give the absence assertion.
Finally, zero in the support ensures positive probability of the
favourable events used in Proposition~\ref{prop:spectrum-new}.

For the direct uniform law, put $b=1+w$ and define
\[
 \eta_n=\frac{b/b_n-1}{w},\qquad M'_\varpi=\diag(1+w\eta_n).
\]
These variables are iid on $[0,1]$, with density
\[
 \rho_\eta(t)=\frac{1+w}{(1+wt)^2}\one_{[0,1]}(t).
\]
It is bounded, has zero in its support, and gives
$\PP\{\eta_n\ge1/2\}=1/(2+w)>0$.
Since $B_\varpi=b(M'_\varpi)^{-1}$,
\[
 B_\varpi^{1/2} C B_\varpi^{1/2}
 =b(M'_\varpi)^{-1/2} C(M'_\varpi)^{-1/2}.
\]
Apply the first assertion with disorder width $w$, and multiply
the spectral energy by $b$. This proves the second assertion.
\end{proof}

\begingroup
\small\raggedright
\let\savedthebibliography\thebibliography
\renewcommand{\thebibliography}[1]{%
  \savedthebibliography{#1}\setlength{\itemsep}{0.5pt}\setlength{\parskip}{0pt}}
\bibliographystyle{amsplain}
\bibliography{Random_Resonator_Crystals}
\endgroup
\end{document}